\documentclass[11pt,titlepage,a4paper]{myarticle}

\usepackage[T1]{fontenc} 
\usepackage{comment} 

\usepackage{amsmath, mathrsfs, amsfonts, amssymb, amsthm, mathtools, graphicx, color, ucs, xparse, tikz, lmodern, physics, varioref, tensor, bm}
\usepackage[nosort]{cite}
\usepackage{youngtab}
\usepackage{float,multirow}
\newcommand{\ybar}[1]{\overline{\vphantom{1}#1}}

\newcommand{\centerresize}[2]{%
  \makebox[\textwidth][c]{\resizebox{#1\textwidth}{!}{#2}}}

\newcommand{\R}{\mathbb{R}}

\newcommand{\Z}{\mathbb{Z}}

\newcommand\mperiod[1][\rlap]{#1{\,.}}	
\newcommand\mcomma[1][\rlap]{#1{\,,}}

\newenvironment{eq}
    {\begin{equation}
    \begin{aligned}
    }
    { 
    \end{aligned}
    \end{equation}
    \ignorespacesafterend
    }

\usepackage{pifont}
\newcommand{\cmark}{{$\bullet$}}%
\newcommand{\xmark}{{\ding{53}}}%

\RequirePackage[colorlinks=true
,urlcolor=blue
,anchorcolor=blue
,citecolor=blue
,filecolor=blue
,linkcolor=blue
,menucolor=blue
,linktocpage=true
,pdfproducer=medialab
,pdfa=true
]{hyperref}

\usepackage{cleveref}
\usepackage{tabularx}

\begin{document} 

\preprint{{\tt IFT-UAM/CSIC-26-73}\\}

\title{Non-supersymmetric dualities beyond the gauge algebra}
\author{Bernardo Fraiman${}^{1}$, Vittorio Larotonda${}^{3,4}$, Ling Lin${}^{3,4}$, Salvatore Raucci${}^{1,2}$
     \oneaddress{
     ${}^1$Instituto de F\'{i}sica Te\'{o}rica IFT-UAM/CSIC, C/ Nicol\'{a}s Cabrera 13--15, Campus de Cantoblanco, 28049 Madrid, Spain \\
     ${}^2$Departamento de F\'{i}sica Te\'{o}rica, Universidad Aut\'{o}noma de Madrid, Cantoblanco, 28049 Madrid, Spain\\ 
     ${}^3$Dipartimento di Fisica e Astronomia, Universit\`{a} di Bologna, via Irnerio 46, Bologna, Italy\\
     ${}^4$INFN, Sezione di Bologna, viale Berti Pichat 6/2, Bologna, Italy\\
      {~}
}}

\Abstract{We provide a testing ground for dualities of non-supersymmetric type 0 orientifolds by finding the global forms of their gauge groups. As an example, we consider the Bergman--Gaberdiel proposal for a duality of theories with $\mathfrak{so}(32) \oplus \mathfrak{so}(32)$ gauge algebra. On the orientifold side, we perform a scan of brane states, similar to the case of the D0-brane in type I string theory, and identify spinorial states that constrain the gauge group.
Comparing with the bosonic string side, where the gauge group is determined by the internal momentum lattice as in heterotic strings, our analysis shows that the gauge groups match, hence supporting the duality beyond the perturbative string spectrum. 
}

\maketitle
\setcounter{page}{1}

\setcounter{tocdepth}{2}
\tableofcontents

\section{Introduction}

String dualities are a window into the physics beyond perturbation theory and are one of the most compelling pieces of evidence for the quantum-gravitational consistency of string theory.
This non-perturbative knowledge of string theories comes at a cost: dualities are well understood only in very controlled setups, where various amounts of spacetime supersymmetries protect the physics. For instance, in ten dimensions, string dualities unify all spacetime-supersymmetric string theories in the framework of M-theory, but the story is less clear for non-supersymmetric strings.\footnote{See~\cite{Mourad:2017rrl,Basile:2021vxh,Angelantonj:2024tns,Raucci:2024fnp,Leone:2025mwo,Dudas:2025ubq} for recent reviews on non-supersymmetric string theories.} There have been several attempts to shed light on the matter, and there are some proposed dualities that involve these theories~\cite{Bergman:1997rf,Blum:1997cs,Blum:1997gw,Bergman:1999km,Blumenhagen:1999ad,Angelantonj:2007ts,Faraggi:2007tj,Acharya:2022shu,Larotonda:2024thv,Fraiman:2025yrx,Baykara:2026gem,Altavista:2026evd,Baykara:2026vdc,Altavista:2026brr,Dasgupta:2026maq,Basile:2026trt,Kamal:2026msr}. In this work, we focus on the Bergman--Gaberdiel proposal~\cite{Bergman:1997rf} for an explicit example.

Specifically, we test the proposed duality by comparing the gauge group, not merely the gauge algebra, on both sides of the duality.
This is motivated by the strong-weak duality between the type I and heterotic theories with $\mathfrak{so}(32)$ gauge symmetry \cite{Dabholkar:1995ep, Hull:1995nu, Polchinski:1995df}, which historically was also (in part) established by matching the gauge groups.
A crucial role in this matching is occupied by non-perturbative states: while the perturbative group in type I is $O(32)/\Z_2$~\cite{Witten:1998cd}, it becomes $Spin(32)/\Z_2$ once D-branes are included, hence matching the heterotic gauge group \cite{Sen:1998tt, Sen:1998ki, Sen:1999mg, Witten:2023snr, Larotonda:2024thv}.

A natural question is whether a matching gauge group is a general feature of string dualities. 
In field-theoretic strong-weak dualities, the gauge group is not necessarily duality-invariant.
For example, the Montonen--Olive duality~\cite{Goddard:1976qe,Montonen:1977sn,Osborn:1979tq} of four-dimensional $\mathcal{N}=4$ gauge theories is an electric-magnetic duality that exchanges a given gauge group at coupling $g\to 0$ with the GNO-dual group at $g\to\infty$. It exchanges W bosons with magnetic monopoles. On the other hand, string dualities exchange perturbative string descriptions: in the dual frame, a solitonic string becomes weakly-coupled with the same worldsheet description. 
Since the gauge sector is part of the spectrum of the weakly-coupled ``fundamental'' string, we expect, in a string duality, the spacetime gauge symmetry to remain the same. 
In particular, this should hold at the level of its global topology---with or without spacetime supersymmetry.

To verify this for the Bergman--Gaberdiel duality \cite{Bergman:1997rf} between the type 0B$/\Omega$ orientifold and a particular $T^{16}$ compactification of the bosonic closed string, we follow a similar approach as in the type I-heterotic duality and compute the spacetime gauge excitations by quantizing the open-string spectrum on the stable D-branes of type 0 orientifold theories, following the procedure in \cite{Larotonda:2024thv}.
When comparing the gauge group of type 0B$/\Omega$ with that of the bosonic string, which is determined by a momentum lattice as in heterotic strings, we find agreement between the global structures of the gauge groups.

This paper is organized as follows. In section~\ref{sec:particle_vs_string}, we review the type I-heterotic duality, motivating how non-perturbative D-brane states are necessary to match the gauge group in a strong-weak string duality.
In section~\ref{sec:nonsusy_dualities}, we briefly review the type 0 orientifolds and discuss two cases in which we determine the gauge groups via brane quantization.
In section~\ref{sec:BG_group}, we use the result for type 0B$/\Omega$ to test the purported duality with the bosonic string.
Section~\ref{sec:discussion} contains some concluding thoughts. We include three appendices: in appendices~\ref{app:0A_branes} and~\ref{app:0B_branes}, we discuss the branes of the type 0A and 0B orientifolds that we use in the main text, also correcting some typos present in the literature for the type 0B$/\Omega$ case; in appendix~\ref{app:bosonic_on_T16}, we include details on the spectrum of bosonic string theory on the special $T^{16}$ that is used in the Bergman--Gaberdiel duality.

\section{Gauge group under type I/heterotic duality}\label{sec:particle_vs_string}

The arguably best understood instance of strong-weak string duality is between type I and the heterotic $Spin(32)/\Z_2$ theory.
In this section, we review this supersymmetric duality along the lines of~\cite{Witten:1998cd,Witten:2023snr}, as our analysis for the non-supersymmetric theories in section~\ref{sec:nonsusy_dualities} will mirror the supersymmetric case.

Type I string theory is an orientifold projection~\cite{Sagnotti:1987tw,Pradisi:1988xd,Horava:1989vt,Bianchi:1990yu,Bianchi:1990tb,Bianchi:1991eu} of type IIB string theory by worldsheet parity $\Omega$ (preserving spacetime SUSY), whose tadpole cancellation condition requires the presence of spacetime-filling D9-branes that host the $\mathfrak{so}(32)$ gauge symmetry.
The excitations of the perturbative open string carry Chan--Paton charges at the two endpoints on the D9s, each transforming in vector representations of $\mathfrak{so}(32)$.
For the 10d gauge symmetry, these give rise to irreducible representations that appear in the tensor product decomposition of even powers of the vector representation of $\mathfrak{so}(32)$.
In particular, this yields representations that transform trivially under the center $\Z_2 \times \Z_2$ of the universal covering group $Spin(32)$.
Hence, the spacetime spectrum arising from the perturbative type I string is consistent with the gauge group $O(32)/\Z_2$\footnote{Open strings have two endpoints, and the generator $-1$ of $O(32)$ acts trivially on all states.}. 

However, type I string theory also has non-perturbative states: D-branes. 
Their dynamical excitations can carry representations that break the center of the gauge group, thus affecting the global structure differently than the weakly-coupled fundamental string.
In fact, a non-BPS stable D0-brane in type I provides the states living in the spinorial representation (but \emph{not} the co-spinors) of $Spin(32)$, which are precisely the states required to establish the duality to the heterotic theory~\cite{Sen:1998tt,Sen:1998ki,Witten:1998cd,Sen:1999mg}.

The argument is as follows. In the D0-D9 system, there are massless Majorana fermions originating from the Ramond sector of the open string, propagating on the D0-brane worldvolume. Taking the Chan--Paton factors into account, there are 32 such zero modes $\chi_i,\, i=1,\dots,32$, transforming in the vector of the $\mathfrak{so}(32)$ spacetime gauge algebra, whose action is
\begin{equation}
    \frac{i}{2}\sum_{i=1}^{32} \int dt \chi_i \frac{d}{dt}\chi_i\,.
\end{equation}
The canonical quantization of these modes, 
\begin{equation}
    \{\chi_i,\chi_j\}=\delta_{ij}\,,
\end{equation}
corresponds to the Clifford algebra over the vector of $\mathfrak{so}(32)$ and leads to a Fock space whose lowest level contains $2^{16}$ states which decompose into spinor and co-spinor of the gauge algebra.\footnote{More generally, the Hilbert space decomposes into representations of the Clifford algebra, which in turn decompose into (irreducible components of tensor products of) the spinor and co-spinor representation of the underlying $\mathfrak{so}(32)$ Lie algebra.}
Either of the spinors already break the $O(1) \subset O(32)$ of the spacetime gauge symmetry.
Crucially, a single D0-brane carries a \emph{worldvolume} $O(1) \cong \Z_2$ gauge symmetry.
Its action on the D0 Hilbert space is generated by the chirality matrix $\bar{\chi}= \prod_{i=1}^{32} \chi_i$, which acts as one of the spinorial $\Z_2$ of the center of $Spin(32)$.
By convention, this $\Z_2$ acts non-trivially on the co-spinor representations.
Therefore, imposing gauge invariance on the D0 projects out the co-spinor, and the physical Hilbert space sees a spacetime $Spin(32)/\Z_2$ gauge group, where the quotiented $\Z_2$ is (by convention) the co-spinorial factor of the center of $Spin(32)$.

D0-branes naturally appear as particles in spacetime, but the excitations of other non-perturbative objects can also carry non-trivial representations under the spacetime $\mathfrak{so}(32)$ symmetry, which a priori can be different than those from the D0s.
In~\cite{Larotonda:2024thv}, the D0-analysis above was generalized to the cases of D1- and D5-branes in type I string theory, which are the other non-tachyonic branes allowed by K-theory~\cite{Witten:1998cd} (in type I, these happen to be also BPS branes).
It turns out that their excitations are consistent with the group structure $Spin(32)/\Z_2$ dictated by the D0s; in particular, the arguments for a single D1-brane are identical to the D0-discussion above, because the worldsheet gauge symmetry ($O(1) \cong \Z_2$) and fermion representations under the spacetime $\mathfrak{so}(32)$ symmetry are identical.\footnote{More generally, a stack of D1-/D5-branes host $\mathfrak{so}(n)$ and $\mathfrak{sp}(n)$ worldvolume gauge algebras respectively.
Inferring the representations under the center of the spacetime symmetry from the quantization is more subtle~\cite{Larotonda:2024thv}.
We will encounter the analogous arguments in section~\ref{sec:0A_D4_quantization} for D4-branes in the non-supersymmetric type 0A$/\Omega$ theory.}
While not adding new information about the gauge spectrum in comparison to the D0s,\footnote{In section~\ref{sec:type0AOmega} we will see that, in general, different non-tachyonic branes can give rise to different gauge representations, hence it is crucial to take into account all of them to deduce the correct gauge groups for non-supersymmetric models.} the D1-brane is crucial for the string duality, in that it precisely is the object that is non-perturbative at weak string coupling, but at strong coupling becomes a weakly-coupled string identical to the fundamental heterotic string \cite{Dabholkar:1995ep,Hull:1995nu,Polchinski:1995df}.

In the duality frame of the latter, the global form of the gauge group is actually fixed by the consistency of its perturbative worldsheet formulation. In the bosonic formulation, the internal momentum lattice must be even self-dual, which at rank 16 is either the root lattice of $\mathfrak{e}_8 \oplus \mathfrak{e}_8$ or the unique (up to lattice isomorphisms) even self-dual extension of the $\mathfrak{so}(32)$-root lattice $D_{16}$,
\begin{eq}
    \Gamma_{16}=D_{16}^+ \mcomma
\end{eq}
which is the (co-)character lattice of the group $Spin(32)/\Z_2$.
Given our motivations, we shall focus on the latter.

Explicitly, the root lattice of $\mathfrak{so}(32)$ (which is the same as its coroot lattice) is 
\begin{eq}
    D_{16}=\left\{(x_1,\dots x_{16})\in \Z^{16} \ \middle| \  \sum_i x_i = 0 \quad (\text{mod }2)\right\}\mperiod
\end{eq}
Let $D_{16}^*$ be the dual lattice (i.e., elements of $\R^{16}$ such that their inner product with any element of $D_{16}$ is an integer), which is also the (co-)weight lattice of $\mathfrak{so}(32)$.
The quotient $D_{16}^*/D_{16} \cong \Z_2 \times \Z_2$ decomposes $D_{16}^*$ into four cosets, denoted by the (equivalence class of the) shift vectors $\vec{o}$, $\vec{v}$, $\vec{s}$, and $\vec{c}$, given by
\begin{eq}\label{eq:equivalence_classes_D16*}
    \vec{o}=(0,\dots,0) \mcomma \qquad \vec{v}=(1,0,\dots,0) \mcomma\qquad \vec{s}= \left(\frac{1}{2},\dots \frac{1}{2}\right)\mcomma \qquad \vec{c}= \left(\frac{1}{2},\dots \frac{1}{2},-\frac{1}{2}\right) \mperiod
\end{eq}
In terms of the $\mathfrak{so}(32)$ representations, these vectors are weight vectors of the trivial representation ${\bf 1} \equiv {\bf O}$, the vector representation ${\bf V}$, the spinor representation ${\bf S}$, and the cospinor representation ${\bf C}$.
The root lattice is associated with $\vec{o}$, and the other three each define a different self-dual character lattice,
\begin{eq}
D_{16} \cup (D_{16}+\vec{v})\,,\quad 
D_{16} \cup (D_{16}+\vec{s})\,,\quad
D_{16} \cup (D_{16}+\vec{c})\,,
\end{eq}
corresponding to the group topologies $SO(32)$, $Spin(32)/\Z_2^{(s)}$ and $Spin(32)/\Z_2^{(c)}$, respectively, where $\Z_2^{(s/c)}$ are the spinorial factors of the center.
Among these three self-dual lattices, only the ones corresponding to $\vec{s}$ and $\vec{c}$ are even.
As they are isomorphic, we are free to pick, say, the one corresponding to $\vec{s}$:
\begin{align}
    D_{16}^+ := D_{16} \cup (D_{16} + \vec{s}) \, .
\end{align}

Since vectors in the momentum lattice correspond to the weights of the spacetime excitations under the $\mathfrak{so}(32)$ gauge symmetry, they all fall into the cosets of $\vec{o}$ or $\vec{s}$. Therefore, the gauge group is the quotient of $Spin(32)$ by the $\Z_2$ that acts trivially on both $\vec{o}$ and $\vec{s}$, or in terms of representations, leaves the representations ${\bf O}$ and ${\bf S}$ (and tensor products thereof) invariant.
This precisely matches the structure of type I string theory.

\section{The gauge group of type 0 orientifolds}\label{sec:nonsusy_dualities}

We want to extend this line of arguments for establishing strong-weak dualities to the proposed non-supersymmetric duality between type 0B$/\Omega$ and the bosonic closed string compactified on $T^{16}$.
To this end, we determine in this section the gauge group of the orientifold theory from the brane spectra.
Because of their similarities, we also perform the brane scan for the type 0A$/\Omega$ orientifold for completeness.

Before presenting our results, we briefly review the orientifold construction~\cite{Sagnotti:1995ga,Sagnotti:1996qj} for these theories; see also~\cite{Angelantonj:2002ct}.

\subsection{Type 0 orientifolds}

In light-cone gauge, type 0 theories are two-dimensional $(1,1)$ theories with $8$ bosons and $8$ Majorana fermions. As in type II theories, there are two variants depending on the GSO projection: type 0A and type 0B. Their one-loop partition functions are
\begin{eq}\label{eq:T_0AB}
    \mathcal{T}_{0A}&=\int_{\mathcal{F}}\frac{d^2\tau}{\tau_2^2}\frac{O_8\bar{O}_8+V_8\bar{V}_8+S_8\bar{C}_8+C_8\bar{S}_8}{\tau_2^4\eta^8\bar{\eta}^8}\mcomma \\
    \mathcal{T}_{0B}&=\int_{\mathcal{F}}\frac{d^2\tau}{\tau_2^2}\frac{O_8\bar{O}_8+V_8\bar{V}_8+S_8\bar{S}_8+C_8\bar{C}_8}{\tau_2^4\eta^8\bar{\eta}^8}\mcomma
\end{eq}
where we used the expression in terms of $\mathfrak{so}(8)$ characters as in~\cite{Angelantonj:2002ct}.
The NS-NS spectrum contains a tachyon from the NS ground state on the left and on the right, and then a graviton, a 2-form field, and a dilaton from the $\mathbf{8}_v\otimes \mathbf{8}_v$ sector. The R-R spectra of type 0A and 0B are two copies of the R-R spectra of type IIA and IIB. There are no NS-R sectors.

Type 0 strings have orientifold descendants~\cite{Bianchi:1990yu}; in particular, there are four consistent ten-dimensional orientifolds, some of which will be the focus of this work. One of these is an orientifold of type 0A string theory, and the other three are orientifolds of type 0B string theory. The various alternatives originate from the existence of non-trivial automorphisms in the parent theory that generate signs in the Klein bottle amplitude~\cite{Fioravanti:1993hf,Pradisi:1995pp}. In particular, for the type 0A orientifold,
\begin{eq}\label{eq:K_0A}
    \mathcal{K}=\frac{1}{2}(O_8+V_8)\mcomma
\end{eq}
while for the three type 0B orientifolds,
\begin{eq}\label{eq:K_0B}
    \mathcal{K}_1&=\frac{1}{2}(O_8+V_8-S_8-C_8)\mcomma \\
    \mathcal{K}_2&=\frac{1}{2}(O_8+V_8+S_8+C_8)\mcomma \\
    \mathcal{K}_3&=\frac{1}{2}(-O_8+V_8+S_8-C_8)\mperiod
\end{eq}
In the above expressions, we have suppressed the integrals and some common parts of the integrands, only keeping track of the relevant combinations of characters; see~\cite{Angelantonj:2002ct}.

Equivalently, there are several options for the orientifold operator in the two theories: 
\begin{itemize}
    \item[-] It can be $\Omega$, worldsheet parity, in which case one obtains the 0A orientifold and the 0B orientifold denoted by $\mathcal{K}_1$;
    \item[-] It can be $\Omega (-1)^{F_L}$, where $F_L$ is the left-moving spacetime fermion number. In this case, one obtains the same orientifold as $\Omega$ from type 0A and the model with $\mathcal{K}_2$ from type 0B;
    \item[-] It can be $\Omega (-1)^{f_L}$, where $f_L$ is the left-moving worldsheet fermion number. In type 0A, this is an order-four orientifold and reproduces a type 0B orientifold~\cite{Bergman:1999km}. In type 0B, this gives the model with $\mathcal{K}_3$.
\end{itemize}

Let us consider two particular cases in more detail, which correspond to the orientifold action $\Omega$ in both type 0A and type 0B. 

Type 0A$/\Omega$ has the torus amplitude that is (half of that) in~\eqref{eq:T_0AB} and the Klein bottle amplitude in~\eqref{eq:K_0A}; the annulus and M\"{o}bius amplitudes are
\begin{eq}
    \mathcal{A}_{99}&=\frac{1}{2}(n_1^2 + n_2^2)(O_8+V_8)-n_1 n_2(S_8+C_8)\mcomma \\
    \mathcal{M}_{9}&=-\frac{1}{2}\left[(n_1+n_2)\hat{V}_8-(n_1-n_2)\hat{O}_8\right]\mperiod
\end{eq}
The Chan--Paton charges $n_1$ and $n_2$ imply the presence of an $\mathfrak{so}(n_1)\times\mathfrak{so}(n_2)$ gauge algebra. The light spectrum consists of a tachyon, a graviton, a dilaton, a 1-form, and a 3-form from the projected closed spectrum, and gauge vectors, tachyons in the $\left(\mathbf{\frac{n_1(n_1+1)}{2}},\mathbf{1}\right)$ and $\left(\mathbf{1},\mathbf{\frac{n_2(n_2-1)}{2}}\right)$, and Majorana fermions in the $(\mathbf{n_1},\mathbf{n_2})$ from the open sector. Note that $n_1$ and $n_2$ can also be zero, which means that the open sector is not necessary. Imposing cancellation of the dilaton tadpole, which is not required by consistency, restricts $n_1+n_2=32$. A special symmetric case that will be relevant later corresponds to $n_1=n_2=16$, which gives the algebra $\mathfrak{so}(16)\times\mathfrak{so}(16)$.

Type 0B$/\Omega$ is analogous: the torus amplitude is half of that in~\eqref{eq:T_0AB}, the Klein bottle amplitude is $\mathcal{K}_1$ in~\eqref{eq:K_0B}, and the annulus and M\"{o}bius amplitudes are 
\begin{eq}\label{eq:type0B_orientifold_open_sector}
    \mathcal{A}_{99,1}&=\frac{1}{2}(n_1^2+n_2^2+n_3^2 + n_4^2)V_8+(n_1 n_2+n_3 n_4)O_8\\
    &-(n_1 n_3+n_2 n_4)S_8-(n_1n_4+n_2n_3)C_8\mcomma\\
    \mathcal{M}_{9,1}&=-\frac{1}{2}(n_1+n_2+n_3+n_4)\hat{V}_8\mperiod
\end{eq}
Note that in this case there are four types of Chan--Paton charges, corresponding to different types of D9-branes and antibranes. The cancellation of R-R tadpoles implies that $n_1=n_2\equiv n$ and $n_3=n_4\equiv m$, thus giving the gauge algebra $\mathfrak{so}(n)^2\times\mathfrak{so}(m)^2$. The light spectrum contains a tachyon, a graviton, a dilaton, and two R-R 2-forms from the projected closed spectrum, together with gauge vectors, bifundamental tachyons and fermions from the open sector. 
Once again, the open sector is not required by consistency (one can set $n=m=0$), and imposing cancellation of the dilaton tadpole gives the further constraint $n+m=32$. A special case that will be relevant later is when $n=0$ and $m=32$.

As in type I string theory, type 0 orientifolds have ``charged and uncharged'' D-branes;\footnote{The nomenclature here follows \cite{Dudas:2001wd}, whereby ``(un-)charged'' refers to continuous R-R $p$-form fields.
A more precise characterization uses K-theory charges \cite{Kaidi:2019tyf}, and we will list these for the relevant branes below.
} see~\cite{Dudas:2001wd,Kaidi:2019tyf} for a full classification. These can provide additional states charged under spinorial representations of the orthogonal gauge groups, and their inclusion determines the gauge group in the same way as D-branes determine the group in type I string theory; see section~\ref{sec:particle_vs_string}.

Performing the analysis of~\cite{Witten:2023snr,Larotonda:2024thv} in the case of type 0 orientifolds is an interesting topic by itself, independent of the implications for dualities. We will discuss the case of the $\Omega$ orientifold of type 0A in section~\ref{sec:type0AOmega}, relying on appendix~\ref{app:0A_branes} for details on D-branes in this model. Similarly, for the case of the $\Omega$ orientifold of type 0B, we will present our results in section~\ref{sec:Bergman-Gaberdiel}, relying on appendix~\ref{app:0B_branes} for details on D-branes in this model.

Analogously to the analysis of type I, we are interested in the branes without open-string tachyons that cannot decay to the vacuum.
A summary of the brane content in the two orientifolds of interest can be found in \cite[Tables 9 and 10]{Angelantonj:2002ct} and~\cite{Kaidi:2019tyf}. The tachyon-free branes, together with their K-theory charges, are:
\begin{itemize}
    \item[-] In type 0A$/\Omega$: $\Z$-charged D$p$-branes with $p$ even, a $\Z_2$-charged D0-brane,\footnote{Here and in the following, ``$\Z_2$-charged'' means that the brane has a $\Z_2$ charge in K-theory.} a $\Z_2$-charged D1-brane.
    \item[-] In type 0B$/\Omega$: 2 types of $\Z$-charged D$1$-branes, 2 types of $\Z$-charged D$5$-branes, 2 types of $\Z_2$-charged D0-branes.\footnote{There are some typos in \cite[Table 10]{Angelantonj:2002ct} that originate from a typo in the discussion after eq.~(6.22) in~\cite{Dudas:2001wd}, which would otherwise prevent the existence of a single D0-brane. We will discuss this in the relevant appendix. We thank Justin Kaidi and Yuji Tachikawa for illuminating discussions on this point.}
\end{itemize}

\subsubsection*{Center cosets of $Spin(2n)$}

To set up the subsequent computation of the gauge groups in these orientifold theories, let us first introduce some notational conventions.
In general, to identify the group $G/Z$ with $G$ simply connected, one determines $Z$ by looking at the central elements of $G$ that act trivially on the physical states. The center of $Spin(2n)$ groups---the relevant ones for all the examples that we consider---is ${\cal Z}(Spin(2n))=\Z_2\times \Z_2$.
We denote its elements by $\left\{1,z_v,z_s,z_c\right\}$, with the conditions $z_i^2 = 1$ and $z_v=z_s z_c = z_c z_s$. 

Under the center, the transformation properties of irreducible representations $\mathbf{R}$ of $Spin(2n)$ fall into equivalence classes $[\mathbf{R}]$ labelled by the Pontryagin dual $\widehat{\Z_2 \times \Z_2} \cong \Z_2 \times \Z_2$ (i.e., irreducible representations of the center):
\begin{align}
    o = [{\bf O}] \, , \quad v = [{\bf V}] \, , \quad s = [{\bf S}] \, , \quad c = [{\bf C}] \, ,
\end{align}
with $s+c = v$.
Note that this is consistent with the explicit parametrization of the glue vectors \eqref{eq:equivalence_classes_D16*} for $Spin(32)$ in that the shift of the root lattice by any weight vector $\vec{i}$ of the indicated representations is the corresponding coset labelled by $i \in D_{16}^*/D_{16} = \Z_2 \times \Z_2$.
We use the convention that $z_i$ acts as the identity on the class $i$;\footnote{Note that there is another common convention in the literature, for which $z_i$ is the Pontryagin-dual of $i$, i.e., acts non-trivially on the representation $i$.} see table~\ref{tab:actions_of_z2s}.
To determine the gauge group $G/Z$, we first identify the representations ${\bf R}$ from quantizing the excitations on the various branes, and then find the largest set of central elements $z_i$ under which all cosets $[{\bf R}]$ transform trivially.

\begin{table}[!ht]
    \centering
    \begin{tabular}{c|c|c|c}
    irrep & $z_v$ & $z_s$ & $z_c$ \\ \hline
    $o = [{\bf O}]$ & $+$ & $+$ & $+$ \\
    $v = [{\bf V}]$ & $+$ & $-$ & $-$ \\
    $s = [{\bf S}]$ & $-$ & $+$ & $-$ \\
    $c = [{\bf C}]$ & $-$ & $-$ & $+$
    \end{tabular}
    \caption{Representation classes of $Spin(2n)$ and action of the central elements on them.}
    \label{tab:actions_of_z2s}
\end{table}

\subsection{Type \texorpdfstring{0A$/\Omega$}{0A/Omega}}\label{sec:type0AOmega}

While there is no known ten-dimensional dual for this model, there are recent proposals that obtain type 0 orientifolds from M-theory on $\Z_2$ quotients of the wedge sum of two circles, $S^1\vee S^1$~\cite{Altavista:2026evd,Baykara:2026vdc}. Finding the global topology of the gauge groups in these orientifolds can thus serve as a test of the emerging duality web.

To this end, we consider the branes that are free of open-string tachyons\footnote{In this theory, there is still a closed-string tachyon and an open-string tachyon from the background D9-branes. These might play a role in the dynamical realization of the duality, but for our purposes, we will simply neglect their presence. While they do play a role in determining the background, we assume they do not change the spectrum of tachyon-free branes. It would be interesting to explore this issue in more detail.} to determine the states that fix the gauge group topology. From appendix~\ref{app:0A_branes}, these are $\Z$-charged D$p$-branes with $p$ even, a single $\Z_2$-charged D1-brane, and an additional type of D0-brane with $m=\bar{m}=1$ and $d_2=0$ in the conventions of appendix~\ref{app:0A_branes}.

For concreteness, we perform the brane-mode quantization in the type 0A orientifold model with $n_1=n_2=16$, and therefore the center is two copies of the center of $Spin(16)$. 
First, we determine which central elements act trivially on the perturbative D9-D9-open-string states, which are listed in table~\ref{tab:0A_open_spectrum}.
In fact, the only states that transform non-trivially under the center are the fermions in the bivector $({\bf 16}, {\bf 16})$ of the two gauge factors, which corresponds to the class $(v,v)$ in terms of the center of the universal cover $Spin(16)\times Spin(16)$. To leave it invariant, a central element $(g_1,g_2)$ of $Spin(16)\times Spin(16)$ must have $g_1$ and $g_2$ that are either both $1$ or $z_v$, or both $z_s$ or $z_c$:
\begin{eq}
    \{(1,1),(1,z_v),(z_v,1),(z_v,z_v),(z_s,z_s),(z_s,z_c),(z_c,z_s),(z_c,z_c)\}\mperiod
\end{eq}
This is a $\Z_2^{3}$ subgroup whose generators can be chosen to be
\begin{eq}\label{eq:0A-center-generators-perturbative}
    (z_s,z_s) \, , \quad (1,z_v) \, , \quad (z_v,1)\mperiod
\end{eq}
The gauge group of type 0A$/\Omega$, taking into account only perturbative states, would then be\footnote{Actually, the perturbative states transform with $O$ groups rather than $SO$ in the numerator. However, instanton effects will reduce them to $SO$ groups, as in type I string theory~\cite{Witten:1998cd}.} 
\begin{eq}
    \frac{Spin(16) \times Spin(16)}{\Z_2^3} = \frac{SO(16)\times SO(16)}{\Z_2}\mperiod
\end{eq}

In the following, we will discuss, case by case, how non-perturbative (tachyon-free) branes modify this structure.

\subsubsection{D1 quantization}

First, we consider a single $\Z_2$-charged D1-brane that is tachyon-free, i.e., a D1 with $d_1=0$ and $d_2=1$. The worldvolume gauge symmetry is $O(d_2=1)\simeq \Z_2$. The analysis of~\cite{Larotonda:2024thv} proceeds by quantizing the states that arise from worldvolume fields on this D1. 
For generic values of $n_1$ and $n_2$, the light spectrum can be found in table~\ref{tab:0A_D1_spectrum} from appendix~\ref{app:0A_branes} and in the discussion accompanying it. 
Relevant to the current discussion is only a simpler version of the table, showing the states that are charged under the spacetime gauge symmetry $\mathfrak{so}(n_1) \times \mathfrak{so}(n_2)$, which can therefore affect the global form of the gauge group; see table~\ref{tab:0A/Omega_charged_D1_spectrum}.\footnote{The scalar modes of table~\ref{tab:0A/Omega_charged_D1_spectrum} are not present in table~\ref{tab:0A_D1_spectrum} because they are massive. See the discussion around table~\ref{tab:0A_D1_spectrum}.}
\begin{table}[!ht]
\centering
\renewcommand{\arraystretch}{1.2}
\begin{tabular}{l|c|c|c}
Worldvolume state  & $\mathfrak{so}(8)_T$ & $\mathfrak{so}(n_1) \times \mathfrak{so}(n_2)$ & worldvolume $O(1)$  \\
\hline
Fermion $\psi$ & $\mathbf{1}$ & $\left(\mathbf{n_1},\mathbf{1}\right)$ & odd \\
Scalar $X$ & $\mathbf{16}$ & $\left(\mathbf{1},\mathbf{n_2}\right)$ & odd \\
\end{tabular}
\caption{Open-string spectrum charged under the gauge symmetry from a single D1-brane in the orientifold of type 0A string theory with generic $n_1$ and $n_2$ and with Chan--Paton charges $d_1=0$ and $d_2=1$. $\mathfrak{so}(8)_T$ is the transverse $\mathfrak{so}(8)$. Here, $\mathbf{16}=\mathbf{8}_s\oplus \mathbf{8}_c$.}
\label{tab:0A/Omega_charged_D1_spectrum}
\end{table}

Hence, for the relevant case of $n_1 = n_2 = 16$, we have 16 fermions $\psi_i$ forming the vector representation ${\bf 16}$ of $\mathfrak{so}(n_1 = 16)$ and a singlet under $\mathfrak{so}(n_2 = 16)$, transforming non-trivially under the worldvolume $\Z_2$ gauge symmetry.
By quantizing their zero modes,
\begin{equation}
    \{\psi_i,\psi_j\} = \delta_{ij}\mcomma
\end{equation}
where $i,j = 1, ..., 16$ are the vector indices of the \emph{first} $\mathfrak{so}(16)$ factor, we obtain a Clifford algebra over the vector representation of $\mathfrak{so}(n_1)$.
For $n_1 = 4k$, the Hilbert space that furnishes a representation of this algebra decomposes into (tensor products of)
\begin{align}
    ({\bf S}, {\bf 1}) \oplus ({\bf C}, {\bf 1})
\end{align}
for the $\mathfrak{so}(16) \oplus \mathfrak{so}(16)$ Lie algebra.
The $({\bf S}, {\bf 1})$ subspace is spanned by the Fock states with \emph{even} numbers of creation operators arising from $\psi_i$, while $({\bf C}, {\bf 1})$ is spanned by Fock states with \emph{odd} numbers of $\psi_i$-creation operators.

This is not the full story, as we must project onto the states invariant under the worldvolume $\Z_2$ gauge symmetry.
Since the fermion fields $\psi_i$ are odd under this gauge symmetry, the gauge invariant sector can only contain Fock states with even numbers of creation operators.
Hence, demanding $O(1)$ gauge invariance projects out the $({\bf C},{\bf 1})$ states and leaves only the $({\bf S}, {\bf 1})$ representation.\footnote{Equivalently, the $O(1)=\Z_2$ gauge symmetry is generated by the chirality matrix $\bar{\psi}=\psi_1 \dots \psi_{16}$, and therefore, just as for type I, we only get the spinor (and not the co-spinor) of the first factor of $\mathfrak{so}(16)\times \mathfrak{so}(16)$.} In terms of the center cosets, we are left with
\begin{align}
    (s,o) \, .
\end{align}
Thus, the D1-brane excitations are only invariant under the generators $(z_s, z_s)$ and $(z_s, z_c)$ out of those in \eqref{eq:0A-center-generators-perturbative} that generate the perturbatively invariant $\Z_2^3$.
This means that the unbroken center reduces to the $\Z_2\times \Z_2$ given by
\begin{eq}\label{eq:0A_center_after_D1s}
    (g_1, g_2) \in \{(1,1),(z_s,z_s),(z_s,z_c),(1,z_v)\}\mperiod
\end{eq}

Note that the worldvolume also contains scalar fields transforming in the vector representations of $\mathfrak{so}(n_2 = 16)$.
Quantizing these, i.e., imposing $[X_a,X_b]=\delta_{ab}$, where $a,b$ are the vector indices of $\mathfrak{so}(n_2 = 16)$, the resulting Fock space contains only symmetric tensor products of the vector representation of $\mathfrak{so}(n_2 = 16)$ and singlet representation of $\mathfrak{so}(n_1 = 16)$.
In fact, since $X_a$ are also odd under the $O(1) = \Z_2$ gauge symmetry,\footnote{In principle, we should also include the $\mathfrak{so}(8)_T$ representation of the scalars, so we have fields $X_{a,\beta}$ with $\beta$ being the $\mathfrak{so}(8)_T$ indices.
The quantization of these yields the direct sum of tensor products of $\mathfrak{so}(n_2 = 16)$ and $\mathfrak{so}(8)_T$ representations, respectively.
In principle, projecting to the $O(1)$ gauge invariant sector correlates the center representations under $Spin(n_2 = 16)$ and $Spin(8)_T$; however, as we are only interested in the $Spin(n_2 = 16)$, this correlation is irrelevant, and hence we will ignore the $\mathfrak{so}(k)_T$ transformation properties in the following.\label{footnote:ignore_soT-indices}} gauge invariance projects onto the sector containing only even rank tensors of $\mathfrak{so}(n_2 = 16)$ (and singlets under the first factor), thus always invariant under the center of $Spin(16)^2$.

In summary, the D1-excitations, together with the bivector from D9-D9-open-string states, leave invariant only the central subgroup \eqref{eq:0A_center_after_D1s}.
Therefore, we would have the gauge group topology
\begin{align}
    \frac{Spin(16) \times Spin(16)}{\Z_2 \times \Z_2} \, .
\end{align}

\subsubsection{D0 quantization}\label{sec:0A_D0}

Consider now D0-branes. The light spectrum can be found in table~\ref{tab:0A_D0_D8_spectrum} from appendix~\ref{app:0A_branes} and in the discussion accompanying it. Here, we include simpler versions of the table, showing states that can potentially constrain the gauge group.
Note that we have two qualitatively different types depending on the values of $d_{1,2}$ and $m$.

Take a single D0 of the first type with $m=0$, $d_1=1$, and $d_2=0$, which is $\Z$-charged; see table~\ref{tab:0A/Omega_charged_D0_spectrum}.\footnote{As for the D1, the scalars in table~\ref{tab:0A/Omega_charged_D0_spectrum} are massive.}
There is an $O(1) = \Z_2$ gauge symmetry on the worldvolume, under which both the fermions and scalars are charged non-trivially.
\begin{table}[!ht]
\centering
\renewcommand{\arraystretch}{1.2}
\begin{tabular}{l|c|c|c}
Worldvolume state  & $\mathfrak{so}(9)_T$ & $\mathfrak{so}(n_1) \oplus \mathfrak{so}(n_2)$ & worldvolume $O(1)$ \\
\hline
Fermion & $\mathbf{1}$ & $\left(\mathbf{1},\mathbf{n_2}\right)$ & odd \\
Scalar & $\mathbf{16}_{s'}$ & $\left(\mathbf{n_1}, \mathbf{1}\right)$ & odd \\
\end{tabular}
\caption{Open-string spectrum charged under the gauge symmetry from a single D0-brane in the orientifold of type 0A string theory with generic $n_1$ and $n_2$ and with Chan--Paton charges $d_1=1$, $d_2 = 0$, and $m = 0$. $\mathfrak{so}(9)_T$ is the transverse $\mathfrak{so}(9)$ and $\mathbf{16}_{s'}$ is its irreducible Dirac-spinor representation.}
\label{tab:0A/Omega_charged_D0_spectrum}
\end{table}
As in the D1-case, the scalars, which are charged only under the $\mathfrak{so}(n_1)$ factor, give rise to center-invariant states.
The interesting states come from quantizing the fermions as in the D1 case, which are now spinors of the \emph{second} factor of $\mathfrak{so}(16)\times \mathfrak{so}(16)$ (see the first row of the table). 
The quantization proceeds analogously to the D1-case, and we find---after imposing $O(1)$ gauge invariance---only the coset of $(o,s)$ under the center of $Spin(n_1) \times Spin(n_2)$ (again, this holds only for $n_1, n_2$ divisible by 4, which is all of the cases we will discuss).

Hence, out of the central elements \eqref{eq:0A_center_after_D1s} preserved by the D1-brane, the ones acting trivially on the D0s are reduced to the $\Z_2$ given by
\begin{eq}
    \{(1,1), (z_s,z_s)\}\mperiod
\end{eq}

The other type of $\Z_2$-charged tachyon-free D0-brane has $m=1$ and $d_1=d_2=0$; see table~\ref{tab:0A/Omega_uncharged_D0_spectrum}. 
\begin{table}[!ht]
\centering
\renewcommand{\arraystretch}{1.2}
\begin{tabular}{l|c|c|c}
Worldvolume state  & $\mathfrak{so}(9)_T$ & $\mathfrak{so}(n_1) \oplus \mathfrak{so}(n_2)$  & worldvolume $\mathfrak{u}(1)$ \\
\hline
Fermion $\psi^{\bar{q}}_a$ & $\mathbf{1}$ & $\left(\mathbf{1},\mathbf{n_2}\right)$ & $\bar{q}$ \\
Fermion $\chi^q_i$ & $\mathbf{1}$ & $\left(\mathbf{n_1},\mathbf{1}\right)$ & $q$ \\
Scalar & $\mathbf{16}_{s'}$ & $\left(\mathbf{n_1}, \mathbf{1}\right)$ & $\bar{q}$ \\
Scalar & $\mathbf{16}_{s'}$ & $\left(\mathbf{1},\mathbf{n_2}\right)$ & $q$ \\
\end{tabular}
\caption{Open-string spectrum charged under the gauge symmetry from a single D0-brane in the orientifold of type 0A string theory with generic $n_1$ and $n_2$ and with Chan--Paton charges $d_1=d_2 = 0$, and $m=1$. $\mathfrak{so}(9)_T$ is the transverse $\mathfrak{so}(9)$ and $\mathbf{16}_{s'}$ is its irreducible Dirac-spinor representation. $q$ and $\bar{q}$ denote opposite charges under the worldvolume $\mathfrak{u}(1)$.}
\label{tab:0A/Omega_uncharged_D0_spectrum}
\end{table}
The worldvolume gauge symmetry requires the quantization of $\mathfrak{u}(1)$ representations, specifically the fermions in the fundamental representation of $\mathfrak{so}(n_{1,2})$ with unit charge with respect to the $\mathfrak{u}(1)$ factor.
The canonical quantization of these states amounts to imposing
\begin{equation}\label{eq:quantization_two_fermions}
    \{\psi_a^{\bar{q}}, \bar{\psi}_b^{\bar{q}}\} = \delta_{ab} \mcomma \quad \{\chi^q_i, \bar{\chi}^q_j\} = \delta_{ij} \mcomma \quad \{\psi_a^{\bar{q}}, \chi_i^q\} = 0 \, .
\end{equation}
where $a$ and $i$ are the $\mathfrak{so}(n_{1})$ and $\mathfrak{so}(n_2)$ vector indices, respectively.
The label $q$ indicates that the fundamental fields under consideration transform under the $\mathfrak{u}(1)$ gauge factor. The field and its complex conjugate carry opposite unit charges. 
Ignoring the $\mathfrak{u}(1)$-charge for the moment, these anti-commutation relations correspond to the Clifford algebra
\begin{align}
    Cl(n_1) \widehat{\otimes} Cl(n_2) \cong Cl(n_1 + n_2) \, ,
\end{align}
where $\widehat{\otimes}$ denotes the graded tensor product.
Therefore, the Fock space carries the representation ${\bf S}_{n_1+n_2} \oplus {\bf C}_{n_1+n_2}$ under $\mathfrak{so}(n_1+n_2)$, which, under the branching $\mathfrak{so}(n_1 + n_2) \supset \mathfrak{so}(n_1) \oplus \mathfrak{so}(n_2)$, decomposes as
\begin{align}
\begin{split}
    {\bf S}_{n_1 + n_2} & \rightarrow ({\bf S}_{n_1}, {\bf S}_{n_2}) \oplus ({\bf C}_{n_1}, {\bf C}_{n_2}) \mcomma \\
    {\bf C}_{n_1 + n_2} & \rightarrow ({\bf S}_{n_1}, {\bf C}_{n_2} ) \oplus ({\bf C}_{n_1} , {\bf S}_{n_2}) \mperiod
\end{split}
\end{align}
That is, the quantization of the fermions yields the $\mathfrak{so}(n_1) \oplus \mathfrak{so}(n_2)$ representations
\begin{align}
    ({\bf S}, {\bf S}) \oplus ({\bf C}, {\bf C}) \oplus ({\bf S}, {\bf C}) \oplus ({\bf C}, {\bf S})\,,
\end{align}
and tensor products thereof.
Observe that we do not obtain representations in the coset of, e.g., $({\bf 1}, {\bf S})$ or $({\bf 1}, {\bf C})$.
This means that the ``simultaneous'' quantization of the $\psi$ and $\chi$ fermions automatically leads to a $Spin(n_1) \times Spin(n_2)$ spectrum invariant under the $\Z_2$ generated by $(z_v, z_v)$.\footnote{Viewing the origin of these states as coming from the branching $\mathfrak{so}(n_1 + n_2) \supset \mathfrak{so}(n_1) \oplus \mathfrak{so}(n_2)$, this is actually no surprise since at the level of groups we have $Spin(n_1 + n_2) \supset [Spin(n_1) \times Spin(n_2)]/\Z_2$, where the $\Z_2$ is precisely generated by $(z_v, z_v)$.}

To get the physical degrees of freedom, we must further project to the $\mathfrak{u}(1)$-invariant sector by selecting the zero-charge states. 
Under a single $\mathfrak{so}$-factor, states in the ${\bf S}$/${\bf C}$ subspace are generated by even/odd numbers of creation operators.
Thus, states in the $({\bf S}, {\bf S})$ and $({\bf C}, {\bf C})$ will have in total an even number of creation operators (from either $\psi$ or $\chi$), while $({\bf S}, {\bf C})$ and $({\bf C}, {\bf S})$ will have in total an odd number.
Since both $\psi$ and $\chi$ have the same modulus of $\mathfrak{u}(1)$ charge, it is clear that only the subspace associated with $({\bf S}, {\bf S}) \oplus ({\bf C}, {\bf C})$ contains charge-0 states.
Hence, the projection onto this subspace acts as $(z_s, z_s)$ (or, together with $(z_v, z_v)$, it is equivalent to $(z_c, z_c)$).

In summary, the \emph{second} type of D0-brane ($m=1$ and $d_1=d_2=0$) by itself would actually indicate that the spacetime gauge group is
\begin{align}
    \frac{Spin(16) \times Spin(16)}{\Z_2 \times \Z_2} \, ,
\end{align}
with the two $\Z_2$'s generated by $(z_v, z_v)$ and $(z_s, z_s)$, respectively.
However, due to the presence of the D1- and the other type of D0-brane (with $m=0=d_2$ and $d_1=1$), there are excitations transforming non-trivially under $(z_v, z_v)$, and hence the gauge group topology compatible with all of these brane spectra is
\begin{align}\label{eq:gauge_group_0A-after-D0}
    \frac{Spin(16) \times Spin(16)}{\Z_2} \, ,
\end{align}
with the $\Z_2$ generated by $(z_s, z_s)$.

\subsubsection{D2 quantization}

The D2 case is analogous to the D0 one; see table~\ref{tab:0A_D2_spectrum}. The discussion exactly parallels that of section~\ref{sec:0A_D0}, with no consequences for the center, and we will not repeat it here.

\subsubsection{D4 quantization}\label{sec:0A_D4_quantization}

Consider D4-branes. The light spectrum can be found in table~\ref{tab:0A_D4_spectrum} from appendix~\ref{app:0A_branes} and in the discussion accompanying it. 

The relevant tachyon-free D4-branes require $m=0$ and $d_2=0$, which leads to the reduced set of states in table~\ref{tab:0A/Omega_charged_D4_spectrum_restricted}, where $d_1=2k$.\footnote{As for the D0 and D1, the scalar modes are not present in table~\ref{tab:0A_D4_spectrum} because they are massive. See the discussion around table~\ref{tab:0A_D4_spectrum}.}
\begin{table}[!ht]
\centering
\renewcommand{\arraystretch}{1.2}
\begin{tabular}{l|c|c|c}
Worldvolume state  & $\mathfrak{so}(5)_T$ & $\mathfrak{so}(n_1) \times \mathfrak{so}(n_2)$  & worldvolume $\mathfrak{sp}(k)$ \\
\hline
Fermion & $\mathbf{1}$ & $\left(\mathbf{1},\mathbf{n_2}\right)$ & $\mathbf{2k}$ \\
Scalar & $\mathbf{4}_{s'}$ & $\left(\mathbf{n_1}, \mathbf{1}\right)$ & $\mathbf{2k}$ \\
\end{tabular}
\caption{Open-string spectrum charged under the gauge symmetry from D4-branes in the orientifold of type 0A string theory with generic $n_1$ and $n_2$ and with Chan--Paton charges $d_1=2k$, $d_2=0$, and $m=0$. $\mathfrak{so}(5)_T$ is the transverse $\mathfrak{so}(5)$.}
\label{tab:0A/Omega_charged_D4_spectrum_restricted}
\end{table}

From table~\ref{tab:0A/Omega_charged_D4_spectrum_restricted}, setting $n_1=n_2=16$, the interesting states come from quantizing the fermions, as in the previous cases, which are spinors of the \emph{second} factor $\mathfrak{so}(n_2=16)$. Hence, we must quantize the $(\mathbf{16},\mathbf{2k})$ of $\mathfrak{so}(16) \oplus \mathfrak{sp}(k)$, where we are not displaying the singlet representation under the first $\mathfrak{so}(16)$ factor. For these algebras, we must follow the analysis of~\cite{Larotonda:2024thv}.
The naive canonical quantization of these objects requires
\begin{equation}\label{eq:D4_naive_quantization}
    \{\psi_{i,\alpha},\psi_{j,\beta}\} = \delta_{ij}\omega_{\alpha\beta}\,,
\end{equation}
where $i,j = 1,\dots,16$ are indices of the vector of $\mathfrak{so}(16)$ and $\alpha,\beta = 1,\dots,2k$ of the fundamental of $\mathfrak{sp}(k)$. This quantization seems to be inconsistent because the invariant symplectic tensor of $\mathfrak{sp}(k)$ is antisymmetric, $\omega_{\alpha\beta} = - \omega_{\beta\alpha}$, while the left-hand side of~\eqref{eq:D4_naive_quantization} is symmetric. However, in five dimensions, we must impose the symplectic-Majorana condition on the fermions, which reduces the degrees of freedom of the pseudo-real representation $(\mathbf{16},\mathbf{2k})$ of $ \mathfrak{so}(16) \oplus \mathfrak{sp}(k)$ to the real representation $(\mathbf{16},\mathbf{k})$ of $\mathfrak{so}(16) \oplus \mathfrak{so}(k)$. Indeed, the special embedding
\begin{equation}
    \mathfrak{sp}(k) \supset \mathfrak{sp}(1) \oplus \mathfrak{so}(k) 
\end{equation}
gives us the $\mathfrak{sp}(1) \cong \mathfrak{su}(2)$ factor that maps the two real parts of the fermion into each other, and the symplectic tensor is given by $\omega = \epsilon \otimes \delta $, where $\epsilon$ is the invariant tensor of $\mathfrak{su}(2)$ and $\delta$ of $\mathfrak{so}(k)$. Therefore, we must quantize the $(\mathbf{16},\mathbf{k})$ of $\mathfrak{so}(16) \oplus \mathfrak{so}(k)$, imposing the now consistent canonical quantization
\begin{equation}
    \{\psi_{i,\alpha},\psi_{j,\beta}\} = \delta_{ij}\delta_{\alpha\beta}\,,
\end{equation}
where this time $\alpha,\beta = 1,\dots,k$ are indices of the vector of $\mathfrak{so}(k)$. We can rearrange the components using capital indices $(i,\alpha) \rightarrow I = 1, \dots, 16k$ and obtain a Clifford algebra over the vector of $\mathfrak{so}(16k)$, consistent with the special embedding
\begin{equation}
    \mathfrak{so}(16k) \supset \mathfrak{so}(16) \oplus \mathfrak{so}(k) \, , \quad {\bf 16k} \rightarrow ({\bf 16},{\bf k}) \,.
\end{equation}
Hence, we obtain $\mathfrak{so}(16k)$ spinors, which decompose into representations of $\mathfrak{so}(16) \oplus \mathfrak{so}(k)$ under the branching rule for the above special embedding.
This can be quite involved to compute in general, and depends on the value of $k$, see~\cite[Section 3.1]{Larotonda:2024thv}.
The upshot is that, once we select the representations that are invariant under the worldvolume gauge symmetry, i.e., the $\mathfrak{so}(k) \subset \mathfrak{sp}(k)$ part, for any value of $k$, the remaining representations fall into the $o$-coset of $\mathfrak{so}(n_2 = 16)$.
Thus, under the full spacetime gauge algebra $\mathfrak{so}(n_1 = 16) \oplus \mathfrak{so}(n_2 = 16)$, the D4-brane states are all in the $(o,o)$ coset, which is trivially invariant under the $\Z_2$ appearing in the quotient \eqref{eq:gauge_group_0A-after-D0}.

\subsubsection{The gauge group}

The quantization of the brane modes implies that the central elements that act trivially are $\{(1,1),(z_s,z_s)\}$. Thus, the global topology of the gauge group of type 0A$/\Omega$ is fixed by non-perturbative D-brane states to be
\begin{eq}
    \frac{Spin(16)\times Spin(16)}{\Z_2}\mperiod
\end{eq}

This can be useful, for instance, in the proposed M-theory derivation of this orientifold in~\cite{Altavista:2026evd,Baykara:2026vdc}, since the above global structure should emerge from the M-theory degrees of freedom giving rise to the type 0A D9-branes.

\subsection{Type \texorpdfstring{0B$/\Omega$}{0B/Omega}}\label{sec:Bergman-Gaberdiel}

We now turn our attention to the type 0B orientifold, setting up a comparison of gauge groups with the proposed dual bosonic string theory.
To compute the gauge group of the orientifold, the first step is to consider the perturbative string spectrum from the D9-branes in table~\ref{tab:0B/Omega_spacetime_spectrum}. Focusing here on the case $n_1=n_2=0$ and $n_3=n_4=32$, we obtain the gauge sector in the $(o,o)$ class and the tachyons in the $(v,v)$ class. 
This implies that the gauge group compatible with the perturbative spectrum of type 0B$/\Omega$ is\footnote{Again, up to the same subtleties with $SO$ and $O$ groups as in the type 0A case.}
\begin{eq}
    \frac{SO(32)\times SO(32)}{\Z_2}\mperiod
\end{eq}
As we have done for the type 0A orientifold, the next step to understand the gauge group topology is to include states from non-tachyonic D-branes.

\subsubsection{D0 quantization}

The first type of brane that we have to consider is actually a special case: a single D0-brane. There are two types of such D0-branes, charged under a K-theory $\Z_2$; see~\cite{Kaidi:2019tyf} and the discussion in appendix~\ref{app:0B_branes}. 
Only one of the two types is relevant to us, as the other does not carry charged excitations under $\mathfrak{so}(n_3 = 32) \oplus \mathfrak{so}(n_4 = 32)$.
Our analysis clarifies an earlier discussion in~\cite{Michishita:1999it}, by making explicit the ``truncations'' on the $\mathfrak{so}(32) \oplus \mathfrak{so}(32)$ representations to be matched in the dual bosonic string.

The light spectrum of the D0 can be found in table~\ref{tab:0B/Omega_D0_spectrum} from appendix~\ref{app:0B_branes}. Here, we include a simpler version of the table, showing the states that can potentially constrain the spacetime gauge group for the special case that we are considering, $n_1=n_2=0$ and $n_3=n_4=32$, taking a single brane with $d_1=1$ and $d_2=0$; see table~\ref{tab:0B/Omega_D0_uncharged_spectrum}.\footnote{Here and in the following cases, there would also be additional scalars charged under the gauge symmetry (see appendix~\ref{app:0B_branes}) but as in type 0A$/\Omega$, they do not impose further constraints on the gauge groups.}

\begin{table}[H]
\centering
\renewcommand{\arraystretch}{1.2}
\begin{tabular}{l|c|c|c}
Worldvolume state  & $\mathfrak{so}(9)_T$ & $\mathfrak{so}(32)\oplus\mathfrak{so}(32)$ & worldvolume $O(1)$ \\
\hline
Fermions $\Psi \equiv \psi + \chi$ & $\mathbf{1} \oplus {\bf 9}_v$ & $(\mathbf{32}, \mathbf{1}) \oplus (\mathbf{1}, \mathbf{32}) $ & odd 
\end{tabular}
\caption{Open-string light spectrum charged under the gauge symmetry from a single D0-brane with $d_1=1$ and $d_2=0$ in the orientifold of type 0B string theory.
There is an $O(1) \cong \Z_2$ gauge symmetry on the D0.
We have set $n_1=n_2=0$ and $n_3=n_4=32$ for the proposed dual setting of the bosonic string.
According to the representation under the $\mathfrak{so}(n_3 = 32) \oplus \mathfrak{so}(n_4 = 32)$ spacetime gauge algebra, the fermion is split such that $\psi$ is in the $({\bf 32},{\bf 1})$, and $\chi$ is in the $({\bf 1}, {\bf 32})$ representation.
}
\label{tab:0B/Omega_D0_uncharged_spectrum}
\end{table}

To quantize the fermions, we must perform the same procedure as for the \emph{second type} of D0s of type 0A$/\Omega$ in section~\ref{sec:0A_D0}.
Namely, canonical quantization imposes
\begin{align}\label{eq:canonical_quantization_D0-0B}
    \{\psi_i, \psi_j\} = \delta_{ij} \, , \quad \{\chi_a, \chi_b\} = \delta_{ab} \, , \quad \{\psi_i, \chi_a\} = 0 \, ,
\end{align}
where $i=1,...,32$ is the vector index of $\mathfrak{so}(n_3 = 32)$, and $a=1,...,32$ that of $\mathfrak{so}(n_4 = 32)$,\footnote{Again, we are ignoring the $\mathfrak{so}(9)_T$ indices, as they will not affect our discussion of the spacetime representations obtained after quantization; see footnote \ref{footnote:ignore_soT-indices}.} analogous to~\eqref{eq:quantization_two_fermions}.
Proceeding as in that case, we obtain a Clifford algebra over the vector of $\mathfrak{so}(64) \supset \mathfrak{so}(32) \oplus \mathfrak{so}(32)$, which---after decomposing the $\mathfrak{so}(64)$ (co-)spinors---gives rise only to the $\mathfrak{so}(32) \oplus \mathfrak{so}(32)$ cosets generated by $(s,s)$, $(c,c)$, $(c,s)$, and $(s,c)$.
In other words, because the simultaneous quantization of $\psi$ and $\chi$ produces states naturally transforming as Dirac spinors of $Spin(64)$, they form representations of $[Spin(32) \times Spin(32)]/\Z_2 \subset Spin(64)$, where the $\Z_2$ is generated by $(z_v, z_v)$.

As usual, we must further impose invariance under the $O(1) = \Z_2$ gauge symmetry acting on the Chan--Paton index on the D0-brane.
Similarly to the arguments for obtaining charge-0 states under a $\mathfrak{u}(1)$ gauge symmetry in section \ref{sec:0A_D0}, the key property is that Fock states in the $s$/$c$ coset of one $\mathfrak{so}$-factor contain an even/odd number of creation operators of the corresponding field.
Since both fields, $\psi$ and $\chi$, are odd under the $\Z_2$ gauge symmetry, projecting onto the gauge invariant sector preserves only states in the cosets $(s,s)$ and $(c,c)$, as these have an even total number of creation operators.
Equivalently, the $\Z_2$ acts on the cosets as $(z_s, z_s)$ (or $(z_c, z_c) = (z_s, z_s) \cdot (z_v, z_v)$), and imposing gauge invariance projects out representations that are not invariant under this central element.

In conclusion, the excitations of the D0-brane of type 0B$/\Omega$ transform faithfully under the spacetime gauge group
\begin{align}
    \frac{Spin(32) \times Spin(32)}{\Z_2 \times \Z_2} \, ,
\end{align}
where $\Z_2 \times \Z_2 = \{(1,1), (z_s,z_s), (z_c,z_c), (z_v,z_v)\}$ is generated by $(z_v, z_v)$ and $(z_s, z_s)$ (or any other combination of two non-trivial elements).

\subsubsection{D1 quantization}\label{subsubsec:D1-in-0B}

Consider now the charged D1-branes. The light spectrum can be found in table~\ref{tab:0B/Omega_D1_spectrum} from appendix~\ref{app:0B_branes}. Here, we include a simpler version of the table, showing the states that can potentially constrain the gauge group for the special case that we are considering, $n_1=n_2=0$ and $n_3=n_4=32$, and taking only branes, i.e., $d_1$, $d_3\neq0$ and $d_2=d_4=0$; see table~\ref{tab:0B/Omega_charged_D1_spectrum}.
\begin{table}[!ht]
\centering
\renewcommand{\arraystretch}{1.2}
\begin{tabular}{l|c|c|c}
Worldvolume state  & $\mathfrak{so}(8)_T$ & $\mathfrak{so}(32)\oplus\mathfrak{so}(32)$ & worldvolume $\mathfrak{so}(d_1)\oplus\mathfrak{so}(d_3)$ \\
\hline
Fermions & $\mathbf{1} \oplus {\bf 8}_v$ & $
     (\mathbf{32}, \mathbf{1}) \oplus  (\mathbf{1}, \mathbf{32})$ & $  (\mathbf{d_1}, \mathbf{1}) $ 
\end{tabular}
\caption{Open-string light spectrum charged under the gauge symmetry from D1-branes in the orientifold of type 0B string theory with $n_1=n_2=0$, $n_3=n_4=32$, and with generic Chan--Paton charges $d_1$ and $d_3$. We are not including antibranes with non-zero $d_2$ and $d_4$.
From the $d_1$ Chan--Paton charges, the fermions are further charged under an $O(1) \cong \Z_2$ worldvolume gauge symmetry.
}
\label{tab:0B/Omega_charged_D1_spectrum}
\end{table}

By keeping $d_3$ generic, we can explicitly see that branes charged only under the corresponding 2-form (i.e., with $d_1=0$) impose no constraint on the gauge group, as there are no $\mathfrak{so}(32) \oplus \mathfrak{so}(32)$-charged states on these 1-branes.
This is consistent with the picture of~\cite{Baykara:2026vdc}, for which the corresponding 2-form can be Higgsed without affecting the spacetime gauge symmetry; see section \ref{sec:BG_group} for further explanation.

On the other hand, branes with $d_1 \neq 0$ host fermions in the vector representations of the two $\mathfrak{so}(32)$ factors, which are odd under a worldvolume $O(1) = \Z_2$ gauge symmetry.
As these are identical to the D0-worldvolume degrees of freedom in terms of the spacetime and worldvolume representations, their quantization proceeds analogously:
the canonical anti-commutation relations of the form \eqref{eq:canonical_quantization_D0-0B} lead to a Clifford algebra associated with $\mathfrak{so}(64) \supset \mathfrak{so}(32) \oplus \mathfrak{so}(32)$, and decomposing the corresponding Dirac spinor representation yields the cosets $(s,s)$, $(c,c)$, $(s,c)$, and $(c,s)$.
Imposing $O(1)$ gauge invariance further projects out $(s,c)$ and $(c,s)$.
The resulting spacetime gauge group is therefore again
\begin{align}\label{eq:gauge_group_0B-after-D1}
    \frac{Spin(32) \times Spin(32)}{\Z_2 \times \Z_2} \, ,
\end{align}
with the denominator being the central subgroup $\Z_2 \times \Z_2 = \{(1,1), (z_s, z_s), (z_c, z_c), (z_v, z_v)\}$.

\subsubsection{D5 quantization}

The last type of tachyon-free branes is given by charged D5-branes. The light spectrum can be found in table~\ref{tab:0B/Omega_D5_spectrum} from appendix~\ref{app:0B_branes}. Here, we include a simpler version of the table, showing the states that can potentially constrain the gauge groups for the special case that we are considering, $n_1=n_2=0$ and $n_3=n_4=32$, and taking only branes, i.e., $d_2=d_4=0$; see table~\ref{tab:0B/Omega_charged_D5_spectrum}.
\begin{table}[!ht]
\centering
\renewcommand{\arraystretch}{1.2}
\begin{tabular}{l|c|c|c}
Worldvolume state  & $\mathfrak{so}(4)_T$ & $\mathfrak{so}(32)\oplus\mathfrak{so}(32)$ & worldvolume $\mathfrak{sp}(d_1/2)\oplus\mathfrak{sp}(d_3/2)$ \\
\hline
Fermions $\psi + \chi$ & $\mathbf{1} \oplus {\bf 4}_v$ & $
     (\mathbf{32}, \mathbf{1}) \oplus  (\mathbf{1}, \mathbf{32})$ & $ (\mathbf{d_1}, \mathbf{1})$ 
\end{tabular}
\caption{Open-string light spectrum charged under the gauge symmetry from D5-branes in the orientifold of type 0B string theory with $n_1=n_2=0$, $n_3=n_4=32$, and with generic Chan--Paton charges $d_1$ and $d_3$ which both have to be even. We are not including antibranes with $d_2$ and $d_4$.}
\label{tab:0B/Omega_charged_D5_spectrum}
\end{table}

Once again, branes corresponding to $d_3$ impose no constraint. Those corresponding to $d_1$ yield fermions in the fundamentals of the two $\mathfrak{so}(32)$ factors and of $\mathfrak{sp}(d_1/2) \equiv \mathfrak{sp}(k)$.
To quantize these, we have to combine the anti-commutation relation \eqref{eq:canonical_quantization_D0-0B} for two $\mathfrak{so}$ factors with the quantization of 6d symplectic Majorana--Weyl fermions carrying $\mathfrak{so} \oplus \mathfrak{sp}$ symmetry indices.
Explicitly, we start from the fields $\psi_{i, \tilde\alpha}$ and $\chi_{a, \tilde\beta}$ in table \ref{tab:0B/Omega_charged_D5_spectrum}, where $i$ and $a$ are the vector indices of $\mathfrak{so}(n_3=32)$ and $\mathfrak{so}(n_4=32)$, respectively, and $\tilde\alpha, \tilde\beta=1,...,d_1 \equiv 2k$ the indices of the fundamental representation ${\bf d_1} = {\bf 2k}$ of $\mathfrak{sp}(d_1/2=k)$.\footnote{Again, we have ignored the $\mathfrak{so}(4)_T$ indices for the same reasons as in footnote \ref{footnote:ignore_soT-indices}.}

As explained in \cite{Larotonda:2024thv} and adapted to 5d symplectic Majorana fermions in section \ref{sec:0A_D4_quantization}, the symplectic Majorana--Weyl condition effectively reduces ${\bf 2k}$ to the vector representation ${\bf k}$ of $\mathfrak{so}(k)$, i.e., we end up having to quantize $\psi_{i, \alpha} \in ({\bf 32}, {\bf 1}, {\bf k})$ and $\chi_{a, \beta} \in ({\bf 1}, {\bf 32}, {\bf k})$ of $\mathfrak{so}(32) \oplus \mathfrak{so}(32) \oplus \mathfrak{so}(k)$.
As in section \ref{sec:0A_D4_quantization}, we can group the indices of each field into the vector index of $\mathfrak{so}(32 k)$, and then impose canonical anti-commutation relations
\begin{align}
    \{ \psi_I, \psi_J\} = \delta_{IJ} \, , \quad \{\chi_A, \chi_B\} = \delta_{AB} \, , \quad \{\psi_I, \chi_A\} = 0 \, ,
\end{align}
where $I \equiv (i, \alpha) = 1,..., 32k = n_3 \times \frac{d_1}{2}$ and $A \equiv (a, \beta) = 1,..., 32k = n_4 \times \frac{d_1}{2}$.
Analogously to the discussions after \eqref{eq:quantization_two_fermions} and \eqref{eq:canonical_quantization_D0-0B}, this produces the Clifford algebra $Cl(32k+32k)$, and thus the (co-)spinor representations ${\bf S}_{64k} \oplus {\bf C}_{64k}$.

To infer the representations under the physical symmetries, we first decompose according to the special embedding $\mathfrak{so}(64k) \supset \mathfrak{so}(64) \oplus \mathfrak{so}(k)$, because this allows us to directly apply invariance under the worldvolume $\mathfrak{so}(k) \subset \mathfrak{sp}(k) = \mathfrak{sp}(d_1/2)$ gauge symmetry.
By the same reasoning as in \cite[Section 3.1]{Larotonda:2024thv} and section \ref{sec:0A_D4_quantization}, gauge invariance leaves only the trivial coset $o$ of $\mathfrak{so}(64)$.
Under the subsequent branching of $\mathfrak{so}(64) \supset \mathfrak{so}(32) \oplus \mathfrak{so}(32)$ into the actual spacetime gauge algebra, this coset decomposes into $(o,o)$ and $(v,v)$, both of which are invariant under the $\Z_2 \times \Z_2$ in the denominator of \eqref{eq:gauge_group_0B-after-D1}.
That is, the D5-brane excitations are compatible with the gauge group derived from D0- and D1-branes.

\subsubsection{The gauge group}

In summary, we see that the spectrum on the stable branes of 0B$/\Omega$ constrains the spacetime gauge group to be
\begin{align}\label{eq:gauge_group_0B}
    \frac{Spin(32) \times Spin(32)}{\Z_2\times \Z_2} \, ,
\end{align}
where the $\Z_2\times \Z_2$ is $\{(1,1), (z_s, z_s),(z_c,z_c), (z_v, z_v)\} \subset {\cal Z}(Spin(32)^2)$.

It is worth highlighting again that the quotient by the $\Z_2^{(v)} := \{(1,1), (z_v, z_v)\}$ subgroup originates from the canonical quantization procedure of fermions in the $({\bf 32},{\bf 1}) \oplus ({\bf 1}, {\bf 32})$ representations of $\mathfrak{so}(32) \oplus \mathfrak{so}(32)$, which produces Fock states naturally living in the weight lattice of $Spin(64) \supset [Spin(32) \times Spin(32)]/\Z_2^{(v)}$.
Meanwhile, the quotient by $(z_s, z_s)$ (or equivalently $(z_c, z_c)$) is implemented by the $O(1) = \Z_2$ gauge symmetry on D0- and D1-branes.

\section{Testing the Bergman--Gaberdiel duality}\label{sec:BG_group}

With a focus on the involved gauge groups, we now turn to the duality proposed in~\cite{Bergman:1997rf} between the 0B$/\Omega$ theory discussed above and bosonic string theory in $D=26$ compactified on a $T^{16}$ at a special $\mathfrak{so}(32)$ enhancement point (see appendix \ref{app:bosonic_on_T16} for details).
The duality is motivated by a suggestive partial matching of the light spectrum and by the possibility of identifying the fundamental bosonic string in the spectrum of the type 0 orientifold, as we now review.

Take type 0B$/\Omega$ with $n_1=n_2=0$ and $n_3=n_4=32$ in~\eqref{eq:type0B_orientifold_open_sector}. In this case, there are no massless fermions (see~\cite[Figure 2]{Altavista:2026evd} for a pictorial explanation), while there are open-string tachyons in the bifundamental  $\left(\mathbf{32},\mathbf{32}\right)$ of $\mathfrak{so}(32)\oplus \mathfrak{so}(32)$. In addition, there is the closed-string sector with a tachyon singlet, the graviton and the dilaton, gauge fields, and two R-R 2-forms $C_2^\pm$.

Consider now bosonic string theory on the $T^{16}$ in question. In the light-cone gauge, there are 8 left-moving bosonic oscillators $\alpha^\mu_n$ and, by fermionizing the 16 internal coordinates, 32 left-moving fermionic oscillators $\psi^{a=1,\dots, 32}_r$, where half-integer $r$ yields massless states, and the corresponding right-moving fields $\tilde\alpha^\mu_n$ and $\tilde\psi_ r^a$. The light spectrum contains the bosonic string theory tachyon, the graviton, the dilaton, the Kalb--Ramond field $B_2$, gauge bosons in the gauge algebra $\mathfrak{so}(32)_L\oplus \mathfrak{so}(32)_R$ coming from the operators $\psi^a_{-\frac{1}{2}}\psi^b_{-\frac{1}{2}}\widetilde\alpha^\mu_{-1}$ and from those with left and right exchanged, tachyons in the $\left(\mathbf{32},\mathbf{32}\right)$ of $\mathfrak{so}(32)\oplus \mathfrak{so}(32)$ coming from the operators $\psi^a_{-\frac{1}{2}}\widetilde\psi^b_{-\frac{1}{2}}$, and Narain scalars in the bi-adjoint of $\mathfrak{so}(32)\oplus \mathfrak{so}(32)$ coming from the operators $\psi^a_{-\frac{1}{2}}\psi^b_{-\frac{1}{2}}\widetilde\psi^c_{-\frac{1}{2}}\widetilde\psi^d_{-\frac{1}{2}}$; see appendix \ref{app:bosonic_on_T16} for more details.

The first observation is that the light spacetime spectrum does not fully match, although it nearly does: the differences are the two 2-forms in the type 0B orientifold, to be compared with a single 2-form in the bosonic theory, and the presence of additional scalars in the bi-adjoint in the bosonic theory. A solution to this mismatch was recently proposed in~\cite{Baykara:2026vdc}. For the 2-forms, the idea is that one of the two R-R forms of the orientifold is Higgsed by a condensate of one of the two types of D1-branes present in the theory, and the corresponding gauge symmetry is spontaneously broken. On the other hand, for the Narain scalars, the proposal of~\cite{Baykara:2026vdc} is that the dynamics of the system removes them from the light spectrum. In fact, a one-loop potential should develop for all these scalars~\cite{Ginsparg:1986wr}, driving the system to the maximal enhancement point\footnote{This had already been noted in~\cite{Bergman:1997rf}.}. Here, the would-be moduli are massive, and at strong coupling they are invisible from the point of view of the light spectrum of type 0B$/\Omega$.

In addition to the quasi-matching of the light spectrum and the above resolutions, the remarkable feature of the duality of~\cite{Bergman:1997rf} is the identification of the fundamental bosonic string with one of the two D1-branes. In fact, type 0B$/\Omega$ has two types of D1-branes, which we encountered in section~\ref{subsubsec:D1-in-0B}. The one with $d_1=0$ and $d_3\neq0$, associated with the Higgsed 2-form, has as its only massless open-string states the scalars describing the brane position. The one with $d_1\neq0$ and $d_3=0$, associated with the 2-form that remains massless, gives rise to the massless scalars denoting the position of the branes from D1-D1 modes, together with 32 left-moving and 32 right-moving fermions in the $\left(\mathbf{32},\mathbf{1}\right)$ and $\left(\mathbf{1},\mathbf{32}\right)$ of $\mathfrak{so}(32)\oplus \mathfrak{so}(32)$. This is precisely the field content of the fundamental bosonic string that we introduced before, with the oscillators $\alpha^\mu_n$ and $\psi^a_r$. Hence, the proposal of~\cite{Bergman:1997rf} is that this D1-brane becomes the fundamental bosonic string (at the $\mathfrak{so}(32)$ point) at strong coupling.

With our brane scan, we can verify the duality also at the level of the gauge group.
On the bosonic string side, this is determined by the internal lattice $\Gamma$ which encode the fermionic states.
For this to be well-defined, one must appropriately sum over spin structures, resulting in a ``GSO projection'' acting on the $\psi^a_r$, which selects $\Gamma$ as an even self-dual sublattice of the weight lattice $D^*_{16} \oplus D^*_{16}$ of $\mathfrak{so}(32) \oplus \mathfrak{so}(32)$.
There are two possible such lattices, but only one has the tachyon in the $({\bf 32}, {\bf 32})$ representation (see appendix \ref{app:bosonic_on_T16}) needed to match the dual 0B$/\Omega$, which is the lattice~\cite{Baykara:2026vdc}
\begin{eq}
    \Gamma_{16,16}&=[D_{16}\oplus D_{16}] \\
    & = \{(p_L,p_R)\in D_{16}^*\oplus D_{16}^* \, | \, p_L-p_R\in D_{16} \}\mperiod
\end{eq}
The associated spacetime representations thus fall only into the classes $(o,o)$, $(v,v)$, $(s,s)$, and $(c,c)$.
This means that the gauge group is
\begin{eq}\label{eq:gauge_group_bosonic_lattice}
    \frac{Spin(32)\times Spin(32)}{\Z_2\times \Z_2}\mcomma
\end{eq}
where $\Z_2\times \Z_2 = \{(1,1), (z_s, z_s), (z_c, z_c), (z_v, z_v)\}$.
This perfectly matches the gauge group \eqref{eq:gauge_group_0B} of 0B$/\Omega$ enforced by the brane spectra, thus further validating the Bergman--Gaberdiel duality beyond the perturbative spectrum.

We close this section with a comparison of this duality with the supersymmetric type-I-heterotic duality.
As reviewed in section \ref{sec:particle_vs_string}, the heterotic worldsheet has similar consistency conditions that project the weight lattice of $\mathfrak{so}(32)$ onto the $\Z_2^{(s)}$-invariant states, i.e., selecting the even self-dual character lattice $D_{16}^+$ of $Spin(32)/\Z_2$.
Under the supersymmetric duality, this projection is identified with the $O(1) \cong \Z_2$ gauge symmetry of the D0- and D1-branes of type I.
In the case of the bosonic $T^{16}$ compactification, the projection by $\Z_2 \times \Z_2$ has two generators, while the D0- and D1-branes on the dual orientifold side have the same gauge symmetry as in type I, namely $O(1) = \Z_2$.
As we have emphasized in section \ref{sec:Bergman-Gaberdiel}, the additional $\Z_2$ arises, in a more subtle fashion, from the quantization of the fermions in the reducible $({\bf 32}, {\bf 1}) \oplus ({\bf 1}, {\bf 32})$ representation under the spacetime gauge symmetry $\mathfrak{so}(32) \oplus \mathfrak{so}(32)$.
Interestingly, if we also include $m$ anti-D9-branes in type I (and, by tadpole cancellation, $32+m$ D9s), we have a (tachyonic) theory with an $\mathfrak{so}(32+m) \oplus \mathfrak{so}(m)$ gauge algebra, and the D1-brane also has the reducible fermion spectrum $({\bf 32+m}, {\bf 1}) \oplus ({\bf 1}, {\bf m})$.
The quantization of these would also have a ``built in'' $\Z_2$ quotient, associated with the branching $Spin(32+2m) \supset [Spin(32+m)\times Spin(m)]/\Z_2$ induced by the quantization procedure.
For $m\equiv 0\pmod{4}$, the $O(1)$ gauge symmetry provides a further $\Z_2$ projection and yields the gauge group $[Spin(32+m) \times Spin(m)]/(\Z_2 \times \Z_2)$; for $m\equiv 2\pmod{4}$, the generator of the $\Z_2$ induced by the quantization procedure is the square of the generator of the gauge $O(1)$---that is, $z_v=z_s^2$ when $m\equiv 2\pmod{4}$---and the gauge group is $[Spin(32+m) \times Spin(m)]/\Z_4$; for odd $m$, the $O(1)$-projection acts as the $\Z_2$ from the branching, leaving the gauge group $ [Spin(32+m)\times Spin(m)]/\Z_2$. Note that in all cases, the quotient is by the diagonal center of the two factors, and leads to a self-dual character lattice.

\section{Discussion}\label{sec:discussion}

We have found the global structure of spacetime gauge symmetries in two orientifolds of non-supersymmetric type 0 strings. 
To fully capture the gauge group topology, the quantization of brane states is crucial. Indeed, these provide additional degrees of freedom that constrain the structure of the gauge groups \cite{Witten:1998cd,Witten:2023snr,Larotonda:2024thv} and refine the naive topology that would arise by taking into account only the perturbative excitations.

We have then put the results for type 0B$/\Omega$ to use by comparing with the gauge group of the bosonic string compactified on $T^{16}$ at an $\mathfrak{so}(32) \oplus \mathfrak{so}(32)$ enhancement point.
Our gauge-group test supports the Bergman--Gaberdiel duality. 
Specifically, the non-perturbative brane states in the 0B$/\Omega$ duality frame precisely constrain the gauge group structure to be the same as that of the bosonic string on the special $T^{16}$ lattice.

It is worthwhile to comment on a recent consideration of another possible duality among non-supersymmetric strings. In~\cite{Larotonda:2024thv}, an analogous brane scan for the Sugimoto model~\cite{Sugimoto:1999tx} led to the argument that the global form of the gauge group for this string theory is $Sp(16)/\Z_2$.\footnote{This further led to a bottom-up analysis pointing towards String Universality \cite{Kumar:2009us, Adams:2010zy, Kim:2019vuc, Kim:2019ths, Cvetic:2020kuw, Montero:2020icj, Hamada:2021bbz, Bedroya:2021fbu, Hamada:2023zol} in the non-supersymmetric landscape.} 
In the same work, the intuition from the Montonen--Olive duality led to the conjecture that the Sugimoto string may be dual to one of the non-supersymmetric heterotic theories---the one with gauge group $Spin(32)$.
This relied on an agreement entirely at the level of the lattices, whereby the heterotic theory is actually realized on the weight lattice of $Spin(33)$ (see \cite{Fraiman:2023cpa}), which is the Langlands dual of $Sp(16)/\Z_2$.
However, if we believe that this duality is also identifying a solitonic string as a fundamental string in a dual frame, then our initial discussion points to a different picture: the gauge group must be preserved under string dualities, which raises the question of which theory could be dual to the Sugimoto model. A naive counting of central charges is inconsistent with a current-algebra realization of $\mathfrak{sp}(16)$ at level one, which would challenge a perturbative heterotic string interpretation.

Note that for the type I-heterotic duality, the gauge group $Spin(32)/\Z_2$ happens to be GNO-self-dual. However, this is a mere coincidence, given that $D_{16}$ is simply-laced, so the algebra is Langlands-self-dual, and the relevant global form $Spin(32)/\Z_2$ has a unimodular character lattice $D_{16}^+$. Neither property is generic beyond the supersymmetric landscape---e.g., the type 0A$/\Omega$ model for which we found the gauge group $[Spin(16) \times Spin(16)]/\Z_2$ has a non-unimodular character lattice---so one cannot learn a general lesson about string dualities from this.

Transcending the primary goal of this work, several related directions remain open. Both type 0 orientifolds that we considered admit Sugimoto-like variants with symplectic-like projections instead of orthogonal ones. For these, an analogous quantization of brane states would yield the global topologies of the gauge groups. To achieve this, one should apply our procedures to the tachyon-free branes of those variants. Moreover, a bottom-up anomaly analysis as in~\cite{Larotonda:2024thv} would provide a further test of non-supersymmetric String Universality and shed light on the landscape of gravitational effective field theories without supersymmetry.  

Obtaining the global form of the gauge groups in all these cases would be interesting, especially given that no M-theory embeddings are known for them. Recently, there has been renewed interest in obtaining non-supersymmetric string theories from M-theory~\cite{Baykara:2026gem,Altavista:2026evd,Baykara:2026vdc,Altavista:2026brr,Dasgupta:2026maq,Basile:2026trt,Kamal:2026msr}.
In this context, the exotic gauge groups and the possible dualities in the non-supersymmetric landscape are a critical testing ground for the proposals. The primary example of this is the $Sp(16)/\Z_2$ Sugimoto model, but also the 0A$/\Omega$ orientifold we studied here provides an interesting case. These string theories do not have an M-theory embedding in the proposals of the above works, and obtaining the non-simply laced groups in ten dimensions is a key challenge that still evades our understanding of dualities.

From a more general perspective, constraining non-perturbative string dualities is our path to go beyond the protecting principle of supersymmetry in quantum gravity. In the absence of dynamical protection, topological mechanisms are our trusted guiding principles, and the matching of gauge groups that we have explored in this paper can be a window into the consistency of non-supersymmetric quantum gravity.

\section*{Acknowledgements}

We thank Matilda Delgado, Emilian Dudas, Ben Heidenreich, Justin Kaidi, H\'{e}ctor Parra de Freitas, Augusto Sagnotti, Shigeki Sugimoto, Yuji Tachikawa, and Cumrun Vafa for useful discussions.
We are further grateful to Emilian Dudas, H\'{e}ctor Parra de Freitas, and Cumrun Vafa for comments on an earlier version of the manuscript.
B.F.~is supported by a Juan de la Cierva contract (JDC2023-050850-I) from Spain’s Ministry of Science, Innovation and Universities. S.R.~is supported by the ERC Starting Grant QGuide101042568 - StG 2021.~The work of B.F.~and S.R.~is supported by the grants CEX2020-001007-S, PID2021-123017NB-I00, and PID2024-156043NB-I00, funded by MCIN\slash AEI\slash10.13039\slash501100011033, and ERDF A way of making Europe. This work was performed in part at the Aspen Center for Physics, which is supported by a grant from the Simons Foundation (1161654, Troyer). V.L.~would like to thank Kavli IPMU for the hospitality during the early stages of this work.

\appendix

\section{Branes of the type 0A orientifold}\label{app:0A_branes}

In this and in the next appendix, we use the following convention for the various representations: the Young tableaux $\text{\tiny\yng(1)}$ denote the vector representations for $\mathfrak{so}(n)$ algebras and the fundamental representations for $\mathfrak{sp}(n)$ or $\mathfrak{u}(n)$ algebras. The symbols $\text{\tiny\yng(2)}$ and $\text{\tiny\yng(1,1)}$ are used for the symmetric and antisymmetric tensor products of vector representations of $\mathfrak{so}(n)$ or of fundamental representations of $\mathfrak{sp}(n)$ or $\mathfrak{u}(n)$.

In this appendix, we derive the worldvolume spectra of the branes of the type 0A orientifold using the open-string amplitudes from~\cite{Dudas:2001wd}.

The combination that contains the spectra of interest is the sum of the annulus and the M\"{o}bius strip amplitudes,
\begin{equation}
    \mathcal{A}+\mathcal{M}\,.
\end{equation}
The spectra can be extracted directly from the one-loop amplitudes written in terms of $\mathfrak{so}(n)$ characters, using the fact that the character $O_p$ starts with a scalar, the string vacuum, $V_p$ starts with a vector, and $S_p$ and $C_p$ start with spinors of opposite chiralities. Then, the coefficients multiplying the characters are the degeneracy of a given state, and contain the information of which representation of the gauge group it belongs to.

For example, take the spacetime-filling D9-branes in the 0A$/\Omega$ model. The annulus amplitude reads
\begin{equation}\label{eq:0A_D9_D9_annulus}
    \mathcal{A}_{99} = \frac{n_1^2 + n_2^2}{2} \ (O_{8}+ V_{8}) - n_1 n_2 (S_{8}+ C_{8})\,,
\end{equation}
and corresponds to string states stretching between two D9-branes. The states coming from the D9-O9 sector are given by the M\"{o}bius amplitude,
\begin{equation}\label{eq:0A_D9_O9_moebius}
    \mathcal{M}_9 = -\, \frac{n_1+n_2}{2} \, {\hat V}_8 + \frac{n_1 - n_2}{2} \,{\hat O}_8\,.
\end{equation}
From the above amplitudes, we can extract that the spacetime gauge boson transforms in the adjoint representation of the gauge group. In fact, the degeneracy of the vector character $V_8$ is
\begin{equation}\label{eq:0A_spacetime_vector}
    \frac{n_1^2 + n_2^2}{2} - \frac{n_1+n_2}{2} = \frac{n_1(n_1-1)}{2} + \frac{n_2(n_2-1)}{2}\,,
\end{equation}
which corresponds to antisymmetric representations. This implies that the gauge symmetry is $\mathfrak{so}(n_1) \times \mathfrak{so}(n_2)$.
Chiral fermions from $S_8$ and $C_8$ transform in the bifundamental of the spacetime gauge algebra, having degeneracy $n_1n_2$.
The $O_8$ character also signals the presence of two open-string tachyons, with degeneracies
\begin{equation}\label{eq:spacetime_tachyon}
    \frac{n_1^2 + n_2^2}{2} + \frac{n_1 - n_2}{2}  =\frac{n_1(n_1+1)}{2} + \frac{n_2(n_2-1)}{2}\,.
\end{equation}
One transforms in the symmetric representation of the first gauge factor, while being a singlet of the second, and the other is a singlet of the first and transforms in the antisymmetric (adjoint) representation of the second $\mathfrak{so}(n_2)$.
We summarize the ten-dimensional spectrum arising from the open sector in table~\ref{tab:0A_open_spectrum}.
\begin{table}[!ht]
\centering
\renewcommand{\arraystretch}{1.2}
\begin{tabular}{l|c|c|c}
State & character & $\mathfrak{so}(n_1)$ & $\mathfrak{so}(n_2)$ \\
\hline
Gauge vector & $V_8$ & $\mathbf{Adj}$ & $\mathbf{1}$ \\
Gauge vector & $V_8$ & $\mathbf{1}$ & $\mathbf{Adj}$ \\
Fermion & $S_8$ & \tiny\yng(1) & \tiny\yng(1) \\
Fermion & $C_8$ & \tiny\yng(1) & \tiny\yng(1) \\
Tachyon & $O_8$ & \tiny\yng(2) & $\mathbf{1}$ \\
Tachyon & $O_8$ & $\mathbf{1}$ & $\mathbf{Adj}$
\end{tabular}
\caption{Open-string light spectrum for D9-branes in the type 0A$/\Omega$.}
\label{tab:0A_open_spectrum}
\end{table}

After this preliminary discussion, let us now analyze D-branes. In general, a D$p$-brane breaks the ten-dimensional Lorentz group into a longitudinal and transverse component:
\begin{equation}
    \mathfrak{so}(1,9) \longrightarrow \mathfrak{so}(1,p)_L \times \mathfrak{so}(9-p)_T\,.
\end{equation}
Therefore, in brane amplitudes, we have to distinguish between the characters of the worldvolume Lorentz group, e.g., $V_{p-1}$, and those associated with the transverse directions, e.g., $V_{9-p}$.

Type 0A$/\Omega$ contains charged D$p$-branes with $p=0,2,4,6,8$. These have odd dimensions, and therefore the spinorial characters of the would-be different chiralities, $S$ and $C$, can be replaced by a single character related to the Dirac spinor:
\begin{equation}
    S' = \frac{1}{2}\left(S+C\right)\,.
\end{equation}
In this context, one finds two different types of branes with a given dimension. The first type is denoted by D$p_1$-branes, together with their antibranes; the degeneracies of these branes and antibranes are denoted by $d_1$ and $d_2$. The second type comes from a bound state of a brane and an antibrane in the parent theory, and $m$ such branes generate a $U(m)$ gauge factor. 
In addition to these branes, type 0A$/\Omega$ has uncharged D$p$-branes\footnote{Here and in the following, we will not consider D-instantons.} with $p=1,3,5,7$, some of which have a K-theory $\Z_2$ charge. 

We now discuss the various branes in detail.

\subsection{Charged branes in the 0A orientifold}

For $p=0,2,4,6,8$, the D$p$-D$p$ annulus amplitude is
\begin{equation}\label{eq:0A_Dp_Dp_annulus_p_even}
\begin{split}
    \mathcal{A}_{pp} =& \left(\frac{d_1^2+d_2^2}{2}+m {\bar m}\right)
    \left(O_{p-1}V_{9-p}+ V_{p-1}O_{9-p}\right) + \\ &+ \left(d_1 d_2+\frac{m^2+\bar m^2}{2}\right) \left(O_{p-1}O_{9-p}+ V_{p-1} V_{9-p}\right) 
    - (d_1+d_2)\left(m+\bar m\right) S'_{p-1} S'_{9-p}\,.
 \end{split}
\end{equation}

\subsubsection{\texorpdfstring{$p=0,4,8$}{p=0,4,8}}

Let us first focus on the cases $p=0$, $4$, and $8$. The M\"{o}bius amplitude is
\begin{equation}\label{eq:0A_Dp_O9_moebius_p048}
    \mathcal M_p = \frac{d_1+d_2}{2} \ \epsilon \ ({\hat O}_{p-1}{\hat V}_{9-p}- {\hat V}_{p-1}{\hat O}_{9-p}) -\, \frac{m+\bar m}{2} \ \epsilon \ ({\hat O}_{p-1}{\hat O}_{9-p}+ {\hat V}_{p-1}{\hat V}_{9-p})\,,
\end{equation}
where $\epsilon = +1$ for $p=0,8$ and $\epsilon = -1$ for $p=4$.

The worldvolume gauge boson arises from the $V_{p-1}O_{9-p}$ character, whose multiplicity is
\begin{equation}
    \frac{d_1^2+d_2^2}{2}+m {\bar m} - \frac{d_1+d_2}{2} \ \epsilon = \frac{d_1(d_1 - \epsilon)}{2} +  \frac{d_2(d_2 - \epsilon)}{2} + m {\bar m}\,.
\end{equation}
Therefore, if $p=0,8$, the adjoint representation corresponds to the antisymmetric for the first two factors, fixing the gauge structure to be $\mathfrak{so}(d_1)\oplus \mathfrak{so}(d_2) \oplus \mathfrak{u}(m)$. Note that $m\bar{m}$ is the multiplicity of the adjoint of $U(m)$.
For $p=4$, we find a symmetric representation for the first two gauge bosons, implying that the gauge group is $\mathfrak{sp}(d_1/2)\oplus \mathfrak{sp}(d_2/2) \oplus \mathfrak{u}(m)$.

The $O_{p-1}V_{9-p}$ character corresponds to worldvolume bosons that transform as a vector of the transverse Lorentz group: they are the position of the brane. These come in a symmetric representation for the first two factors when $p=0,8$ and in an antisymmetric one when $p=4$. With respect to $\mathfrak{u}(m)$, we find a boson in the adjoint.
The first state from the combination $V_{p-1} V_{9-p}$ is massive, and we will not consider it.
From the combination $O_{p-1} O_{9-p}$, the worldvolume theory contains tachyons with degeneracy
\begin{equation}
    \left(d_1 d_2+\frac{m^2+\bar m^2}{2}\right) - \epsilon \frac{m+\bar m}{2}\,,
\end{equation}
thus yielding a bifundamental representation of $\mathfrak{so}(d_1)\oplus \mathfrak{so}(d_2)$, an antisymmetric representation with its conjugate with respect to $\mathfrak{u}(m)$ for $p=0,8$, and a bifundamental representation of $\mathfrak{sp}(d_1/2)\oplus \mathfrak{sp}(d_2/2)$ and a symmetric representation of $\mathfrak{u}(m)$ with its conjugate for $p=4$.
Finally, the worldvolume spinors in $S_{p-1}'$ are also spinors with respect to the transverse Lorentz group and transform in the bifundamental of the first or second factor and of the $\mathfrak{u}(m)$ $+$ conjugate part of the gauge algebra.

There are additional states coming from the D$p$-D9 sector, whose (longitudinal) annulus amplitude is
\begin{equation}\label{eq:0A_Dp_D9_annulus}
\begin{split}
    {\cal A}_{p9} =& \bigl[n_1(d_1+d_2) + n_1 \bar{m} + n_2 m\bigr](O_{p-1} + V_{p-1}) S'_{9-p} \\
    &- \bigl[n_2(d_1+d_2) + n_1 m + n_2 \bar{m} \bigr]S'_{p-1} (O_{9-p} + V_{9-p})\,.
\end{split}
\end{equation}
While the sectors that mix $S$ and $V$ are massive, there are several fermions arising from $S'_{p-1} O_{9-p}$. The first two are in the bifundamental of $\mathfrak{so}(n_2)$ and of the first and second factors of the worldvolume gauge algebra, respectively. In addition, there is a fermion in the bifundamental of $\mathfrak{so}(n_1)\times \mathfrak{u}(m)$ and in the fundamental-antifundamental of $\mathfrak{so}(n_2)\times \mathfrak{u}(m)$. Finally, we have other scalars in the same representations exchanging the first and second factors. In general these states are massive for $p<5$, massless for $p=5$ and tachyonic for $p>5$. We summarize all these results in table~\ref{tab:0A_D4_spectrum} for $p=4$ and in table~\ref{tab:0A_D0_D8_spectrum} for $p=0,8$.

\begin{table}[!ht]
\centering
\textbf{D4}
\renewcommand{\arraystretch}{1.2}
\centerresize{}{
\begin{tabular}{l|c|c|c}
Worldvolume state  & $\mathfrak{so}(5)_T$ & $\mathfrak{so}(n_1) \oplus \mathfrak{so}(n_2)$  & $\mathfrak{sp}(d_1/2)\oplus \mathfrak{sp}(d_2/2) \oplus \mathfrak{u}(m)$ \\
\hline
Gauge vector  & $\mathbf{1}$ & $(\mathbf{1}, \mathbf{1})$  & $(\mathbf{Adj},  \mathbf{1}, \mathbf{1}) \oplus(\mathbf{1}, \mathbf{Adj}, \mathbf{1}) \oplus (\mathbf{1},  \mathbf{1}, \mathbf{Adj})$ \\
Scalar & $\mathbf{5}_{v}$ & $(\mathbf{1}, \mathbf{1})$ & $\left(\text{\tiny{\yng(1,1)}},  \mathbf{1}, \mathbf{1}\right) \oplus\left(\mathbf{1}, \text{\tiny{\yng(1,1)}},\mathbf{1}\right)\oplus (\mathbf{1},  \mathbf{1}, \mathbf{Adj})$\\
Tachyon & $\mathbf{1}$ & $(\mathbf{1}, \mathbf{1})$ & $ \left(\text{\tiny{\yng(1)}},\text{\tiny{\yng(1)}}, \mathbf{1} \right) \oplus \left(\mathbf{1},  \mathbf{1}, \text{\tiny{\yng(2)}}\right)\oplus \left(\mathbf{1},  \mathbf{1}, \ybar{\text{\tiny{\yng(2)}}}\right)$ \\
Fermion & $\mathbf{4}_{s'}$ & $(\mathbf{1}, \mathbf{1})$ & $\left(\text{\tiny{\yng(1)}}, \mathbf{1},\text{\tiny{\yng(1)}} \right) \oplus \left(\mathbf{1},\text{\tiny{\yng(1)}},\text{\tiny{\yng(1)}}\right) \oplus \left(\text{\tiny{\yng(1)}}, \mathbf{1},\ybar{\text{\tiny{\yng(1)}}} \right) \oplus \left(\mathbf{1},\text{\tiny{\yng(1)}},\ybar{\text{\tiny{\yng(1)}}} \right)$ \\
Fermion & $\mathbf{1}$ & $\left(\mathbf{1},\text{\tiny\yng(1)}\right)$ & $\left(\text{\tiny\yng(1)}, \mathbf{1},\mathbf{1}\right) \oplus \left( \mathbf{1},\text{\tiny{\yng(1)}},\mathbf{1}\right)\oplus\left(\mathbf{1},\mathbf{1},\ybar{\text{\tiny\yng(1)}}\right)$ \\
Fermion & $\mathbf{1}$ & $\left(\text{\tiny\yng(1)},\mathbf{1}\right)$ & $\left(\mathbf{1},\mathbf{1},\text{\tiny\yng(1)}\right)$ 
\end{tabular}}
\caption{Open-string light spectrum for D4-branes in the type 0A orientifold. The bar over a Young tableaux stands for the conjugate representation. $\mathbf{4}_{s'}$ is the irreducible Dirac spinor of $\mathfrak{so}(5)$ and $\mathbf{5}_{v}$ is the vector.}
\label{tab:0A_D4_spectrum}
\end{table}
\begin{table}[!ht]
\centering
\renewcommand{\arraystretch}{1.2}
\textbf{D$p$ with $p=0,8$}
\centerresize{}{
\begin{tabular}{l|c|c|c}
Worldvolume state  & $\mathfrak{so}(9-p)_T$ & $\mathfrak{so}(n_1) \oplus \mathfrak{so}(n_2)$  & $\mathfrak{so}(d_1) \oplus \mathfrak{so}(d_2) \oplus \mathfrak{u}(m)$ \\
\hline
Gauge vector  & $\mathbf{1}$ & $(\mathbf{1}, \mathbf{1})$  & $(\mathbf{Adj},  \mathbf{1}, \mathbf{1}) \oplus(\mathbf{1}, \mathbf{Adj}, \mathbf{1}) \oplus (\mathbf{1},  \mathbf{1}, \mathbf{Adj})$ \\
Scalar & Vector & $(\mathbf{1}, \mathbf{1})$ & $\left(\text{\tiny{\yng(2)}},  \mathbf{1}, \mathbf{1}\right) \oplus\left(\mathbf{1}, \text{\tiny{\yng(2)}},\mathbf{1}\right)\oplus (\mathbf{1},  \mathbf{1}, \mathbf{Adj})$\\
Tachyon & $\mathbf{1}$ & $(\mathbf{1}, \mathbf{1})$ & $ \left(\text{\tiny{\yng(1)}},\text{\tiny{\yng(1)}}, \mathbf{1} \right) \oplus \left(\mathbf{1},  \mathbf{1}, \text{\tiny{\yng(1,1)}}\right)\oplus \left(\mathbf{1},  \mathbf{1}, \ybar{\text{\tiny{\yng(1,1)}}}\right)$ \\
Fermion & Spinor & $(\mathbf{1}, \mathbf{1})$ & $\left(\text{\tiny{\yng(1)}}, \mathbf{1},\text{\tiny{\yng(1)}}\right) \oplus \left(\mathbf{1},\text{\tiny{\yng(1)}},\text{\tiny{\yng(1)}}\right) \oplus \left(\text{\tiny{\yng(1)}}, \mathbf{1},\ybar{\text{\tiny{\yng(1)}}}\right) \oplus \left(\mathbf{1},\text{\tiny{\yng(1)}},\ybar{\text{\tiny{\yng(1)}}}\right)$ \\
Fermion & $\mathbf{1}$ & $\left(\mathbf{1},\text{\tiny\yng(1)}\right)$ & $\left(\text{\tiny\yng(1)}, \mathbf{1},\mathbf{1}\right) \oplus \left( \mathbf{1},\text{\tiny{\yng(1)}},\mathbf{1}\right)\oplus\left(\mathbf{1},\mathbf{1},\ybar{\text{\tiny\yng(1)}}\right)$ \\
Fermion & $\mathbf{1}$ & $\left(\text{\tiny\yng(1)},\mathbf{1}\right)$ & $\left(\mathbf{1},\mathbf{1},\text{\tiny\yng(1)}\right)$ \\
Scalar & Spinor & $\left(\text{\tiny\yng(1)}, \mathbf{1}\right)$ & $\left(\text{\tiny\yng(1)}, \mathbf{1},\mathbf{1}\right) \oplus \left( \mathbf{1},\text{\tiny{\yng(1)}},\mathbf{1}\right) \oplus \left(\mathbf{1},\mathbf{1},\ybar{\text{\tiny\yng(1)}}\right)$ \\
Scalar & Spinor & $\left(\mathbf{1},\text{\tiny\yng(1)}\right)$ & $\left(\mathbf{1},\mathbf{1},\text{\tiny\yng(1)}\right)$ \\
\end{tabular}}
\caption{Open-string light spectrum for D0- and D8-branes in the type 0A orientifold. The last two rows give massive scalars for the D0-brane while yielding tachyonic states for the D8-brane.}
\label{tab:0A_D0_D8_spectrum}
\end{table}

\subsubsection{\texorpdfstring{$p=2,6$}{p=2,6}}

Let us perform the analogous computation for $p=2$ and $6$. The D$p$-D$p$ and D$p$-D9 annulus amplitudes are the same as in~\eqref{eq:0A_Dp_Dp_annulus_p_even} and~\eqref{eq:0A_Dp_D9_annulus}. The M\"{o}bius amplitude reads
\begin{equation}\label{eq:0A_Dp_O9_moebius_p26}
\begin{split}
    \mathcal{M}_p =& \frac{d_1+d_2}{2} \,\epsilon\,({\hat O}_{p-1}{\hat V}_{9-p}- {\hat V}_{p-1}{\hat O}_{9-p})\\
 &-\,\frac{m+\bar m}{2} \,\epsilon\, ({\hat O}_{p-1}{\hat O}_{9-p}+ {\hat V}_{p-1}{\hat V}_{9-p})\,,
\end{split}
\end{equation}
with $\epsilon = 1$ for $p=2$ and $\epsilon = -1$ for $p=6$.

The computation proceeds as for the other D$p$-branes: the gauge vectors come in the antisymmetric for $p=2$ and in the symmetric for $p=6$, forcing the gauge structure to be $\mathfrak{so}(d_1) \oplus \mathfrak{so}(d_2) \oplus \mathfrak{u}(m)$ in the former case and $\mathfrak{sp}(d_1/2) \oplus \mathfrak{sp}(d_2/2) \oplus \mathfrak{u}(m)$ in the latter case. The other degrees of freedom are analogous to the ones computed in the previous section, with the D2 mapping to the D0 and D8 and the D6 mapping to the D4. We show the explicit results in tables~\ref{tab:0A_D2_spectrum} and~\ref{tab:0A_D6_spectrum}.

\begin{table}[!ht]
\centering
\renewcommand{\arraystretch}{1.2}
\textbf{D2}
\centerresize{}{
\begin{tabular}{l|c|c|c}
Worldvolume state  & $\mathfrak{so}(7)_T$ & $\mathfrak{so}(n_1) \oplus \mathfrak{so}(n_2)$  & $\mathfrak{so}(d_1) \oplus \mathfrak{so}(d_2) \oplus \mathfrak{u}(m)$ \\
\hline
Gauge vector  & $\mathbf{1}$ & $(\mathbf{1}, \mathbf{1})$  & $(\mathbf{Adj},  \mathbf{1}, \mathbf{1}) \oplus(\mathbf{1}, \mathbf{Adj}, \mathbf{1}) \oplus (\mathbf{1},  \mathbf{1}, \mathbf{Adj})$ \\
Scalar & $\mathbf{7}_{v}$ & $(\mathbf{1}, \mathbf{1})$ & $\left(\text{\tiny{\yng(2)}},  \mathbf{1}, \mathbf{1}\right) \oplus\left(\mathbf{1}, \text{\tiny{\yng(2)}},\mathbf{1}\right)\oplus (\mathbf{1},  \mathbf{1}, \mathbf{Adj})$\\
Tachyon & $\mathbf{1}$ & $(\mathbf{1}, \mathbf{1})$ & $ \left(\text{\tiny{\yng(1)}},\text{\tiny{\yng(1)}}, \mathbf{1} \right) \oplus \left(\mathbf{1},  \mathbf{1}, \text{\tiny{\yng(1,1)}}\right)\oplus \left(\mathbf{1},  \mathbf{1}, \ybar{\text{\tiny{\yng(1,1)}}}\right)$ \\
Fermion & $\mathbf{8}_{s'}$ & $(\mathbf{1}, \mathbf{1})$ & $\left(\text{\tiny{\yng(1)}}, \mathbf{1},\text{\tiny{\yng(1)}} \right) \oplus \left(\mathbf{1},\text{\tiny{\yng(1)}},\text{\tiny{\yng(1)}}\right) \oplus \left(\text{\tiny{\yng(1)}}, \mathbf{1},\ybar{\text{\tiny{\yng(1)}}}\right) \oplus \left(\mathbf{1},\text{\tiny{\yng(1)}},\ybar{\text{\tiny{\yng(1)}}} \right)$ \\
Fermion & $\mathbf{1}$ & $\left(\mathbf{1},\text{\tiny\yng(1)}\right)$ & $\left(\text{\tiny\yng(1)}, \mathbf{1},\mathbf{1}\right) \oplus \left( \mathbf{1},\text{\tiny{\yng(1)}},\mathbf{1}\right)\oplus\left(\mathbf{1},\mathbf{1},\ybar{\text{\tiny\yng(1)}}\right)$ \\
Fermion & $\mathbf{1}$ & $\left(\text{\tiny\yng(1)},\mathbf{1}\right)$ & $\left(\mathbf{1},\mathbf{1},\text{\tiny\yng(1)}\right)$ 
\end{tabular}}
\caption{Open-string light spectrum for D2-branes in the type 0A orientifold. The bar over a Young tableaux stands for the conjugate representation, $\mathbf{8}_{s'}$ is the irreducible Dirac spinor of $\mathfrak{so}(7)$, and $\mathbf{7}_v$ is the vector.}
\label{tab:0A_D2_spectrum}
\end{table}
\begin{table}[!ht]
\centering
\textbf{D6}
\renewcommand{\arraystretch}{1.2}
\centerresize{}{
\begin{tabular}{l|c|c|c}
Worldvolume state  & $\mathfrak{so}(3)_T$ & $\mathfrak{so}(n_1) \oplus \mathfrak{so}(n_2)$  & $\mathfrak{sp}(d_1/2) \oplus \mathfrak{sp}(d_2/2) \oplus \mathfrak{u}(m)$ \\
\hline
Gauge vector  & $\mathbf{1}$ & $(\mathbf{1}, \mathbf{1})$  & $(\mathbf{Adj},  \mathbf{1}, \mathbf{1}) \oplus(\mathbf{1}, \mathbf{Adj}, \mathbf{1}) \oplus (\mathbf{1},  \mathbf{1}, \mathbf{Adj})$ \\
Scalar & $\mathbf{3}_{v}$ & $(\mathbf{1}, \mathbf{1})$ & $\left(\text{\tiny{\yng(1,1)}},  \mathbf{1}, \mathbf{1}\right) \oplus\left(\mathbf{1}, \text{\tiny{\yng(1,1)}},\mathbf{1}\right)\oplus (\mathbf{1},  \mathbf{1}, \mathbf{Adj})$\\
Tachyon & $\mathbf{1}$ & $(\mathbf{1}, \mathbf{1})$ & $ \left(\text{\tiny{\yng(1)}},\text{\tiny{\yng(1)}}, \mathbf{1} \right) \oplus \left(\mathbf{1},  \mathbf{1}, \text{\tiny{\yng(2)}}\right)\oplus \left(\mathbf{1},  \mathbf{1}, \ybar{\text{\tiny{\yng(2)}}}\right)$ \\
Fermion & $\mathbf{2}_{s'}$ & $(\mathbf{1}, \mathbf{1})$ & $\left(\text{\tiny{\yng(1)}}, \mathbf{1},\text{\tiny{\yng(1)}}\right) \oplus \left(\mathbf{1},\text{\tiny{\yng(1)}},\text{\tiny{\yng(1)}}\right) \oplus \left(\text{\tiny{\yng(1)}}, \mathbf{1},\ybar{\text{\tiny{\yng(1)}}} \right) \oplus \left(\mathbf{1},\text{\tiny{\yng(1)}},\ybar{\text{\tiny{\yng(1)}}}\right)$ \\
Fermion & $\mathbf{1}$ & $\left(\mathbf{1},\text{\tiny\yng(1)}\right)$ & $\left(\text{\tiny\yng(1)}, \mathbf{1},\mathbf{1}\right) \oplus \left( \mathbf{1},\text{\tiny{\yng(1)}},\mathbf{1}\right)\oplus\left(\mathbf{1},\mathbf{1},\ybar{\text{\tiny\yng(1)}}\right)$ \\
Fermion & $\mathbf{1}$ & $\left(\text{\tiny\yng(1)},\mathbf{1}\right)$ & $\left(\mathbf{1},\mathbf{1},\text{\tiny\yng(1)}\right)$ \\
Tachyon & $\mathbf{2}_{s'}$ & $\left(\text{\tiny\yng(1)}, \mathbf{1}\right)$ & $\left(\text{\tiny\yng(1)}, \mathbf{1},\mathbf{1}\right) \oplus \left( \mathbf{1},\text{\tiny{\yng(1)}},\mathbf{1}\right) \oplus \left(\mathbf{1},\mathbf{1},\ybar{\text{\tiny\yng(1)}}\right)$ \\
Tachyon & $\mathbf{2}_{s'}$ & $\left(\mathbf{1},\text{\tiny\yng(1)}\right)$ & $\left(\mathbf{1},\mathbf{1},\text{\tiny\yng(1)}\right)$ \\
\end{tabular}}
\caption{Open-string light spectrum for D6-branes in the type 0A orientifold. $\mathbf{2}_{s'}$ is the irreducible Dirac spinor of $\mathfrak{so}(3)$ and $\mathbf{3}_{v}$ is the vector.}
\label{tab:0A_D6_spectrum}
\end{table}

\subsection{Uncharged branes in the 0A orientifold}

In the type 0A$/\Omega$ orientifold, there are uncharged D$p$-branes with $p=1,3,5,7$.
The annulus amplitude for the D$p$-D$p$ sector is
\begin{equation}\label{eq:0A_Dp_Dp_annulus_p_odd}
    \mathcal{A}_{pp} = \frac{d_1^2 + d_2^2}{2} \ (O_{p-1}+ V_{p-1})(O_{9-p}+V_{9-p}) - \, d_1 d_2 \, (S_{p-1}+ C_{p-1}) (S_{9-p}+ C_{9-p})\,.
\end{equation}
The M\"{o}bius amplitude for strings stretching between D$p$-branes and O9-planes is
\begin{equation}\label{eq:0A_Dp_O9_moebius_p_odd}
\begin{split} 
    \mathcal{M}_p =&  \frac{d_1 - \epsilon' d_2}{2} \, \epsilon\,({\hat O}_{p-1}{\hat O}_{9-p} + {\hat V}_{p-1}{\hat V}_{9-p})  \\
    & + \frac{d_1 + \epsilon' d_2}{2} \, \epsilon \, ({\hat O}_{p-1}{\hat V}_{9-p} - {\hat V}_{p-1}{\hat O}_{9-p}) \,.
\end{split}
\end{equation}
The parameters $(\epsilon, \epsilon')$ are $(+1,+1)$ for the D1 branes, $(+1,-1)$ for the D3 branes, $(-1,+1)$ for the D5 branes, and $(-1,-1)$ for the D7 and D$(-1)$ branes.
Finally, the D$p$-D9 annulus amplitude for uncharged branes reads
\begin{equation}\label{eq:0A_Dp_D9_annulus_p_odd}
\begin{split}
    \mathcal{A}_{p9} =& (n_1d_1+n_2d_2) \ (O_{p-1} + V_{p-1}) (S_{9-p}+ C_{9-p})  \\ 
    &-\,(n_1d_2+n_2d_1) (S_{p-1}+C_{p-1}) (O_{9-p}+ V_{9-p})\,.
\end{split}
\end{equation}

Again, combining the annulus in~\eqref{eq:0A_Dp_Dp_annulus_p_odd} with the M\"{o}bius in~\eqref{eq:0A_Dp_O9_moebius_p_odd} and extracting the coefficient of the $V_{p-1}O_{9-p}$ combination, we can determine the worldvolume gauge structure. This turns out to be
\begin{equation}\label{eq:0A_gauge_vector_degeneracies_p_odd}
\begin{split}
    \frac{d_1^2 + d_2^2}{2} - \frac{d_1 + \epsilon' d_2}{2} \, \epsilon = \frac{d_1(d_1 - \epsilon)}{2} + \frac{d_2(d_2 - \epsilon \epsilon')}{2}\,,
\end{split}
\end{equation}
and the corresponding degeneracies and gauge structure for each uncharged brane are collected in table~\ref{tab:0A_gauge_structures_uncharged_branes}.
\begin{table}[!ht]
\centering
\renewcommand{\arraystretch}{1.4}
\centerresize{1,1}{
\begin{tabular}{c|c|c|c|c|c}
     & D1 & D3 & D5 & D7 & D$(-1)$ \\
    \hline
    $(\epsilon,\epsilon')$ & $(1,1)$ & $(1,-1)$ & $(-1,1)$ & $(-1,-1)$ & $(-1,-1)$ \\
    vector degeneracy & $\frac{d_1(d_1 - 1)}{2} + \frac{d_2(d_2 - 1)}{2}$ & $\frac{d_1(d_1 - 1)}{2} + \frac{d_2(d_2 + 1)}{2}$ & $\frac{d_1(d_1 + 1)}{2} + \frac{d_2(d_2 + 1)}{2}$ & $\frac{d_1(d_1 + 1)}{2} + \frac{d_2(d_2 - 1)}{2}$  & $\frac{d_1(d_1 + 1)}{2} + \frac{d_2(d_2 - 1)}{2}$ \\
    gauge structure & $\mathfrak{so}(d_1) \oplus \mathfrak{so}(d_2)$ & $\mathfrak{so}(d_1) \oplus \mathfrak{sp}(d_2/2)$ & $\mathfrak{sp}(d_1/2) \oplus \mathfrak{sp}(d_2/2)$ & $\mathfrak{sp}(d_1/2) \oplus \mathfrak{so}(d_2)$ & $\mathfrak{sp}(d_1/2) \oplus \mathfrak{so}(d_2)$
\end{tabular}}
\caption{Gauge structures of the uncharged D$p$-branes in type 0A$/\Omega$.}
\label{tab:0A_gauge_structures_uncharged_branes}
\end{table}

The tachyons arising from $O_{p-1}O_{9-p}$ come with a multiplicity of
\begin{equation}
    \frac{d_1^2 + d_2^2}{2} + \frac{d_1 - \epsilon' d_2}{2} \, \epsilon = \frac{d_1(d_1 + \epsilon)}{2} + \frac{d_2(d_2 - \epsilon \epsilon')}{2}\,.
\end{equation}
Thus, these states are always in the adjoint of the second factor of the gauge group, while they are in the symmetric representation when the first factor is $\mathfrak{so}(d_1)$ and in the antisymmetric representation when it is $\mathfrak{sp}(d_1/2)$.
An analogous argument holds for the coordinates of the branes, coming from the $O_{p-1}V_{9-p}$ sector. The prefactor is
\begin{equation}
    \frac{d_1^2 + d_2^2}{2} + \frac{d_1 + \epsilon' d_2}{2} \, \epsilon = \frac{d_1(d_1 + \epsilon)}{2} + \frac{d_2(d_2 + \epsilon \epsilon')}{2}\,,
\end{equation}
which stands for a symmetric representation each time the corresponding factor is $\mathfrak{so}(d_{1,2})$ and an antisymmetric representation when the corresponding factor is $\mathfrak{sp}(d_{1,2}/2)$.
The fermions in the annulus amplitude in~\eqref{eq:0A_Dp_Dp_annulus_p_odd} are non-chiral; indeed, the two chiralities $(S+C)$ are summed, and the fermions transform in the bifundamental representation of the gauge algebra. This is not surprising since the type 0A theory and its descendants are non-chiral, and only the sum of the spinorial characters appears in the amplitudes.
Two more fermions arise from the D$p$-D9 system from~\eqref{eq:0A_Dp_D9_annulus_p_odd}. They are singlets with respect to the transverse Lorentz group and transform in the bifundamental of the first spacetime and second worldvolume gauge algebra factors and of the second and first factors of the spacetime and worldvolume gauge groups, respectively.
Finally, additional bosons from $O_{p-1}(S_{9-p}+C_{9-p})$ appear in the bifundamental representations of the first factors of the spacetime and worldvolume algebras, as well as the second factors. Again, these have a positive mass for $p<5$, while being massless and tachyonic for $p=5$ and $p>5$, respectively.

The results of the above discussion are summarized in tables~\ref{tab:0A_D1_spectrum}, \ref{tab:0A_D3_spectrum}, \ref{tab:0A_D5_spectrum}, and~\ref{tab:0A_D7_spectrum}.

\begin{table}[H]
\centering
\textbf{D1}
\renewcommand{\arraystretch}{1.2}
\centerresize{0.85}{
\begin{tabular}{l|c|c|c}
Worldvolume state  & $\mathfrak{so}(8)_T$ & $\mathfrak{so}(n_1) \oplus \mathfrak{so}(n_2)$  & $\mathfrak{so}(d_1) \oplus \mathfrak{so}(d_2)$ \\
\hline
Gauge vector  & $\mathbf{1}$ & $(\mathbf{1}, \mathbf{1})$  & $(\mathbf{Adj},  \mathbf{1}) \oplus(\mathbf{1}, \mathbf{Adj})$ \\
Scalar & $\mathbf{8}_{v}$ & $(\mathbf{1}, \mathbf{1})$ & $\left(\text{\tiny{\yng(2)}},  \mathbf{1}\right) \oplus \left(\mathbf{1}, \text{\tiny{\yng(2)}}\right)$\\
Tachyon & $\mathbf{1}$ & $(\mathbf{1}, \mathbf{1})$ & $ \left(\text{\tiny{\yng(2)}}, \mathbf{1} \right) \oplus \left(\mathbf{1},\mathbf{Adj}\right)$ \\
Fermion & $\mathbf{16}$ & $(\mathbf{1}, \mathbf{1})$ & $\left(\text{\tiny{\yng(1)}},\text{\tiny{\yng(1)}}\right)$ \\
Fermion & $\mathbf{1}$ & $\left(\text{\tiny\yng(1)},\mathbf{1}\right)$ & $\left(\mathbf{1},\text{\tiny\yng(1)}\right)$ \\
Fermion & $\mathbf{1}$ & $\left(\mathbf{1},\text{\tiny\yng(1)}\right)$  & $\left(\text{\tiny\yng(1)},\mathbf{1}\right)$\\
\end{tabular}}
\caption{Open-string light spectrum for D1-branes in type 0A$/\Omega$. Here, $\mathbf{16}=\mathbf{8}_s\oplus \mathbf{8}_c$.}
\label{tab:0A_D1_spectrum}
\end{table}
\begin{table}[H]
\centering
\textbf{D3}
\renewcommand{\arraystretch}{1.2}
\centerresize{0.85}{
\begin{tabular}{l|c|c|c}
Worldvolume state  & $\mathfrak{so}(6)_T$ & $\mathfrak{so}(n_1) \oplus \mathfrak{so}(n_2)$  & $\mathfrak{so}(d_1) \times \mathfrak{sp}(d_2/2)$ \\
\hline
Gauge vector  & $\mathbf{1}$ & $(\mathbf{1}, \mathbf{1})$  & $(\mathbf{Adj},  \mathbf{1}) \oplus(\mathbf{1}, \mathbf{Adj})$ \\
Scalar & $\mathbf{6}_{v}$ & $(\mathbf{1}, \mathbf{1})$ & $\left(\text{\tiny{\yng(2)}},  \mathbf{1}\right) \oplus \left(\mathbf{1}, \text{\tiny{\yng(1,1)}}\right)$\\
Tachyon & $\mathbf{1}$ & $(\mathbf{1}, \mathbf{1})$ & $ \left(\text{\tiny{\yng(2)}}, \mathbf{1} \right) \oplus \left(\mathbf{1},\mathbf{Adj}\right)$ \\
Fermion & $\mathbf{8}$ & $(\mathbf{1}, \mathbf{1})$ & $\left(\text{\tiny{\yng(1)}},\text{\tiny{\yng(1)}}\right)$ \\
Fermion & $\mathbf{1}$ & $\left(\text{\tiny\yng(1)},\mathbf{1}\right)$ & $\left(\mathbf{1},\text{\tiny\yng(1)}\right)$ \\
Fermion & $\mathbf{1}$ & $\left(\mathbf{1},\text{\tiny\yng(1)}\right)$  & $\left(\text{\tiny\yng(1)},\mathbf{1}\right)$\\
\end{tabular}}
\caption{Open-string light spectrum for D3-branes in type 0A$/\Omega$. Here, $\mathbf{8} = \mathbf{4}_s\oplus\mathbf{4}_c$.}
\label{tab:0A_D3_spectrum}
\end{table}
\begin{table}[H]
\centering
\textbf{D5}
\renewcommand{\arraystretch}{1.2}
\centerresize{0.85}{
\begin{tabular}{l|c|c|c}
Worldvolume state  & $\mathfrak{so}(4)_T$ & $\mathfrak{so}(n_1) \oplus \mathfrak{so}(n_2)$  & $\mathfrak{sp}(d_1/2) \times \mathfrak{sp}(d_2/2)$ \\
\hline
Gauge vector  & $\mathbf{1}$ & $(\mathbf{1}, \mathbf{1})$  & $(\mathbf{Adj},  \mathbf{1}) \oplus(\mathbf{1}, \mathbf{Adj})$ \\
Scalar & $\mathbf{4}_{v}$ & $(\mathbf{1}, \mathbf{1})$ & $\left(\text{\tiny{\yng(1,1)}},  \mathbf{1}\right) \oplus \left(\mathbf{1}, \text{\tiny{\yng(1,1)}}\right)$\\
Tachyon & $\mathbf{1}$ & $(\mathbf{1}, \mathbf{1})$ & $ \left(\text{\tiny{\yng(1,1)}}, \mathbf{1} \right) \oplus \left(\mathbf{1},\mathbf{Adj}\right)$ \\
Fermion & $\mathbf{4}$ & $(\mathbf{1}, \mathbf{1})$ & $\left(\text{\tiny{\yng(1)}},\text{\tiny{\yng(1)}}\right)$ \\
Fermion & $\mathbf{1}$ & $\left(\text{\tiny\yng(1)},\mathbf{1}\right)$ & $\left(\mathbf{1},\text{\tiny\yng(1)}\right)$ \\
Fermion & $\mathbf{1}$ & $\left(\mathbf{1},\text{\tiny\yng(1)}\right)$  & $\left(\text{\tiny\yng(1)},\mathbf{1}\right)$\\
Scalar & $\mathbf{4}$ & $\left(\mathbf{1},\text{\tiny\yng(1)}\right)$  & $\left(\mathbf{1},\text{\tiny\yng(1)}\right)$\\
Scalar & $\mathbf{4}$ & $\left(\text{\tiny\yng(1)},\mathbf{1}\right)$  & $\left(\text{\tiny\yng(1)},\mathbf{1}\right)$\\
\end{tabular}}
\caption{Open-string light spectrum for D5-branes in type 0A$/\Omega$. Here, $\mathbf{4} = \mathbf{2}_s\oplus\mathbf{2}_c$.}
\label{tab:0A_D5_spectrum}
\end{table}
\begin{table}[H]
\centering
\textbf{D7}
\renewcommand{\arraystretch}{1.2}
\centerresize{0.85}{
\begin{tabular}{l|c|c|c}
Worldvolume state  & $\mathfrak{so}(2)_T$ & $\mathfrak{so}(n_1) \oplus \mathfrak{so}(n_2)$  & $\mathfrak{sp}(d_1/2) \times \mathfrak{so}(d_2)$ \\
\hline
Gauge vector  & $\mathbf{1}$ & $(\mathbf{1}, \mathbf{1})$  & $(\mathbf{Adj},  \mathbf{1}) \oplus(\mathbf{1}, \mathbf{Adj})$ \\
Scalar & $\mathbf{2}_{v}$ & $(\mathbf{1}, \mathbf{1})$ & $\left(\text{\tiny{\yng(1,1)}},  \mathbf{1}\right) \oplus (\mathbf{1}, \text{\tiny{\yng(2)}})$\\
Tachyon & $\mathbf{1}$ & $(\mathbf{1}, \mathbf{1})$ & $ \left(\text{\tiny{\yng(1,1)}}, \mathbf{1} \right) \oplus (\mathbf{1},\mathbf{Adj})$ \\
Fermion & $\mathbf{2}$ & $(\mathbf{1}, \mathbf{1})$ & $\left(\text{\tiny{\yng(1)}},\text{\tiny{\yng(1)}}\right)$ \\
Fermion & $\mathbf{1}$ & $\left(\text{\tiny\yng(1)},\mathbf{1}\right)$ & $\left(\mathbf{1},\text{\tiny\yng(1)}\right)$ \\
Fermion & $\mathbf{1}$ & $\left(\mathbf{1},\text{\tiny\yng(1)}\right)$  & $\left(\text{\tiny\yng(1)},\mathbf{1}\right)$\\
Tachyon & $\mathbf{2}$ & $\left(\mathbf{1},\text{\tiny\yng(1)}\right)$  & $\left(\mathbf{1},\text{\tiny\yng(1)}\right)$\\
Tachyon & $\mathbf{2}$ & $\left(\text{\tiny\yng(1)},\mathbf{1}\right)$  & $\left(\text{\tiny\yng(1)},\mathbf{1}\right)$\\
\end{tabular}}
\caption{Open-string light spectrum for D7-branes in type 0A$/\Omega$. Here, $\mathbf{2} = \mathbf{1}_s\oplus\mathbf{1}_c$.}
\label{tab:0A_D7_spectrum}
\end{table}

\section{Branes of the type 0B orientifold}\label{app:0B_branes}

There are three consistent ten-dimensional orientifold models starting from type 0B string theory~\cite{Sagnotti:1996qj,Bianchi:1991eu}. These correspond to the operators $\Omega$, $\Omega\otimes(-1)^{F_L}$, and $\Omega\otimes(-1)^{f_L}$, where $F_L$ is the left-moving spacetime fermion number and $f_L$ is the left-moving worldsheet fermion number.
We focus on the first case, 0B$/\Omega$, which is the relevant one for this work.

The Klein bottle amplitude reads
\begin{equation}\label{eq:0B/Omega_Klein}
    \mathcal{K} = \frac{1}{2} (O_8+V_8-S_8-C_8)\,.
\end{equation}
In order to cancel the dilaton tadpole, one requires the presence of D9-branes, which yield the following open string amplitudes:
\begin{equation}\label{eq:0B/Omega_D9_amplitudes}
    \begin{split}
        \mathcal{A}_{99} &= \frac{n_1^2 + n_2^2 + n_3^2 + n_4^2}{2} \, V_8 + (n_1n_2+n_3n_4)\, O_8  \\
        & -\, (n_1n_3+n_2n_4) \, S_8 - (n_1n_4+n_2n_3) \, C_8 \,,\\
        \mathcal{M}_9 &= -\, \frac{1}{2} \, (n_1+n_2+n_3+n_4) \,\hat{V}_8\,.
    \end{split}
\end{equation}
R-R tadpole cancellation forces $n_1=n_2$ and $n_3=n_4$, while the dilaton tadpole is canceled for $n_1+n_3=32$.
Combining the annulus and M\"{o}bius amplitudes, the $V_8$ character comes with a multiplicity of
\begin{equation}
    \frac{n_i(n_i-1)}{2} \quad i \in \{1,2,3,4\}\,,
\end{equation}
implying an $\mathfrak{so}(n_1)^2\oplus\mathfrak{so}(n_3)^2$ spacetime algebra. While we are interested in the case $n_3 = n_4 = 32$, the following arguments apply more generally.

From the amplitudes in~\eqref{eq:0B/Omega_D9_amplitudes}, there are tachyons and spinors of both chiralities in the bivector representation of the spacetime gauge algebra. The spacetime chiral massless spectrum is summarized in table~\ref{tab:0B/Omega_spacetime_spectrum}.
\begin{table}[!ht]
\centering
\renewcommand{\arraystretch}{1.2}
\centerresize{}{
\begin{tabular}{l|c|c}
State  & spacetime character & $\mathfrak{so}(n_1)\oplus\mathfrak{so}(n_2)\oplus\mathfrak{so}(n_3)\oplus\mathfrak{so}(n_4)$ \\
\hline
Gauge vector  & $V_8$ & $(\mathbf{Adj}, \mathbf{1}, \mathbf{1}, \mathbf{1}) \oplus (\mathbf{1}, \mathbf{Adj}, \mathbf{1},  \mathbf{1})\oplus(\mathbf{1}, \mathbf{1},\mathbf{Adj}, \mathbf{1})\oplus(\mathbf{1}, \mathbf{1}, \mathbf{1},\mathbf{Adj})$ \\
Tachyon & $O_8$ & $(\text{\tiny$\yng(1)$},\text{\tiny$\yng(1)$},\mathbf{1},\mathbf{1}) \oplus (\mathbf{1},\mathbf{1},\text{\tiny$\yng(1)$},\text{\tiny$\yng(1)$})$ \\
Spinor $(+)$ & $S_8$ & $(\text{\tiny$\yng(1)$},\mathbf{1},\text{\tiny$\yng(1)$},\mathbf{1})\oplus(\mathbf{1},\text{\tiny$\yng(1)$},\mathbf{1},\text{\tiny$\yng(1)$})$\\
Spinor $(-)$ & $C_8$ & $(\text{\tiny$\yng(1)$},\mathbf{1},\mathbf{1},\text{\tiny$\yng(1)$})\oplus(\mathbf{1},\text{\tiny$\yng(1)$},\text{\tiny$\yng(1)$},\mathbf{1})$
\end{tabular}}
\caption{Open-string light spectrum for the 0B$/\Omega$ theory.}
\label{tab:0B/Omega_spacetime_spectrum}
\end{table}

\subsection{Charged branes in the 0B orientifold}
The presence of two R-R 2-forms in the type 0B$/\Omega$ theory implies the existence of charged D1- and D5-branes.

The D$p$-D$p$ annulus amplitude is the same in both cases, and it reads
\begin{equation}\label{eq:0B/Omega_D1_D5_App}
\begin{split}
    \mathcal{A}_{pp} &= \frac{d_1^2 + d_2^2 + d_3^2 + d_4^2}{2} (V_{p-1} O_{9-p}+O_{p-1}V_{9-p}) + (d_1d_2+d_3d_4) (O_{p-1} O_{9-p}+  V_{p-1}V_{9-p}) \\
    & -\, (d_1d_3+d_2d_4) (S_{p-1} S_{9-p}+ C_{p-1} C_{9-p}) -  (d_1d_4+d_2d_3) (S_{p-1} C_{9-p}+ C_{p-1} S_{9-p})\,.
\end{split}
\end{equation}
The M\"{o}bius strip amplitude changes between the D1s and D5s:
\begin{equation}\label{eq:0B/Omega_D1_D5_Mp}
    \mathcal{M}_p = \frac{d_1+d_2+d_3+d_4}{2} \ \epsilon \,({\hat V}_{p-1}{\hat O}_{9-p}-{\hat O}_{p-1}{\hat V}_{9-p})\,,
\end{equation}
with $\epsilon = +1$ for the D5-branes and $\epsilon=-1$ for the D1-branes.
The D5-D9 amplitude is
\begin{equation}\label{eq:0B/Omega_D5_Ap9}
    \begin{split}
        \mathcal{A}_{59} &= (n_1 d_1 + n_2 d_2 + n_3 d_3 + n_4 d_4) \ (O_{4} C_4+ V_{4}S_{4}) \\
        &+\, (n_1 d_2 + n_2 d_1 + n_3 d_4 + n_4 d_3) \ (O_{4} S_4+ V_{4}C_{4}) \\ 
        &-\,  (n_1 d_4 + n_2 d_3 + n_3 d_2 + n_4 d_1) \ (C_{4} O_4+ S_{4}V_{4}) \\
        &- \, (n_1 d_3 + n_2 d_4 + n_3 d_1 + n_4 d_2) \ (S_{4} O_4+ C_{4} V_{4})\,,
    \end{split}
\end{equation}
while the D1-D9 is basically the same, exchanging $S$ and $C$:
\begin{equation}\label{eq:0B/Omega_D1_Ap9}
    \begin{split}
        \mathcal{A}_{19} &= (n_1 d_1 + n_2 d_2 + n_3 d_3 + n_4 d_4) \ (O_{0} S_8+ V_{0}C_{8})  \\
        &+\, (n_1 d_2 + n_2 d_1 + n_3 d_4 + n_4 d_3) \ (O_{0} C_8+ V_{0}S_{8}) \\ 
        &-\,  (n_1 d_4 + n_2 d_3 + n_3 d_2 + n_4 d_1) \ (S_{0} O_8+ C_{0}V_{8}) \\
        &- \, (n_1 d_3 + n_2 d_4 + n_3 d_1 + n_4 d_2) \ (C_{0} O_8+ S_{0} V_{8})\,.
    \end{split}
\end{equation}

\subsubsection{\texorpdfstring{$p=1$}{p=1}}

We can read the light spectrum of the D1-brane in the 0B$/\Omega$ theory from the amplitudes in~\eqref{eq:0B/Omega_D1_D5_App}, \eqref{eq:0B/Omega_D1_D5_Mp}, and~\eqref{eq:0B/Omega_D1_Ap9}, setting $\epsilon = -1$.
For the vector boson arising from $V_{p-1}O_{9-p}$, we find the multiplicity
\begin{equation}\label{eq:0B/Omega_D1_vector_degeneracy}
    \sum_{i=1}^4\frac{d_i(d_i-1)}{2}\mcomma 
\end{equation}
implying an $\mathfrak{so}(d_1)\oplus\mathfrak{so}(d_2)\oplus\mathfrak{so}(d_3)\oplus\mathfrak{so}(d_4)$ gauge symmetry for a stack of such branes.
The positions of the branes, encoded in $O_{p-1}V_{9-p}$, are in the symmetric representation of the worldvolume gauge algebra.
The degeneracy of the tachyons from $O_{p-1}O_{9-p}$ is
\begin{equation}\label{eq:0B/Omega_D1_tachyon_degeneracy}
    d_1d_2 + d_3 d_4\,,
\end{equation}
hence configurations with $d_2=d_4=0$ are tachyon-free.
From the fermionic characters, $S_{p-1}S_{9-p}$, $C_{p-1}C_{9-p}$, $S_{p-1}C_{9-p}$, and $C_{p-1}S_{9-p}$, there are chiral fermions in the bivector representations of the gauge algebra factors. The degrees of freedom are collected in the first half of table~\ref{tab:0B/Omega_D1_spectrum}.
Additional massless spinors from the D1-D9 amplitude in~\eqref{eq:0B/Omega_D1_Ap9}, in various bivector representations of the spacetime and worldvolume gauge algebra factors. These are listed in the second part of table~\ref{tab:0B/Omega_D1_spectrum}.
\begin{table}[!ht]
\centering
\textbf{D1}
\renewcommand{\arraystretch}{1.2}
\centerresize{}{
\begin{tabular}{l|c|c|c}
Worldvolume state  & $\mathfrak{so}(8)_T$ & $\mathfrak{so}(n_1)\oplus\mathfrak{so}(n_2)\oplus\mathfrak{so}(n_3)\oplus\mathfrak{so}(n_4)$ & $\mathfrak{so}(d_1)\oplus\mathfrak{so}(d_2)\oplus\mathfrak{so}(d_3)\oplus\mathfrak{so}(d_4)$ \\
\hline
Gauge vector  & $\mathbf{1}$ & $(\mathbf{1}, \mathbf{1}, \mathbf{1}, \mathbf{1})$ & $(\mathbf{Adj}, \mathbf{1}, \mathbf{1}, \mathbf{1}) \oplus (\mathbf{1}, \mathbf{Adj}, \mathbf{1},  \mathbf{1})\oplus(\mathbf{1}, \mathbf{1},\mathbf{Adj}, \mathbf{1})\oplus(\mathbf{1}, \mathbf{1}, \mathbf{1},\mathbf{Adj})$ \\
Scalar & $\mathbf{8}_v$ & $(\mathbf{1}, \mathbf{1}, \mathbf{1}, \mathbf{1})$ & $(\text{\tiny$\yng(2)$},\mathbf{1}, \mathbf{1},\mathbf{1}) \oplus (\mathbf{1},\text{\tiny$\yng(2)$}, \mathbf{1},\mathbf{1}) \oplus(\mathbf{1}, \mathbf{1}, \text{\tiny$\yng(2)$}, \mathbf{1}) \oplus(\mathbf{1}, \mathbf{1},\mathbf{1}, \text{\tiny$\yng(2)$})$ \\
Tachyon & $\mathbf{1}$ & $(\mathbf{1}, \mathbf{1}, \mathbf{1}, \mathbf{1})$ & $(\text{\tiny$\yng(1)$},\text{\tiny$\yng(1)$},\mathbf{1},\mathbf{1})\oplus (\mathbf{1},\mathbf{1},\text{\tiny$\yng(1)$},\text{\tiny$\yng(1)$})$\\
Spinor $(+)$ & $\mathbf{8}_{s}$ & $(\mathbf{1}, \mathbf{1}, \mathbf{1}, \mathbf{1})$ & $(\text{\tiny$\yng(1)$},\mathbf{1},\text{\tiny$\yng(1)$},\mathbf{1})\oplus(\mathbf{1},\text{\tiny$\yng(1)$},\mathbf{1},\text{\tiny$\yng(1)$})$ \\
Spinor $(-)$ & $\mathbf{8}_{c}$ & $(\mathbf{1}, \mathbf{1}, \mathbf{1}, \mathbf{1})$ & $(\text{\tiny$\yng(1)$},\mathbf{1},\text{\tiny$\yng(1)$},\mathbf{1})\oplus(\mathbf{1},\text{\tiny$\yng(1)$},\mathbf{1},\text{\tiny$\yng(1)$})$ \\
Spinor $(+)$ & $\mathbf{8}_{c}$ & $(\mathbf{1}, \mathbf{1}, \mathbf{1}, \mathbf{1})$ & $(\text{\tiny$\yng(1)$},\mathbf{1},\mathbf{1},\text{\tiny$\yng(1)$})\oplus(\mathbf{1},\text{\tiny$\yng(1)$},\text{\tiny$\yng(1)$},\mathbf{1})$ \\
Spinor $(-)$ & $\mathbf{8}_{s}$ & $(\mathbf{1}, \mathbf{1}, \mathbf{1}, \mathbf{1})$ & $(\text{\tiny$\yng(1)$},\mathbf{1},\mathbf{1},\text{\tiny$\yng(1)$})\oplus(\mathbf{1},\text{\tiny$\yng(1)$},\text{\tiny$\yng(1)$},\mathbf{1})$ \\
\hline
\hline
Fermion ($+$) & $\mathbf{1}$ & $\begin{array}{l}
    (\text{\tiny$\yng(1)$}, \mathbf{1}, \mathbf{1}, \mathbf{1})\\(\mathbf{1}, \text{\tiny$\yng(1)$}, \mathbf{1}, \mathbf{1})\\(\mathbf{1}, \mathbf{1}, \text{\tiny$\yng(1)$}, \mathbf{1})\\(\mathbf{1}, \mathbf{1}, \mathbf{1}, \text{\tiny$\yng(1)$}) \end{array}$ & $\begin{array}{l} (\mathbf{1}, \mathbf{1}, \mathbf{1}, \text{\tiny$\yng(1)$}) \\ (\mathbf{1}, \mathbf{1}, \text{\tiny$\yng(1)$}, \mathbf{1}) \\ (\mathbf{1}, \text{\tiny$\yng(1)$}, \mathbf{1}, \mathbf{1}) \\ (\text{\tiny$\yng(1)$}, \mathbf{1}, \mathbf{1}, \mathbf{1}) \end{array}$ \\
\hline
Fermion ($-$) & $\mathbf{8}_{v}$ & $\begin{array}{l}
    (\text{\tiny$\yng(1)$}, \mathbf{1}, \mathbf{1}, \mathbf{1})\\(\mathbf{1}, \text{\tiny$\yng(1)$}, \mathbf{1}, \mathbf{1})\\(\mathbf{1}, \mathbf{1}, \text{\tiny$\yng(1)$}, \mathbf{1})\\(\mathbf{1}, \mathbf{1}, \mathbf{1}, \text{\tiny$\yng(1)$}) \end{array}$ & $\begin{array}{l} (\mathbf{1}, \mathbf{1}, \mathbf{1}, \text{\tiny$\yng(1)$}) \\ (\mathbf{1}, \mathbf{1}, \text{\tiny$\yng(1)$}, \mathbf{1}) \\ (\mathbf{1}, \text{\tiny$\yng(1)$}, \mathbf{1}, \mathbf{1}) \\ (\text{\tiny$\yng(1)$}, \mathbf{1}, \mathbf{1}, \mathbf{1}) \end{array}$ \\
\hline
Fermion ($-$) & $\mathbf{1}$ & $\begin{array}{l}
    (\text{\tiny$\yng(1)$}, \mathbf{1}, \mathbf{1}, \mathbf{1})\\(\mathbf{1}, \text{\tiny$\yng(1)$}, \mathbf{1}, \mathbf{1})\\(\mathbf{1}, \mathbf{1}, \text{\tiny$\yng(1)$}, \mathbf{1})\\(\mathbf{1}, \mathbf{1}, \mathbf{1}, \text{\tiny$\yng(1)$}) \end{array}$ & $\begin{array}{l} (\mathbf{1}, \mathbf{1}, \text{\tiny$\yng(1)$}, \mathbf{1}) \\ (\mathbf{1}, \mathbf{1}, \mathbf{1}, \text{\tiny$\yng(1)$}) \\ (\text{\tiny$\yng(1)$}, \mathbf{1}, \mathbf{1}, \mathbf{1}) \\ (\mathbf{1}, \text{\tiny$\yng(1)$}, \mathbf{1}, \mathbf{1}) \end{array}$ \\
\hline
Fermion ($+$) & $\mathbf{8}_{v}$ & $\begin{array}{l}
    (\text{\tiny$\yng(1)$}, \mathbf{1}, \mathbf{1}, \mathbf{1})\\(\mathbf{1}, \text{\tiny$\yng(1)$}, \mathbf{1}, \mathbf{1})\\(\mathbf{1}, \mathbf{1}, \text{\tiny$\yng(1)$}, \mathbf{1})\\(\mathbf{1}, \mathbf{1}, \mathbf{1}, \text{\tiny$\yng(1)$}) \end{array}$ & $\begin{array}{l} (\mathbf{1}, \mathbf{1}, \text{\tiny$\yng(1)$}, \mathbf{1}) \\ (\mathbf{1}, \mathbf{1}, \mathbf{1}, \text{\tiny$\yng(1)$}) \\ (\text{\tiny$\yng(1)$}, \mathbf{1}, \mathbf{1}, \mathbf{1}) \\ (\mathbf{1}, \text{\tiny$\yng(1)$}, \mathbf{1}, \mathbf{1}) \end{array}$ \\
\end{tabular}}
\caption{Open-string light spectrum for D1-branes in type 0B$/\Omega$.}
\label{tab:0B/Omega_D1_spectrum}
\end{table}

\subsubsection{\texorpdfstring{$p=5$}{p=5}}

As for the D1, the light spectrum of the D5-brane in the type 0B$/\Omega$ orientifold can be deduced from the amplitudes in~\eqref{eq:0B/Omega_D1_D5_App}, \eqref{eq:0B/Omega_D1_D5_Mp}, and~\eqref{eq:0B/Omega_D5_Ap9}, letting $\epsilon = +1$.
In this case, the gauge algebra structure is $\mathfrak{sp}(d_1/2)\oplus\mathfrak{sp}(d_2/2)\oplus\mathfrak{sp}(d_3/2)\oplus\mathfrak{sp}(d_4/2)$ since the vector multiplicity is
\begin{equation}\label{eq:0B/Omega_D5_vector_degeneracy}
    \sum_{i=1}^4\frac{d_i(d_i+1)}{2}\,.
\end{equation}
Scalars denoting the positions of the branes are in the antisymmetric representation of the worldvolume gauge algebra, while the degeneracy of the tachyons is
\begin{equation}\label{eq:0B/Omega_D5_tachyon_degeneracy}
    d_1d_2 + d_3 d_4\,,
\end{equation}
implying that non-tachyonic configurations are, for instance, those with $d_2=d_4=0$.
In addition, there are the same chiral fermions in the bivector representations of the worldvolume gauge algebra as in the D1 case.
From the D5-D9 amplitude in~\eqref{eq:0B/Omega_D5_Ap9}, there are additional massless scalars and fermions with flipped chirality with respect to the analogous degrees of freedom on the D1 worldvolume, transforming in the bivector representations of spacetime and worldvolume gauge algebras.
These results are collected in table~\ref{tab:0B/Omega_D5_spectrum}.
\begin{table}[!ht]
\centering
\textbf{D5}
\renewcommand{\arraystretch}{1.2}
\centerresize{}{
\begin{tabular}{l|c|c|c}
Worldvolume state  & $\mathfrak{so}(4)_T$ & $\mathfrak{so}(n_1)\oplus\mathfrak{so}(n_2)\oplus\mathfrak{so}(n_3)\oplus\mathfrak{so}(n_4)$ & $\mathfrak{sp}(d_1/2)\oplus\mathfrak{sp}(d_2/2)\oplus\mathfrak{sp}(d_3/2)\oplus\mathfrak{sp}(d_4/2)$ \\
\hline
Gauge vector  & $\mathbf{1}$ & $(\mathbf{1}, \mathbf{1}, \mathbf{1}, \mathbf{1})$ & $(\mathbf{Adj}, \mathbf{1}, \mathbf{1}, \mathbf{1}) \oplus (\mathbf{1}, \mathbf{Adj}, \mathbf{1},  \mathbf{1})\oplus(\mathbf{1}, \mathbf{1},\mathbf{Adj}, \mathbf{1})\oplus(\mathbf{1}, \mathbf{1}, \mathbf{1},\mathbf{Adj})$ \\
Scalar & $\mathbf{4}_{v}$ & $(\mathbf{1}, \mathbf{1}, \mathbf{1}, \mathbf{1})$ & $\left(\text{\tiny$\yng(1,1)$},\mathbf{1}, \mathbf{1},\mathbf{1}\right) \oplus \left(\mathbf{1},\text{\tiny$\yng(1,1)$}, \mathbf{1},\mathbf{1}\right) \oplus\left(\mathbf{1}, \mathbf{1}, \text{\tiny$\yng(1,1)$}, \mathbf{1}\right) \oplus\left(\mathbf{1}, \mathbf{1},\mathbf{1}, \text{\tiny$\yng(1,1)$}\right)$ \\
Tachyon & $\mathbf{1}$ & $(\mathbf{1}, \mathbf{1}, \mathbf{1}, \mathbf{1})$ & $(\text{\tiny$\yng(1)$},\text{\tiny$\yng(1)$},\mathbf{1},\mathbf{1})\oplus (\mathbf{1},\mathbf{1},\text{\tiny$\yng(1)$},\text{\tiny$\yng(1)$})$\\
Spinor $(+)$ & $\mathbf{2}_{s}$ & $(\mathbf{1}, \mathbf{1}, \mathbf{1}, \mathbf{1})$ & $(\text{\tiny$\yng(1)$},\mathbf{1},\text{\tiny$\yng(1)$},\mathbf{1})\oplus(\mathbf{1},\text{\tiny$\yng(1)$},\mathbf{1},\text{\tiny$\yng(1)$})$ \\
Spinor $(-)$ & $\mathbf{2}_{c}$ & $(\mathbf{1}, \mathbf{1}, \mathbf{1}, \mathbf{1})$ & $(\text{\tiny$\yng(1)$},\mathbf{1},\text{\tiny$\yng(1)$},\mathbf{1})\oplus(\mathbf{1},\text{\tiny$\yng(1)$},\mathbf{1},\text{\tiny$\yng(1)$})$ \\
Spinor $(+)$ & $\mathbf{2}_{c}$ & $(\mathbf{1}, \mathbf{1}, \mathbf{1}, \mathbf{1})$ & $(\text{\tiny$\yng(1)$},\mathbf{1},\mathbf{1},\text{\tiny$\yng(1)$})\oplus(\mathbf{1},\text{\tiny$\yng(1)$},\text{\tiny$\yng(1)$},\mathbf{1})$ \\
Spinor $(-)$ & $\mathbf{2}_{s}$ & $(\mathbf{1}, \mathbf{1}, \mathbf{1}, \mathbf{1})$ & $(\text{\tiny$\yng(1)$},\mathbf{1},\mathbf{1},\text{\tiny$\yng(1)$})\oplus(\mathbf{1},\text{\tiny$\yng(1)$},\text{\tiny$\yng(1)$},\mathbf{1})$ \\
\hline
\hline
Various scalars & & & \\
\hline
Fermion ($-$) & $\mathbf{1}$ & $\begin{array}{l}
    (\text{\tiny$\yng(1)$}, \mathbf{1}, \mathbf{1}, \mathbf{1})\\(\mathbf{1}, \text{\tiny$\yng(1)$}, \mathbf{1}, \mathbf{1})\\(\mathbf{1}, \mathbf{1}, \text{\tiny$\yng(1)$}, \mathbf{1})\\(\mathbf{1}, \mathbf{1}, \mathbf{1}, \text{\tiny$\yng(1)$}) \end{array}$ & $\begin{array}{l} (\mathbf{1}, \mathbf{1}, \mathbf{1}, \text{\tiny$\yng(1)$}) \\ (\mathbf{1}, \mathbf{1}, \text{\tiny$\yng(1)$}, \mathbf{1}) \\ (\mathbf{1}, \text{\tiny$\yng(1)$}, \mathbf{1}, \mathbf{1}) \\ (\text{\tiny$\yng(1)$}, \mathbf{1}, \mathbf{1}, \mathbf{1}) \end{array}$ \\
\hline
Fermion ($+$) & $\mathbf{4}_{v}$ & $\begin{array}{l}
    (\text{\tiny$\yng(1)$}, \mathbf{1}, \mathbf{1}, \mathbf{1})\\(\mathbf{1}, \text{\tiny$\yng(1)$}, \mathbf{1}, \mathbf{1})\\(\mathbf{1}, \mathbf{1}, \text{\tiny$\yng(1)$}, \mathbf{1})\\(\mathbf{1}, \mathbf{1}, \mathbf{1}, \text{\tiny$\yng(1)$}) \end{array}$ & $\begin{array}{l} (\mathbf{1}, \mathbf{1}, \mathbf{1}, \text{\tiny$\yng(1)$}) \\ (\mathbf{1}, \mathbf{1}, \text{\tiny$\yng(1)$}, \mathbf{1}) \\ (\mathbf{1}, \text{\tiny$\yng(1)$}, \mathbf{1}, \mathbf{1}) \\ (\text{\tiny$\yng(1)$}, \mathbf{1}, \mathbf{1}, \mathbf{1}) \end{array}$ \\
\hline
Fermion ($+$) & $\mathbf{1}$ & $\begin{array}{l}
    (\text{\tiny$\yng(1)$}, \mathbf{1}, \mathbf{1}, \mathbf{1})\\(\mathbf{1}, \text{\tiny$\yng(1)$}, \mathbf{1}, \mathbf{1})\\(\mathbf{1}, \mathbf{1}, \text{\tiny$\yng(1)$}, \mathbf{1})\\(\mathbf{1}, \mathbf{1}, \mathbf{1}, \text{\tiny$\yng(1)$}) \end{array}$ & $\begin{array}{l} (\mathbf{1}, \mathbf{1}, \text{\tiny$\yng(1)$}, \mathbf{1}) \\ (\mathbf{1}, \mathbf{1}, \mathbf{1}, \text{\tiny$\yng(1)$}) \\ (\text{\tiny$\yng(1)$}, \mathbf{1}, \mathbf{1}, \mathbf{1}) \\ (\mathbf{1}, \text{\tiny$\yng(1)$}, \mathbf{1}, \mathbf{1}) \end{array}$ \\
\hline
Fermion ($-$) & $\mathbf{4}_{v}$ & $\begin{array}{l}
    (\text{\tiny$\yng(1)$}, \mathbf{1}, \mathbf{1}, \mathbf{1})\\(\mathbf{1}, \text{\tiny$\yng(1)$}, \mathbf{1}, \mathbf{1})\\(\mathbf{1}, \mathbf{1}, \text{\tiny$\yng(1)$}, \mathbf{1})\\(\mathbf{1}, \mathbf{1}, \mathbf{1}, \text{\tiny$\yng(1)$}) \end{array}$ & $\begin{array}{l} (\mathbf{1}, \mathbf{1}, \text{\tiny$\yng(1)$}, \mathbf{1}) \\ (\mathbf{1}, \mathbf{1}, \mathbf{1}, \text{\tiny$\yng(1)$}) \\ (\text{\tiny$\yng(1)$}, \mathbf{1}, \mathbf{1}, \mathbf{1}) \\ (\mathbf{1}, \text{\tiny$\yng(1)$}, \mathbf{1}, \mathbf{1}) \end{array}$ \\
\end{tabular}}
\caption{Open-string light spectrum for D5-branes in type 0B$/\Omega$.}
\label{tab:0B/Omega_D5_spectrum}
\end{table}

\subsection{Uncharged branes in the 0B orientifold}

The orientifold model 0B$/\Omega$ also contains uncharged D$p$-branes. Although nearly all of them have open-string tachyons, we will discuss their worldvolume spectra in the following for completeness. We analyze the cases with $p=-1,3,7$ first, and then those with $p=0,2,4,6,8$.

\subsubsection{\texorpdfstring{$p=-1,3,7$}{p=-1,3,7}}

The D$(-1)$, D3, and D7 branes have unitary gauge groups. Indeed, their open-string amplitudes read
\begin{equation}\label{eq:0B/Omega_D-1_D3_D7_App}
    \begin{split}
        \mathcal{A}_{pp} =& (d_1 {\bar d}_1 + d_2 {\bar d}_2) (V_{p-1}
        O_{9-p}+O_{p-1}V_{9-p}) \\
        &+\, \left(\frac{d_1^2 + {\bar d}_1^2 + d_2^2 + {\bar d}_2^2}{2}\right) (O_{p-1} O_{9-p}+ V_{p-1}V_{9-p})  \\
        & - \, (d_1 {\bar d}_2+{\bar d}_1 d_2)(S_{p-1} S_{9-p}+ C_{p-1} C_{9-p})  \\
        & - \, (d_1 d_2+{\bar d}_1 {\bar d}_2)(S_{p-1} C_{9-p}+ C_{p-1} S_{9-p})\,
    \end{split}
\end{equation}
and
\begin{equation}\label{eq:0B/Omega_D-1_D3_D7_Mpp}
        \mathcal{M}_p = -\frac{d_1+{\bar d}_1 + d_2 + {\bar d}_2}{2} \, \epsilon\,  ({\hat O}_{p-1}{\hat O}_{9-p}+ {\hat V}_{p-1}{\hat V}_{9-p})\,.
\end{equation}
In this case, $\epsilon = +1$ for D7- and D$(-1)$-branes and $\epsilon = -1$ for D3-branes.

The degeneracy of the vector character $V_{p-1}V_{9-p}$ is
\begin{equation}
    d_1\bar{d_1} + d_2\bar{d_2}\,,
\end{equation}
leading to the gauge algebra $\mathfrak{u}(d_1)\oplus\mathfrak{u}(d_2)$.
Tachyons appear with the prefactor
\begin{equation}
    \frac{d_i(d_i-\epsilon)}{2} \quad i = 1,2\,,
\end{equation}
and thus can be eliminated either for a single D7-brane or for two D7-branes of different types. In fact, each of these branes has a $\Z_2$ charge in K-theory. Despite this, for $p>5$, additional tachyonic states arise from the $\mathcal{A}_{p9}$ amplitudes.
There are fermions in bi-(anti)fundamentals of the gauge algebra and scalars, denoting the positions of the branes, in the adjoint representation.

The D7-D9 spectrum can be derived from the amplitude
\begin{equation}\label{eq:0B/Omega_D7_Ap9}
    \begin{split}
        \mathcal{A}_{79} &= (n_1 d_1 \!+  n_2 {\bar d}_1 +  n_3 d_2  + n_4 {\bar d}_2) (O_{6} S_2 +  V_{6}C_{2}) \\
        &+\, (n_1 {\bar d}_1  +  n_2 d_1  +  n_3 {\bar d}_2  + n_4 d_2) (O_{6} C_2 + V_{6} S_{2}) \\ 
        &- \, (n_1 \bar{d}_2 + n_2 d_2 + n_3 \bar{d}_1 + n_4 d_1) \ (C_{6} O_2+ S_{6}V_{2})  \\
        &-\,  (n_1 d_2 + n_2 \bar{d}_2 + n_3 d_1+  n_4 \bar{d}_1) \ (S_{6} O_2+ C_{6}V_{2}) \,.
    \end{split}
\end{equation}
It contains fermions in the bifundamental or in the fundamental-antifundamental representations of a spacetime and a worldvolume gauge algebra factor as well as tachyons in analogous representations. The results for the D7-branes are collected in table~\ref{tab:0B/Omega_D7_spectrum}.
\begin{table}[H]
\centering
\textbf{D7}
\renewcommand{\arraystretch}{1.2}
\centerresize{}{
\begin{tabular}{l|c|c|c}
Worldvolume state  & $\mathfrak{so}(2)_T$ & $\mathfrak{so}(n_1)\oplus\mathfrak{so}(n_2)\oplus\mathfrak{so}(n_3)\oplus\mathfrak{so}(n_4)$ & $\mathfrak{u}(d_1)\oplus\mathfrak{u}(d_2)$ \\
\hline
Gauge vector  & $\mathbf{1}$ & $(\mathbf{1}, \mathbf{1}, \mathbf{1}, \mathbf{1})$ &  $(\mathbf{Adj}, \mathbf{1}) \oplus (\mathbf{1},\mathbf{Adj})$ \\
Scalar & $\mathbf{2}_{v}$ & $(\mathbf{1}, \mathbf{1}, \mathbf{1}, \mathbf{1})$ & $(\mathbf{Adj}, \mathbf{1}) \oplus (\mathbf{1},\mathbf{Adj})$\\
Tachyon & $\mathbf{1}$ & $(\mathbf{1}, \mathbf{1}, \mathbf{1}, \mathbf{1})$ & $(\text{\tiny$\yng(2)$}, \mathbf{1}) \oplus (\mathbf{1}, \text{\tiny$\yng(2)$})$\\
Spinor $(+)$ & $\mathbf{1}_{s}$ & $(\mathbf{1}, \mathbf{1}, \mathbf{1}, \mathbf{1})$ & $(\text{\tiny$\yng(1)$},\text{\tiny$\ybar{\yng(1)}$})$ \\
Spinor $(-)$ & $\mathbf{1}_{c}$ & $(\mathbf{1}, \mathbf{1}, \mathbf{1}, \mathbf{1})$ & $(\text{\tiny$\ybar{\yng(1)}$},\text{\tiny$\yng(1)$})$ \\
Spinor $(+)$ & $\mathbf{1}_{c}$ & $(\mathbf{1}, \mathbf{1}, \mathbf{1}, \mathbf{1})$ & $(\text{\tiny$\yng(1)$},\text{\tiny$\yng(1)$})$ \\
Spinor $(-)$ & $\mathbf{1}_{s}$ & $(\mathbf{1}, \mathbf{1}, \mathbf{1}, \mathbf{1})$ & $(\text{\tiny$\ybar{\yng(1)}$},\text{\tiny$\ybar{\yng(1)}$})$ \\
\hline
\hline
Various tachyons & & & \\
\hline
Fermion ($-$) & $\mathbf{1}$ & $\begin{array}{l}
    (\text{\tiny$\yng(1)$}, \mathbf{1}, \mathbf{1}, \mathbf{1})\\(\mathbf{1}, \text{\tiny$\yng(1)$}, \mathbf{1}, \mathbf{1})\\(\mathbf{1}, \mathbf{1}, \text{\tiny$\yng(1)$}, \mathbf{1})\\(\mathbf{1}, \mathbf{1}, \mathbf{1}, \text{\tiny$\yng(1)$}) \end{array}$ & $\begin{array}{l} (\mathbf{1}, \text{\tiny$\ybar{\yng(1)}$}) \\ (\mathbf{1}, \text{\tiny$\yng(1)$}) \\ (\text{\tiny$\ybar{\yng(1)}$},\mathbf{1}) \\ (\text{\tiny$\yng(1)$},\mathbf{1}) \end{array}$ \\
\hline
Fermion ($+$) & $\mathbf{2}_{v}$ & $\begin{array}{l}
    (\text{\tiny$\yng(1)$}, \mathbf{1}, \mathbf{1}, \mathbf{1})\\(\mathbf{1}, \text{\tiny$\yng(1)$}, \mathbf{1}, \mathbf{1})\\(\mathbf{1}, \mathbf{1}, \text{\tiny$\yng(1)$}, \mathbf{1})\\(\mathbf{1}, \mathbf{1}, \mathbf{1}, \text{\tiny$\yng(1)$}) \end{array}$ & $\begin{array}{l} (\mathbf{1}, \text{\tiny$\ybar{\yng(1)}$}) \\ (\mathbf{1}, \text{\tiny$\yng(1)$}) \\ (\text{\tiny$\ybar{\yng(1)}$},\mathbf{1}) \\ (\text{\tiny$\yng(1)$},\mathbf{1}) \end{array}$ \\
\hline
Fermion ($+$) & $\mathbf{1}$ & $\begin{array}{l}
    (\text{\tiny$\yng(1)$}, \mathbf{1}, \mathbf{1}, \mathbf{1})\\(\mathbf{1}, \text{\tiny$\yng(1)$}, \mathbf{1}, \mathbf{1})\\(\mathbf{1}, \mathbf{1}, \text{\tiny$\yng(1)$}, \mathbf{1})\\(\mathbf{1}, \mathbf{1}, \mathbf{1}, \text{\tiny$\yng(1)$}) \end{array}$ & $\begin{array}{l} (\mathbf{1}, \text{\tiny$\yng(1)$}) \\ (\mathbf{1}, \text{\tiny$\ybar{\yng(1)}$})  \\ (\text{\tiny$\yng(1)$},\mathbf{1}) \\ (\text{\tiny$\ybar{\yng(1)}$},\mathbf{1}) \end{array}$ \\
\hline
Fermion ($-$) & $\mathbf{2}_{v}$ & $\begin{array}{l}
    (\text{\tiny$\yng(1)$}, \mathbf{1}, \mathbf{1}, \mathbf{1})\\(\mathbf{1}, \text{\tiny$\yng(1)$}, \mathbf{1}, \mathbf{1})\\(\mathbf{1}, \mathbf{1}, \text{\tiny$\yng(1)$}, \mathbf{1})\\(\mathbf{1}, \mathbf{1}, \mathbf{1}, \text{\tiny$\yng(1)$}) \end{array}$ & $\begin{array}{l} (\mathbf{1}, \text{\tiny$\yng(1)$}) \\ (\mathbf{1}, \text{\tiny$\ybar{\yng(1)}$})  \\ (\text{\tiny$\yng(1)$},\mathbf{1}) \\ (\text{\tiny$\ybar{\yng(1)}$},\mathbf{1}) \end{array}$\\
\end{tabular}}
\caption{Open-string light spectrum for D7-branes in type 0B$/\Omega$.}
\label{tab:0B/Omega_D7_spectrum}
\end{table}

The annulus amplitude of the D3-D9 brane system is analogous to that in~\eqref{eq:0B/Omega_D7_Ap9}, exchanging $S$ and $C$. Hence, the resulting spectrum is the same as for the D7 case, with fermions having opposite chiralities and scalars from $\mathcal{A}_{39}$ amplitude being massive. The spectrum of D3-branes can be found in table~\ref{tab:0B/Omega_D3_spectrum}.
\begin{table}[!ht]
\centering
\textbf{D3}
\renewcommand{\arraystretch}{1.2}
\centerresize{}{
\begin{tabular}{l|c|c|c}
Worldvolume state  & $\mathfrak{so}(6)_T$ & $\mathfrak{so}(n_1)\oplus\mathfrak{so}(n_2)\oplus\mathfrak{so}(n_3)\oplus\mathfrak{so}(n_4)$ & $\mathfrak{u}(d_1)\oplus\mathfrak{u}(d_2)$ \\
\hline
Gauge vector  & $\mathbf{1}$ & $(\mathbf{1}, \mathbf{1}, \mathbf{1}, \mathbf{1})$ &  $(\mathbf{Adj}, \mathbf{1}) \oplus (\mathbf{1},\mathbf{Adj})$ \\
Scalar & $\mathbf{6}_{v}$ & $(\mathbf{1}, \mathbf{1}, \mathbf{1}, \mathbf{1})$ & $(\mathbf{Adj}, \mathbf{1}) \oplus (\mathbf{1},\mathbf{Adj})$\\
Tachyon & $\mathbf{1}$ & $(\mathbf{1}, \mathbf{1}, \mathbf{1}, \mathbf{1})$ & $  \left(\text{\tiny$\yng(1,1)$}, \mathbf{1}\right) \oplus  \left(\mathbf{1}, \text{\tiny$\yng(1,1)$}\right)$\\
Spinor $(+)$ & $\mathbf{4}_{s}$ & $(\mathbf{1}, \mathbf{1}, \mathbf{1}, \mathbf{1})$ & $(\text{\tiny$\yng(1)$},\text{\tiny$\ybar{\yng(1)}$})$ \\
Spinor $(-)$ & $\mathbf{4}_{c}$ & $(\mathbf{1}, \mathbf{1}, \mathbf{1}, \mathbf{1})$ & $(\text{\tiny$\ybar{\yng(1)}$},\text{\tiny$\yng(1)$})$ \\
Spinor $(+)$ & $\mathbf{4}_{c}$ & $(\mathbf{1}, \mathbf{1}, \mathbf{1}, \mathbf{1})$ & $(\text{\tiny$\yng(1)$},\text{\tiny$\yng(1)$})$ \\
Spinor $(-)$ & $\mathbf{4}_{s}$ & $(\mathbf{1}, \mathbf{1}, \mathbf{1}, \mathbf{1})$ & $(\text{\tiny$\ybar{\yng(1)}$},\text{\tiny$\ybar{\yng(1)}$})$ \\
\hline
\hline
Fermion ($+$) & $\mathbf{1}$ & $\begin{array}{l}
    (\text{\tiny$\yng(1)$}, \mathbf{1}, \mathbf{1}, \mathbf{1})\\(\mathbf{1}, \text{\tiny$\yng(1)$}, \mathbf{1}, \mathbf{1})\\(\mathbf{1}, \mathbf{1}, \text{\tiny$\yng(1)$}, \mathbf{1})\\(\mathbf{1}, \mathbf{1}, \mathbf{1}, \text{\tiny$\yng(1)$}) \end{array}$ & $\begin{array}{l} (\mathbf{1}, \text{\tiny$\ybar{\yng(1)}$}) \\ (\mathbf{1}, \text{\tiny$\yng(1)$}) \\ (\text{\tiny$\ybar{\yng(1)}$},\mathbf{1}) \\ (\text{\tiny$\yng(1)$},\mathbf{1}) \end{array}$ \\
\hline
Fermion ($-$) & $\mathbf{6}_{v}$ & $\begin{array}{l}
    (\text{\tiny$\yng(1)$}, \mathbf{1}, \mathbf{1}, \mathbf{1})\\(\mathbf{1}, \text{\tiny$\yng(1)$}, \mathbf{1}, \mathbf{1})\\(\mathbf{1}, \mathbf{1}, \text{\tiny$\yng(1)$}, \mathbf{1})\\(\mathbf{1}, \mathbf{1}, \mathbf{1}, \text{\tiny$\yng(1)$}) \end{array}$ & $\begin{array}{l} (\mathbf{1}, \text{\tiny$\ybar{\yng(1)}$}) \\ (\mathbf{1}, \text{\tiny$\yng(1)$}) \\ (\text{\tiny$\ybar{\yng(1)}$},\mathbf{1}) \\ (\text{\tiny$\yng(1)$},\mathbf{1}) \end{array}$ \\
\hline
Fermion ($-$) & $\mathbf{1}$ & $\begin{array}{l}
    (\text{\tiny$\yng(1)$}, \mathbf{1}, \mathbf{1}, \mathbf{1})\\(\mathbf{1}, \text{\tiny$\yng(1)$}, \mathbf{1}, \mathbf{1})\\(\mathbf{1}, \mathbf{1}, \text{\tiny$\yng(1)$}, \mathbf{1})\\(\mathbf{1}, \mathbf{1}, \mathbf{1}, \text{\tiny$\yng(1)$}) \end{array}$ & $\begin{array}{l} (\mathbf{1}, \text{\tiny$\yng(1)$}) \\ (\mathbf{1}, \text{\tiny$\ybar{\yng(1)}$})  \\ (\text{\tiny$\yng(1)$},\mathbf{1}) \\ (\text{\tiny$\ybar{\yng(1)}$},\mathbf{1}) \end{array}$ \\
\hline
Fermion ($+$) & $\mathbf{6}_{v}$ & $\begin{array}{l}
    (\text{\tiny$\yng(1)$}, \mathbf{1}, \mathbf{1}, \mathbf{1})\\(\mathbf{1}, \text{\tiny$\yng(1)$}, \mathbf{1}, \mathbf{1})\\(\mathbf{1}, \mathbf{1}, \text{\tiny$\yng(1)$}, \mathbf{1})\\(\mathbf{1}, \mathbf{1}, \mathbf{1}, \text{\tiny$\yng(1)$}) \end{array}$ & $\begin{array}{l} (\mathbf{1}, \text{\tiny$\yng(1)$}) \\ (\mathbf{1}, \text{\tiny$\ybar{\yng(1)}$})  \\ (\text{\tiny$\yng(1)$},\mathbf{1}) \\ (\text{\tiny$\ybar{\yng(1)}$},\mathbf{1}) \end{array}$\\
\end{tabular}}
\caption{Open-string light spectrum for D3-branes in type 0B$/\Omega$.}
\label{tab:0B/Omega_D3_spectrum}
\end{table}

\subsubsection{\texorpdfstring{$p=0,2,4,6,8$}{p=0,2,4,6,8}}

This orientifold contains additional uncharged D$p$-branes with $p=0,2,4,6,8$, among which the D0 and the D8 are $\Z_2$-charged. The annulus and M\"{o}bius amplitudes in these cases are
\begin{equation}\label{eq:0B/Omega_Dp_even_App}
    \mathcal{A}_{pp} = \frac{d_1^2 + d_2^2}{2} \ (O_{p-1}+
V_{p-1})(O_{9-p} + V_{9-p}) - 2 d_1 d_2 \ S'_{p-1} S'_{9-p}
\end{equation}
and
\begin{eq}\label{eq:0B/Omega_Dp_even_Mpp}
    {\mathcal M}_p = - \frac{d_1+d_2}{\sqrt{2}} & \left[ \sin{\frac{(p-5) 
\pi}{4}}
({\hat O}_{p-1}{\hat O}_{9-p} + {\hat V}_{p-1}
{\hat V}_{9-p}) \right. \\
& \left. + \cos\frac{(p-5) \pi}{4}
({\hat O}_{p-1}{\hat V}_{9-p} -{\hat V}_{p-1}
{\hat O}_{9-p})  \right]\,.
\end{eq}
The multiplicity of the vector bosons is
\begin{equation}\label{eq:0B/Omega_Dp_even_vector_multiplicity}
    \frac{d_i^2}{2} + \frac{d_i}{\sqrt{2}}\,\cos\frac{(p-5) \pi}{4}\quad \text{with}\quad i = 1,2\,, 
\end{equation}
implying $\mathfrak{sp}(d_1/2)\oplus\mathfrak{sp}(d_2/2)$ worldvolume gauge algebras for the D4- and D6-branes and $\mathfrak{so}(d_1)\oplus\mathfrak{so}(d_2)$ gauge algebras for D0-, D2-, and D8-branes. Note that this corrects an old typo in~\cite{Dudas:2001wd}, after eq.~(6.22) of that paper, which propagated to~\cite[Table 10]{Angelantonj:2002ct} and would have otherwise been in contradiction with the brane classification of~\cite{Kaidi:2019tyf}.
Tachyonic states appear with multiplicity
\begin{equation}
    \frac{d_i^2}{2} - \frac{d_i}{\sqrt{2}}\,\sin\frac{(p-5) \pi}{4}\quad \text{with}\quad i = 1,2\,.
\end{equation}
Hence, they transform in the symmetric representation for D2- and D4-branes and in the antisymmetric representation for D0-, D6-, and D8-branes. Note that the number of tachyons can vanish for D0-branes and D6-branes with $d_i=1$, but for the latter, D6-branes come in pairs (in the same sense as D5-branes in type I string theory come in pairs) and $d_i$ must be even. Hence, we find that a single $\Z_2$-charged D0-brane is free of open-string tachyons; there are actually two types of such branes, with $d_1=1$ and $d_2=0$, or with $d_1=0$ and $d_2=1$, thus matching the K-theory classification of~\cite{Kaidi:2019tyf}.
The scalars representing the positions of the branes are
\begin{equation}
    \frac{d_i^2}{2} - \frac{d_i}{\sqrt{2}} \, \cos\frac{(p-5) \pi}{4}\quad \text{with}\quad i = 1,2\,.
\end{equation}
Since this corresponds to the multiplicity of the vector boson in~\eqref{eq:0B/Omega_Dp_even_vector_multiplicity}, these scalars always transform in the adjoint representation of the appropriate gauge algebras.
In addition, from~\eqref{eq:0B/Omega_Dp_even_App}, there are Dirac fermions in the bifundamental representation.

Finally, the D$p$-D9 amplitude for the uncharged branes with $p$ even is
\begin{equation}\label{eq:0B/Omega_Dp_even_Ap9}
    \begin{split}
        \mathcal{A}_{p9} =& [(n_1+n_2) d_1 + (n_3+n_4) d_2] \ (O_{p-1}+V_{p-1}) \ S'_{9-p} \\
        &- \,  [ (n_3+n_4) d_1 + (n_1+n_2) d_2]  \ S'_{p-1} \ (O_{9-p} + V_{9-p})\ ,
    \end{split}
\end{equation}
and it does not change among the various cases. Thus, this sector of the light spectrum is the same for all the D$p$-branes. It yields Dirac fermions in the bifundamental representation of a spacetime and a worldvolume algebra, and it contains tachyons for $p>5$. These results are collected in tables~\ref{tab:0B/Omega_D0_spectrum}, \ref{tab:0B/Omega_D2_spectrum}, \ref{tab:0B/Omega_D4_spectrum}, \ref{tab:0B/Omega_D6_spectrum}, and~\ref{tab:0B/Omega_D8_spectrum}.

\begin{table}[H]
\centering
\textbf{D0}
\renewcommand{\arraystretch}{1.2}
\centerresize{}{
\begin{tabular}{l|c|c|c}
Worldvolume state  & $\mathfrak{so}(9)_T$ & $\mathfrak{so}(n_1)\oplus\mathfrak{so}(n_2)\oplus\mathfrak{so}(n_3)\oplus\mathfrak{so}(n_4)$ & $\mathfrak{so}(d_1)\oplus\mathfrak{so}(d_2)$ \\
\hline
Gauge vector  & $\mathbf{1}$ & $(\mathbf{1}, \mathbf{1}, \mathbf{1}, \mathbf{1})$ &  $(\mathbf{Adj}, \mathbf{1}) \oplus (\mathbf{1},\mathbf{Adj})$ \\
Scalar & $\mathbf{9}_{v}$ & $(\mathbf{1}, \mathbf{1}, \mathbf{1}, \mathbf{1})$ & $(\mathbf{Adj}, \mathbf{1}) \oplus (\mathbf{1},\mathbf{Adj})$\\
Tachyon & $\mathbf{1}$ & $(\mathbf{1}, \mathbf{1}, \mathbf{1}, \mathbf{1})$ & $  \left(\text{\tiny$\yng(1,1)$}, \mathbf{1}\right) \oplus \left(\mathbf{1}, \text{\tiny$\yng(1,1)$}\right)$\\
Dirac spinor  & $\mathbf{16}_{s'}$ & $(\mathbf{1}, \mathbf{1}, \mathbf{1}, \mathbf{1})$ & $(\text{\tiny$\yng(1)$},\text{\tiny$\yng(1)$})$ \\
\hline
\hline
Dirac fermion & $\mathbf{1}$ & $\begin{array}{l}
    (\text{\tiny$\yng(1)$}, \mathbf{1}, \mathbf{1}, \mathbf{1})\\(\mathbf{1}, \text{\tiny$\yng(1)$}, \mathbf{1}, \mathbf{1})\\(\mathbf{1}, \mathbf{1}, \text{\tiny$\yng(1)$}, \mathbf{1})\\(\mathbf{1}, \mathbf{1}, \mathbf{1}, \text{\tiny$\yng(1)$}) \end{array}$ & $\begin{array}{l} (\mathbf{1}, \text{\tiny$\yng(1)$}) \\ (\mathbf{1}, \text{\tiny$\yng(1)$}) \\ (\text{\tiny$\yng(1)$},\mathbf{1}) \\ (\text{\tiny$\yng(1)$},\mathbf{1}) \end{array}$ \\
\hline
Dirac fermion & $\mathbf{9}_{v}$ & $\begin{array}{l}
    (\text{\tiny$\yng(1)$}, \mathbf{1}, \mathbf{1}, \mathbf{1})\\(\mathbf{1}, \text{\tiny$\yng(1)$}, \mathbf{1}, \mathbf{1})\\(\mathbf{1}, \mathbf{1}, \text{\tiny$\yng(1)$}, \mathbf{1})\\(\mathbf{1}, \mathbf{1}, \mathbf{1}, \text{\tiny$\yng(1)$}) \end{array}$ & $\begin{array}{l} (\mathbf{1}, \text{\tiny$\yng(1)$}) \\ (\mathbf{1}, \text{\tiny$\yng(1)$}) \\ (\text{\tiny$\yng(1)$},\mathbf{1}) \\ (\text{\tiny$\yng(1)$},\mathbf{1}) \end{array}$ \\
\end{tabular}}
\caption{Open-string light spectrum for D0-branes in type 0B$/\Omega$.}
\label{tab:0B/Omega_D0_spectrum}
\end{table}

\begin{table}[H]
\centering
\textbf{D2}
\renewcommand{\arraystretch}{1.2}
\centerresize{}{
\begin{tabular}{l|c|c|c}
Worldvolume state  & $\mathfrak{so}(7)_T$ & $\mathfrak{so}(n_1)\oplus\mathfrak{so}(n_2)\oplus\mathfrak{so}(n_3)\oplus\mathfrak{so}(n_4)$ & $\mathfrak{so}(d_1)\oplus\mathfrak{so}(d_2)$ \\
\hline
Gauge vector  & $\mathbf{1}$ & $(\mathbf{1}, \mathbf{1}, \mathbf{1}, \mathbf{1})$ &  $(\mathbf{Adj}, \mathbf{1}) \oplus (\mathbf{1},\mathbf{Adj})$ \\
Scalar & $\mathbf{7}_{v}$ & $(\mathbf{1}, \mathbf{1}, \mathbf{1}, \mathbf{1})$ & $(\mathbf{Adj}, \mathbf{1}) \oplus (\mathbf{1},\mathbf{Adj})$\\
Tachyon & $\mathbf{1}$ & $(\mathbf{1}, \mathbf{1}, \mathbf{1}, \mathbf{1})$ &  $(\mathbf{Adj}, \mathbf{1}) \oplus (\mathbf{1},\mathbf{Adj})$\\
Dirac spinor  & $\mathbf{8}_{s'}$ & $(\mathbf{1}, \mathbf{1}, \mathbf{1}, \mathbf{1})$ & $(\text{\tiny$\yng(1)$},\text{\tiny$\yng(1)$})$ \\
\hline
\hline
Dirac fermion & $\mathbf{1}$ & $\begin{array}{l}
    (\text{\tiny$\yng(1)$}, \mathbf{1}, \mathbf{1}, \mathbf{1})\\(\mathbf{1}, \text{\tiny$\yng(1)$}, \mathbf{1}, \mathbf{1})\\(\mathbf{1}, \mathbf{1}, \text{\tiny$\yng(1)$}, \mathbf{1})\\(\mathbf{1}, \mathbf{1}, \mathbf{1}, \text{\tiny$\yng(1)$}) \end{array}$ & $\begin{array}{l} (\mathbf{1}, \text{\tiny$\yng(1)$}) \\ (\mathbf{1}, \text{\tiny$\yng(1)$}) \\ (\text{\tiny$\yng(1)$},\mathbf{1}) \\ (\text{\tiny$\yng(1)$},\mathbf{1}) \end{array}$ \\
\hline
Dirac fermion & $\mathbf{7}_{v}$ & $\begin{array}{l}
    (\text{\tiny$\yng(1)$}, \mathbf{1}, \mathbf{1}, \mathbf{1})\\(\mathbf{1}, \text{\tiny$\yng(1)$}, \mathbf{1}, \mathbf{1})\\(\mathbf{1}, \mathbf{1}, \text{\tiny$\yng(1)$}, \mathbf{1})\\(\mathbf{1}, \mathbf{1}, \mathbf{1}, \text{\tiny$\yng(1)$}) \end{array}$ & $\begin{array}{l} (\mathbf{1}, \text{\tiny$\yng(1)$}) \\ (\mathbf{1}, \text{\tiny$\yng(1)$}) \\ (\text{\tiny$\yng(1)$},\mathbf{1}) \\ (\text{\tiny$\yng(1)$},\mathbf{1}) \end{array}$ \\
\end{tabular}}
\caption{Open-string light spectrum for D2-branes in type 0B$/\Omega$.}
\label{tab:0B/Omega_D2_spectrum}
\end{table}
\begin{table}[H]
\centering
\textbf{D4}
\renewcommand{\arraystretch}{1.2}
\centerresize{}{
\begin{tabular}{l|c|c|c}
Worldvolume state  & $\mathfrak{so}(5)_T$ & $\mathfrak{so}(n_1)\oplus\mathfrak{so}(n_2)\oplus\mathfrak{so}(n_3)\oplus\mathfrak{so}(n_4)$ & $\mathfrak{sp}(d_1/2)\oplus\mathfrak{sp}(d_2/2)$ \\
\hline
Gauge vector  & $\mathbf{1}$ & $(\mathbf{1}, \mathbf{1}, \mathbf{1}, \mathbf{1})$ &  $(\mathbf{Adj}, \mathbf{1}) \oplus (\mathbf{1},\mathbf{Adj})$ \\
Scalar & $\mathbf{5}_{v}$ & $(\mathbf{1}, \mathbf{1}, \mathbf{1}, \mathbf{1})$ & $(\mathbf{Adj}, \mathbf{1}) \oplus (\mathbf{1},\mathbf{Adj})$\\
Tachyon & $\mathbf{1}$ & $(\mathbf{1}, \mathbf{1}, \mathbf{1}, \mathbf{1})$ &  $(\text{\tiny$\yng(2)$}, \mathbf{1}) \oplus (\mathbf{1}, \text{\tiny$\yng(2)$})$\\
Dirac spinor  & $\mathbf{4}_{s'}$ & $(\mathbf{1}, \mathbf{1}, \mathbf{1}, \mathbf{1})$ & $(\text{\tiny$\yng(1)$},\text{\tiny$\yng(1)$})$ \\
\hline
\hline
Dirac fermion & $\mathbf{1}$ & $\begin{array}{l}
    (\text{\tiny$\yng(1)$}, \mathbf{1}, \mathbf{1}, \mathbf{1})\\(\mathbf{1}, \text{\tiny$\yng(1)$}, \mathbf{1}, \mathbf{1})\\(\mathbf{1}, \mathbf{1}, \text{\tiny$\yng(1)$}, \mathbf{1})\\(\mathbf{1}, \mathbf{1}, \mathbf{1}, \text{\tiny$\yng(1)$}) \end{array}$ & $\begin{array}{l} (\mathbf{1}, \text{\tiny$\yng(1)$}) \\ (\mathbf{1}, \text{\tiny$\yng(1)$}) \\ (\text{\tiny$\yng(1)$},\mathbf{1}) \\ (\text{\tiny$\yng(1)$},\mathbf{1}) \end{array}$ \\
\hline
Dirac fermion & $\mathbf{5}_{v}$ & $\begin{array}{l}
    (\text{\tiny$\yng(1)$}, \mathbf{1}, \mathbf{1}, \mathbf{1})\\(\mathbf{1}, \text{\tiny$\yng(1)$}, \mathbf{1}, \mathbf{1})\\(\mathbf{1}, \mathbf{1}, \text{\tiny$\yng(1)$}, \mathbf{1})\\(\mathbf{1}, \mathbf{1}, \mathbf{1}, \text{\tiny$\yng(1)$}) \end{array}$ & $\begin{array}{l} (\mathbf{1}, \text{\tiny$\yng(1)$}) \\ (\mathbf{1}, \text{\tiny$\yng(1)$}) \\ (\text{\tiny$\yng(1)$},\mathbf{1}) \\ (\text{\tiny$\yng(1)$},\mathbf{1}) \end{array}$ \\
\end{tabular}}
\caption{Open-string light spectrum for D4-branes in type 0B$/\Omega$.}
\label{tab:0B/Omega_D4_spectrum}
\end{table}
\begin{table}[H]
\centering
\textbf{D6}
\renewcommand{\arraystretch}{1.2}
\centerresize{}{
\begin{tabular}{l|c|c|c}
Worldvolume state  & $\mathfrak{so}(3)_T$ & $\mathfrak{so}(n_1)\oplus\mathfrak{so}(n_2)\oplus\mathfrak{so}(n_3)\oplus\mathfrak{so}(n_4)$ & $\mathfrak{sp}(d_1/2)\oplus\mathfrak{sp}(d_2/2)$ \\
\hline
Gauge vector  & $\mathbf{1}$ & $(\mathbf{1}, \mathbf{1}, \mathbf{1}, \mathbf{1})$ &  $(\mathbf{Adj}, \mathbf{1}) \oplus (\mathbf{1},\mathbf{Adj})$ \\
Scalar & $\mathbf{3}_{v}$ & $(\mathbf{1}, \mathbf{1}, \mathbf{1}, \mathbf{1})$ & $(\mathbf{Adj}, \mathbf{1}) \oplus (\mathbf{1},\mathbf{Adj})$\\
Tachyon & $\mathbf{1}$ & $(\mathbf{1}, \mathbf{1}, \mathbf{1}, \mathbf{1})$ &  $\left(\text{\tiny$\yng(1,1)$}, \mathbf{1}\right) \oplus \left(\mathbf{1}, \text{\tiny$\yng(1,1)$}\right)$\\
Dirac spinor  & $\mathbf{2}_{s'}$ & $(\mathbf{1}, \mathbf{1}, \mathbf{1}, \mathbf{1})$ & $(\text{\tiny$\yng(1)$},\text{\tiny$\yng(1)$})$ \\
\hline
\hline
Various tachyons & & & \\
\hline
Dirac fermion & $\mathbf{1}$ & $\begin{array}{l}
    (\text{\tiny$\yng(1)$}, \mathbf{1}, \mathbf{1}, \mathbf{1})\\(\mathbf{1}, \text{\tiny$\yng(1)$}, \mathbf{1}, \mathbf{1})\\(\mathbf{1}, \mathbf{1}, \text{\tiny$\yng(1)$}, \mathbf{1})\\(\mathbf{1}, \mathbf{1}, \mathbf{1}, \text{\tiny$\yng(1)$}) \end{array}$ & $\begin{array}{l} (\mathbf{1}, \text{\tiny$\yng(1)$}) \\ (\mathbf{1}, \text{\tiny$\yng(1)$}) \\ (\text{\tiny$\yng(1)$},\mathbf{1}) \\ (\text{\tiny$\yng(1)$},\mathbf{1}) \end{array}$ \\
\hline
Dirac fermion & $\mathbf{3}_{v}$ & $\begin{array}{l}
    (\text{\tiny$\yng(1)$}, \mathbf{1}, \mathbf{1}, \mathbf{1})\\(\mathbf{1}, \text{\tiny$\yng(1)$}, \mathbf{1}, \mathbf{1})\\(\mathbf{1}, \mathbf{1}, \text{\tiny$\yng(1)$}, \mathbf{1})\\(\mathbf{1}, \mathbf{1}, \mathbf{1}, \text{\tiny$\yng(1)$}) \end{array}$ & $\begin{array}{l} (\mathbf{1}, \text{\tiny$\yng(1)$}) \\ (\mathbf{1}, \text{\tiny$\yng(1)$}) \\ (\text{\tiny$\yng(1)$},\mathbf{1}) \\ (\text{\tiny$\yng(1)$},\mathbf{1}) \end{array}$ \\
\end{tabular}}
\caption{Open-string light spectrum for D6-branes in type 0B$/\Omega$.}
\label{tab:0B/Omega_D6_spectrum}
\end{table}
\begin{table}[H]
\centering
\textbf{D8}
\renewcommand{\arraystretch}{1.2}
\centerresize{}{
\begin{tabular}{l|c|c|c}
Worldvolume state  & $\mathfrak{so}(1)_T$ & $\mathfrak{so}(n_1)\oplus\mathfrak{so}(n_2)\oplus\mathfrak{so}(n_3)\oplus\mathfrak{so}(n_4)$ & $\mathfrak{so}(d_1)\oplus\mathfrak{so}(d_2)$ \\
\hline
Gauge vector  & $\mathbf{1}$ & $(\mathbf{1}, \mathbf{1}, \mathbf{1}, \mathbf{1})$ &  $(\mathbf{Adj}, \mathbf{1}) \oplus (\mathbf{1},\mathbf{Adj})$ \\
Scalar & $\mathbf{1}_{v}$ & $(\mathbf{1}, \mathbf{1}, \mathbf{1}, \mathbf{1})$ & $(\mathbf{Adj}, \mathbf{1}) \oplus (\mathbf{1},\mathbf{Adj})$\\
Tachyon & $\mathbf{1}$ & $(\mathbf{1}, \mathbf{1}, \mathbf{1}, \mathbf{1})$ & $\left(\text{\tiny$\yng(1,1)$}, \mathbf{1}\right) \oplus \left(\mathbf{1}, \text{\tiny$\yng(1,1)$}\right)$\\
Dirac spinor  & $\mathbf{1}_{s'}$ & $(\mathbf{1}, \mathbf{1}, \mathbf{1}, \mathbf{1})$ & $(\text{\tiny$\yng(1)$},\text{\tiny$\yng(1)$})$ \\
\hline
\hline
Various tachyons & & & \\
\hline
Dirac fermion & $\mathbf{1}$ & $\begin{array}{l}
    (\text{\tiny$\yng(1)$}, \mathbf{1}, \mathbf{1}, \mathbf{1})\\(\mathbf{1}, \text{\tiny$\yng(1)$}, \mathbf{1}, \mathbf{1})\\(\mathbf{1}, \mathbf{1}, \text{\tiny$\yng(1)$}, \mathbf{1})\\(\mathbf{1}, \mathbf{1}, \mathbf{1}, \text{\tiny$\yng(1)$}) \end{array}$ & $\begin{array}{l} (\mathbf{1}, \text{\tiny$\yng(1)$}) \\ (\mathbf{1}, \text{\tiny$\yng(1)$}) \\ (\text{\tiny$\yng(1)$},\mathbf{1}) \\ (\text{\tiny$\yng(1)$},\mathbf{1}) \end{array}$ \\
\hline
Dirac fermion & $\mathbf{1}_{v}$ & $\begin{array}{l}
    (\text{\tiny$\yng(1)$}, \mathbf{1}, \mathbf{1}, \mathbf{1})\\(\mathbf{1}, \text{\tiny$\yng(1)$}, \mathbf{1}, \mathbf{1})\\(\mathbf{1}, \mathbf{1}, \text{\tiny$\yng(1)$}, \mathbf{1})\\(\mathbf{1}, \mathbf{1}, \mathbf{1}, \text{\tiny$\yng(1)$}) \end{array}$ & $\begin{array}{l} (\mathbf{1}, \text{\tiny$\yng(1)$}) \\ (\mathbf{1}, \text{\tiny$\yng(1)$}) \\ (\text{\tiny$\yng(1)$},\mathbf{1}) \\ (\text{\tiny$\yng(1)$},\mathbf{1}) \end{array}$ \\
\end{tabular}}
\caption{Open-string light spectrum for D8-branes in type 0B$/\Omega$.}
\label{tab:0B/Omega_D8_spectrum}
\end{table}

\section{Spectrum of bosonic string on the \texorpdfstring{$T^{16}$}{T16} \texorpdfstring{$SO(32)$}{SO(32)} lattice}\label{app:bosonic_on_T16}
In this appendix, we derive the light spectrum for the bosonic string compactified on a set of special $T^{16}$ giving massless states transforming in the adjoint of $\mathfrak{so}(32)_L \oplus \mathfrak{so}(32)_R$, using the fermionic formulation, identifying the dual states to the ones in the type 0B orientifold \cite{Michishita:1999it}.

Bosonic string theory compactified on a 16-dimensional torus leads to ten-dimensional tachyonic theories whose spectrum contains massless gauge bosons transforming in the adjoint representation of some rank-32 group. For generic values of the geometry of the $T^{16}$, this gauge group is $U(1)^{16}_L \times U(1)^{16}_R$, while for special values, additional states become massless and the gauge group is enhanced to some non-abelian group. 

As discussed in section \ref{sec:Bergman-Gaberdiel}, the Bergman--Gaberdiel duality concerns not only the algebra $\mathfrak{so}(32)_L\oplus\mathfrak{so}(32)_R$ seen by the massless states, but the precise lattice of allowed weights, containing the information about the tachyonic and massive spectrum, encoded in the gauge group $\frac{Spin(32)\times Spin(32)}{\mathbb{Z}_2 \times \mathbb{Z}_2}$. 

The worldsheet fields are 8 bosons $X^{\mu}(z,\bar{z})$ with $\mu=2,\dots 9$ a spacetime Lorentz index
representing the spacetime coordinates of the string in the light-cone gauge; the fermionized compact directions become 32 left- and 32 right-moving fermions $\psi^{A}(z)$ and $\widetilde{\psi}^{\widetilde{A}}(\bar{z})$.
For the special set of $T^{16}$ we are interested in, each set of fermions gets common boundary conditions, and we can take $A=1,\dots,32$, $\widetilde A=1,\dots,32$ as indices in the $(\mathbf{32,1})$ and $(\mathbf{1,32})$ fundamental irreps of the gauge algebra $\mathfrak{so}(32)_L \oplus \mathfrak{so}(32)_R$. String left(right)-moving bosonic oscillations are encoded in the operators ${\alpha}^{\mu}_{-n}$($\widetilde\alpha^{\mu}_{-n}$) with $n\in \mathbb{Z}$. Fermionic left(right)-moving oscillation operators are ${\psi}^{A}_{-r}$($\widetilde{\psi}^{A}_{-r}$), where $r$ is integer in the R sector and half-integer in the NS sector. 

For each chiral set of 32 real fermions, the NS sector is built by acting with half-integer modes on the vacuum $\ket{0}$, while the Ramond sector contains fermion zero modes satisfying a Clifford algebra. The Ramond ground states therefore form the Dirac spinor representation of $\mathfrak{so}(32)$, which decomposes into the two chiral spinors $\mathbf{32768_s}\oplus \mathbf{32768_c}$. We denote these ground states by $\ket{a}$ and $\ket{\dot a}$, respectively.

We denote the left-moving NS vacuum, which has $m_L^2=-1$, as $\ket{0}_L$. It is a spacetime boson and a singlet under $\mathfrak{so}(32)_L$.
It has eigenvalue $+1$ under $(-1)^{f_L}$.

The left-moving Ramond sector has $m_L^2 = +1$. In this sector the fermions have zero modes
\begin{align}
\psi^A_{0}\,,\qquad A=1,\dots,32\,,
\end{align}
satisfying the Clifford algebra
\begin{align}
\lbrace \psi^A_0, \psi^B_0 \rbrace = \delta^{AB}\,.
\end{align}
These zero modes act on a degenerate space of Ramond ground states. To describe it explicitly, we introduce the creation/annihilation operators
\begin{equation}
 b^{\dagger}_{i} =  \frac{\psi_0^{2i - 1} + i \psi_0^{2i}}{\sqrt{2}}\,\quad,\quad
b_{i} =  \frac{\psi_0^{2i - 1} - i \psi_0^{2i}}{\sqrt{2}}\,, \qquad i=1,\dots,16\,,
 \end{equation}
so that
\begin{equation}
 \lbrace b_i, b^{\dagger}_j \rbrace = \delta_{ij}\,.
 \end{equation}
Choosing a Clifford vacuum $\ket{\rm low}_L$ annihilated by all the $b_i$, a basis of Ramond ground states is
\begin{align}
\prod_{i=1}^{16}(b^\dagger_i)^{n_i} \ket{\rm low}_L \qquad n_i=0,1\, .
\end{align}
Thus the Ramond ground states form the $2^{16}$-dimensional Dirac spinor representation of $\mathfrak{so}(32)_L$, which decomposes into the two chiral spinor representations according to the parity of $\sum_i n_i$:
\begin{align}
\ket{a}_L &:\quad  \sum_{i=1}n_i \in 2\mathbb{Z}\,, & \mathbf{32768_s}\,,\\
\ket{\dot{a}}_L &:\quad \sum_{i=1}n_i \in 2\mathbb{Z}+1\,, & \mathbf{32768_c} \,.
\end{align}
The left-moving fermion parity acts as
\begin{align}
(-1)^{f_L} \prod_{i=1}^{16}(b_i^\dagger)^{n_i} \ket{\rm low}_L =
(-1)^{\sum_i n_i} \prod_{i=1}^{16}(b_i^\dagger)^{n_i} \ket{\rm low}_L\,.
\end{align}
The right-moving Ramond sector is completely analogous. We denote its chiral spinor ground states by 
\begin{align}
\ket{\tilde{a}}_R\,,\qquad \ket{\dot{\tilde{a}}}_R\,,
\end{align}
transforming in the $\mathbf{32768_s}$ and the $\mathbf{32768_c}$ of $\mathfrak{so}(32)_R$, respectively.

By combining states from a right and a left sector with the same mass, we build closed string states.

In table \ref{tab:bosT16lightspectrum}, we list level-matched states with the lowest masses and indicate their weights under the gauge symmetry $\mathfrak{so}(32)^2$ and little group ($SO(8)$ for massless states, $SO(9)$ for massive states).
\begin{table}[H]\centering\renewcommand{\arraystretch}{1.1}
\centerresize{}{
\begin{tabular}{
@{\,}>{$}c<{$}@{\,}|
@{\,}>{}l<{}@{\,}
@{\,}>{$}r<{$}@{\,}|
@{\,}>{$}r<{$}@{\,}|
@{\,}>{$}c<{$}@{\,}|
@{\,}>{$}c<{$}@{\,}|
@{\,}>{}c<{}@{\,}}
m^2 & \text{Sector} &    & \text{State}\quad\quad\quad\quad\quad\quad\quad\quad & \mathfrak{so}(32)\oplus \mathfrak{so}(32)& \text{little group} & \\ \hline
-1 & NS{\rm -}NS & O_{32} \overline{O}_{32}& \ket{0}_L \ket{0}_R & \mathbf{(1,1)} & \mathbf{1} &\cmark \\ \hline
-\frac12 & NS{\rm -}NS & V_{32}\overline{V}_{32}& {{\psi}}^{{A}}_{\text{-}1/2}\widetilde{\psi}^{\widetilde A}_{\text{-}1/2} \ket{0}_L \ket{0}_R & \mathbf{(32,32)} & \mathbf{1} &\cmark\\ \hline
0 & NS{\rm -}NS & O_{32} \overline{O}_{32} & {\alpha}^{\mu}_{\text{-}1}\widetilde{\alpha}^{\nu}_{\text{-}1}\ket{0}_L \ket{0}_R & \mathbf{(1,1)}& \mathbf{1\oplus 28 \oplus 35_v} &\cmark \\
0 & NS{\rm -}NS & O_{32} \overline{O}_{32} & {{\psi}}^{{A}}_{\text{-}1/2}{{\psi}}^{{B}}_{\text{-}1/2}\widetilde{\alpha}^{\nu}_{\text{-}1}\ket{0}_L \ket{0}_R & \mathbf{(496,1)} &\mathbf{8_v}&\cmark \\
0 & NS{\rm -}NS & O_{32} \overline{O}_{32} & {\alpha}^{\mu}_{\text{-}1}\widetilde{\psi}^{\widetilde A}_{\text{-}1/2}\widetilde{\psi}^{\widetilde B}_{\text{-}1/2} \ket{0}_L \ket{0}_R & \mathbf{(1,496)}&\mathbf{8_v} &\cmark \\
0 & NS{\rm -}NS & O_{32} \overline{O}_{32} & {{\psi}}^{{A}}_{\text{-}1/2}{{\psi}}^{{B}}_{\text{-}1/2}\widetilde{\psi}^{\widetilde A}_{\text{-}1/2}\widetilde{\psi}^{\widetilde B}_{\text{-}1/2} \ket{0}_L \ket{0}_R & \mathbf{(496,496)} & \mathbf{1} &\cmark \\
\hline
\frac12 & NS{\rm -}NS & V_{32} \overline{V}_{32} & (\alpha_{\text{-}1}^{\mu}{{\psi}}_{\text{-}1/2}^{{A}}+{{\psi}}_{\text{-}3/2}^{{A}})(\widetilde{\alpha}_{\text{-}1}^{\nu}{\widetilde{\psi}}_{\text{-}1/2}^{\widetilde{A}}+{\widetilde{\psi}}_{\text{-}3/2}^{\widetilde{A}})\ket{0}_L \ket{0}_R & \mathbf{(32,32)}  & \mathbf{1\oplus 36 \oplus 44} &\cmark \\
\frac12 & NS{\rm -}NS & V_{32} \overline{V}_{32}& {{\psi}}_{\text{-}1/2}^{{A}} {{\psi}}_{\text{-}1/2}^{{B}} {{\psi}}_{\text{-}1/2}^{{C}}(\widetilde{\alpha}_{\text{-}1}^{\mu}{\widetilde{\psi}}_{\text{-}1/2}^{ \widetilde{A}}+{\widetilde{\psi}}_{\text{-}3/2}^{ \widetilde{A}})\ket{0}_L \ket{0}_R & \mathbf{(4960,32)} & \mathbf{9}&\cmark \\
\frac12 & NS{\rm -}NS & V_{32} \overline{V}_{32}& (\alpha_{\text{-}1}^{\mu}{{\psi}}_{\text{-}1/2}^{{A}}+{{\psi}}_{\text{-}3/2}^{{A}}){\widetilde{\psi}}_{\text{-}1/2}^{\widetilde{A}}{\widetilde{\psi}}_{\text{-}1/2}^{\widetilde{B}}\widetilde{\psi}_{\text{-}1/2}^{\widetilde C}\ket{0}_L \ket{0}_R & \mathbf{(32,4960)} & \mathbf{9}& \cmark \\
\frac12 & NS{\rm -}NS & V_{32} \overline{V}_{32}& {{\psi}}_{\text{-}1/2}^{{A}} {{\psi}}_{\text{-}1/2}^{{B}} {{\psi}}_{\text{-}1/2}^{{C}}\widetilde{\psi}_{\text{-}1/2}^{\widetilde A}\widetilde{\psi}_{\text{-}1/2}^{\widetilde B}\widetilde{\psi}_{\text{-}1/2}^{\widetilde C}\ket{0}_L \ket{0}_R & \mathbf{(4960,4960)} & \mathbf{1}&\cmark \\
 \hline
1 & R{\rm -}R & S_{32}\overline{S}_{32} & \ket{a}_L  \ket{\widetilde{a}}_R & \mathbf{(32768_s,32768_s)} & \mathbf{1}&\cmark \\
1 & R{\rm -}R & C_{32}\overline{C}_{32}& \ket{\dot{a}}_L  \ket{\widetilde{\dot{a}}}_R & \mathbf{(32768_c,32768_c)} & \mathbf{1} &\cmark \\
1 & R{\rm -}R & C_{32}\overline{S}_{32}& \ket{\dot{a}}_L  \ket{\widetilde{a}}_R & \mathbf{(32768_c,32768_s)}& \mathbf{1} &\xmark \\
1 & R{\rm -}R & S_{32}\overline{C}_{32}& \ket{a}_L  \ket{\widetilde{\dot{a}}}_R & \mathbf{(32768_s,32768_c)}& \mathbf{1} &\xmark \\ 
\hline 
1 & R{\rm -}NS & S_{32}\overline{O}_{32}& (\widetilde{\alpha}_{\text{-}2}^{\mu}+\widetilde{\alpha}_{\text{-}1}^{\mu}\widetilde{\alpha}_{\text{-}1}^{\nu}) \ket{a}_L \ket{0}_R  &  \mathbf{(32768_s,1)} &\mathbf{44}& \xmark \\
1 & R{\rm -}NS & S_{32}\overline{O}_{32}& (\widetilde{\alpha}_{\text{-}1}^{\mu}\widetilde{\psi}_{\text{-}1/2}^{[\widetilde A}+\widetilde{\psi}_{\text{-}3/2}^{[\widetilde A})\widetilde{\psi}_{\text{-}1/2}^{\widetilde B]} \ket{a}_L \ket{0}_R  &  \mathbf{(32768_s,496)} &\mathbf{9}&\xmark \\
1 & R{\rm -}NS & S_{32}\overline{O}_{32}& \widetilde{\psi}_{\text{-}3/2}^{\widetilde (A}\widetilde{\psi}_{\text{-}1/2}^{\widetilde B)} \ket{a}_L \ket{0}_R  &  \mathbf{(32768_s,527\oplus 1)} & \mathbf{1}& \xmark \\
1 & R{\rm -}NS & S_{32}\overline{O}_{32}& \widetilde{\psi}_{\text{-}1/2}^{\widetilde A}\widetilde{\psi}_{\text{-}1/2}^{\widetilde B}\widetilde{\psi}_{\text{-}1/2}^{\widetilde C}\widetilde{\psi}_{\text{-}1/2}^{\widetilde D} \ket{a}_L \ket{0}_R  &  \mathbf{(32768_s,35960)} & \mathbf{1}&\xmark \\
1 & R{\rm -}NS & C_{32}\overline{O}_{32}& (\widetilde{\alpha}_{\text{-}2}^{\mu}+\widetilde{\alpha}_{\text{-}1}^{\mu}\widetilde{\alpha}_{\text{-}1}^{\nu}) \ket{\dot{a}}_L \ket{0}_R & \mathbf{(32768_c,1)} &\mathbf{44}& \xmark \\
1 & R{\rm -}NS & C_{32}\overline{O}_{32}& (\widetilde{\alpha}_{\text{-}1}^{\mu}\widetilde{\psi}_{\text{-}1/2}^{[\widetilde A}+\widetilde{\psi}_{\text{-}3/2}^{[\widetilde A})\widetilde{\psi}_{\text{-}1/2}^{\widetilde B]}\ket{\dot{a}}_L \ket{0}_R & \mathbf{(32768_c,496)} &\mathbf{9}& \xmark \\
1 & R{\rm -}NS & C_{32}\overline{O}_{32}& \widetilde{\psi}_{\text{-}3/2}^{\widetilde (A}\widetilde{\psi}_{\text{-}1/2}^{\widetilde B)} \ket{\dot{a}}_L \ket{0}_R  &  \mathbf{(32768_c,527 \oplus 1)} & \mathbf{1}&\xmark \\
1 & R{\rm -}NS & C_{32}\overline{O}_{32}& \widetilde{\psi}_{\text{-}1/2}^{\widetilde A}\widetilde{\psi}_{\text{-}1/2}^{\widetilde B}\widetilde{\psi}_{\text{-}1/2}^{\widetilde C}\widetilde{\psi}_{\text{-}1/2}^{\widetilde D} \ket{\dot{a}}_L \ket{0}_R  &  \mathbf{(32768_c,35960)} & \mathbf{1}& \xmark \\
\end{tabular}}\caption{Light spectrum before applying a GSO projection. For each sector, only the first massive level is shown. The NS-R sector is analogous to the R-NS sector. The last column indicates whether the corresponding states survive or not the projection by $\frac{(1+(-1)^f)(1+(-1)^{f'})}{4}$.}\label{tab:bosT16lightspectrum}
\end{table}
The $q$-expansion of the total partition function is, neglecting unphysical states,
\begin{align}
(q \overline{q})^{-1} + 1024 \, (q \overline{q})^{-\frac12}
+ 254016 + 27541504 \, (q \overline{q})^{\frac12} + 11349067024  \, q \overline{q} \,.
\end{align}

In addition to the fermion numbers $f_L$ and $f_R$ defined above, it is useful to introduce operators $f'_L$ and $f'_R$. On NS states, $(-1)^{f'_L}$ and $(-1)^{f_L}$ act in the same way: they count the number of left-moving fermionic oscillators modulo two, and similarly for the right-moving sector. The difference is in the Ramond sector: $(-1)^{f'_L}$ exchanges the signs assigned by $(-1)^{f_L}$ to the two chiral Ramond ground states, $\ket{a}_L$, $\ket{\dot a}_L$, and analogously for the right-moving ground states. We also define
\begin{align}
(-1)^{g_L}=(-1)^{f_L+f'_L},\qquad  (-1)^{g_R}=(-1)^{f_R+f'_R}.
\end{align}
With these definitions, the charges of the sectors are summarized in table \ref{tab:chargesBosT16}.
\begin{table}[H]\centering\renewcommand{\arraystretch}{1.1}
\small{
\begin{tabular}{@{\,}>{}l<{}@{\,}@{\,}>{}r<{}@{\,}|@{\,}>{}c<{}@{\,}@{\,}>{}c<{}@{\,}@{\,}>{}c<{}@{\,}@{\,}>{}c<{}@{\,}@{\,}>{}c<{}@{\,}@{\,}>{}c<{}@{\,}@{\,}>{}c<{}@{\,}@{\,}>{}c<{}@{\,}@{\,}>{}c<{}@{\,}}
Sector&&
$(-1)^{f_L}$ & 
$(-1)^{f_R}$ & 
$(-1)^{f'_L}$ & 
$(-1)^{f'_R}$ &
$(-1)^{f}$ & 
$(-1)^{f'}$ & 
$(-1)^{g_L}$ & 
$(-1)^{g_R}$ &
$(-1)^{g}$ 
\\ \hline
NS{\rm -}NS & $O_{32} \overline{O}_{32}$ & + & + & + & + & + & + & + & + & + \\
NS{\rm -}NS &$V_{32} \overline{V}_{32}$ & $-$ & $-$ & $-$ & $-$ & + & + & + & + & + \\
NS{\rm -}NS & $O_{32} \overline{V}_{32}$ & + & $-$ & + & $-$ & $-$ & $-$ & + & + & + \\
NS{\rm -}NS & $V_{32} \overline{O}_{32}$ & $-$ & + & $-$ & + & $-$ & $-$ & + & + & + \\
\hline
R{\rm -}R & $S_{32} \overline{S}_{32}$ & + & + & $-$ & $-$ & + & + & $-$ & $-$ & + \\
R{\rm -}R & $C_{32} \overline{C}_{32}$ & $-$ & $-$ & + & + & + & + & $-$ & $-$ & + \\
R{\rm -}R & $S_{32} \overline{C}_{32}$ & + & $-$ & $-$ & + & $-$ & $-$ & $-$ & $-$ & +\\
R{\rm -}R  & $C_{32} \overline{S}_{32}$ & $-$ & + & + & $-$ & $-$ & $-$ & $-$ & $-$ & + \\
 \hline
NS{\rm -}R  & $O_{32} \overline{S}_{32}$ & + & + & + & $-$ & + & $-$ & + & $-$ & $-$\\
NS{\rm -}R  &$V_{32} \overline{C}_{32}$ & $-$ & $-$ & $-$ & + & + & $-$ & + & $-$ & $-$ \\
NS{\rm -}R  & $O_{32} \overline{C}_{32}$ & + & $-$ & + & + & $-$ & + & + & $-$ & $-$\\
NS{\rm -}R  &$V_{32} \overline{S}_{32}$ & $-$ & + & $-$ & $-$ & $-$ & + & + & $-$ & $-$\\
 \hline
R{\rm -}NS & $S_{32} \overline{O}_{32}$ & + & + & $-$ & + & + & $-$ & $-$ & + & $-$\\
R{\rm -}NS & $C_{32} \overline{V}_{32}$ & $-$ & $-$ & + & $-$&+&$-$&$-$&+&$-$\\
R{\rm -}NS & $S_{32} \overline{V}_{32}$ & + & $-$ & $-$ & $-$ & $-$ & + & $-$ & + & $-$\\
R{\rm -}NS & $C_{32} \overline{O}_{32}$ & $-$ & + & + & +&$-$&+&$-$&+&$-$\\
\end{tabular}}\caption{Charges of the combinations of $\mathfrak{so}(32)$ characters under the operators $\lbrace (-1)^{f_L},(-1)^{f_R},(-1)^{f'_L},(-1)^{f'_R}\rbrace$ and the corresponding diagonal combinations.}\label{tab:chargesBosT16}
\end{table}
$\lbrace (-1)^{f_L},(-1)^{f_R},(-1)^{f'_L},(-1)^{f'_R}\rbrace$ generate a $\mathbb{Z}_2^4$ Abelian symmetry. Every state of the theory has an associated charge given by one of the 16 elements of this group. Each GSO projection corresponds to the states whose charges are invariant under some $\mathbb{Z}_2^2$ subgroup.

All states in NS-NS and R-R sectors survive the $\frac{1+(-1)^g}{2}$ projection. None of the states in NS-R and R-NS survive it.

The first two rows of each sector survive the $\frac{1+(-1)^f}{2}$ projection, they correspond to the combination of characters $O_{32}\overline{O}_{32}$, $V_{32}\overline{V}_{32}$, $S_{32}\overline{S}_{32}$, $C_{32}\overline{C}_{32}$, $O_{32}\overline{S}_{32}$, $V_{32}\overline{C}_{32}$, $S_{32}\overline{O}_{32}$ and $C_{32}\overline{V}_{32}$.

We list the modular-invariant projections obtained from these $\mathbb Z_2^2$ subgroups in table~\ref{tab:projections_bosT16}.
\begin{table}[H]\centering\renewcommand{\arraystretch}{1.4}\centerresize{}{
\begin{tabular}{c|c|c}
Projector & Characters & $q$-expansion \\ \hline
$\frac{(1+(-1)^f)(1+(-1)^{f'})}{4}$ & $O_{32}\overline{O}_{32} + V_{32}\overline{V}_{32} + S_{32}\overline{S}_{32} + C_{32}\overline{C}_{32}$ & $(q \overline{q})^{-1} + 1024 \, (q \overline{q})^{-\frac12}
+ 254016\,+ $  \\
$\frac{(1+(-1)^{f_L+f_R'})(1+(-1)^{f_L'+f_R})}{4}$ & $O_{32}\overline{O}_{32} + V_{32}\overline{V}_{32} + S_{32}\overline{C}_{32} + C_{32}\overline{S}_{32}$ & $+ 27541504 \, (q \overline{q})^{\frac12} +3828155664  \, q \overline{q}$ \\ \hline
$\frac{(1+(-1)^{f_L})(1+(-1)^{f_R})}{4}$ & $(O_{32}+S_{32})(\overline{O}_{32} + \overline{S}_{32})$ & \multirow{4}{*}{$(q \overline{q})^{-1}
+ 254016 +  5441127696  \, q \overline{q}$} \\
$\frac{(1+(-1)^{f'_L})(1+(-1)^{f'_R})}{4}$ & $(O_{32}+C_{32})(\overline{O}_{32} + \overline{C}_{32})$ &  \\
$\frac{(1+(-1)^{f_L})(1+(-1)^{f'_R})}{4}$ & $(O_{32}+S_{32})(\overline{O}_{32} + \overline{C}_{32})$ &  \\
$\frac{(1+(-1)^{f'_L})(1+(-1)^{f_R})}{4}$ & $(O_{32}+C_{32})(\overline{O}_{32} + \overline{S}_{32})$ & 
\end{tabular}}\caption{Modular invariant $\mathbb{Z}_2^2$ projections generated by the fermion-number operators in table~\ref{tab:chargesBosT16}. The first row is the diagonal invariant relevant for the Bergman--Gaberdiel proposal; the second row differs by exchanging one Ramond chirality, while the separative invariants in the last four rows are shown for comparison.}\label{tab:projections_bosT16}
\end{table}
The projections $\frac{(1+(-1)^f)(1+(-1)^{f'})}{4}$ and $\frac{(1+(-1)^{f_L})(1+(-1)^{f_R})}{4}$ are the ``diagonal'' and ``separative'' modular invariant projections discussed in \cite{Michishita:1999it}. The former gives the gauge group $\frac{Spin(32)\times Spin(32)}{\mathbb{Z}_2\times \mathbb{Z}_2}$ and is the candidate dual of Bergman--Gaberdiel; the latter gives $\frac{Spin(32)}{\mathbb{Z}_2}\times \frac{Spin(32)}{\mathbb{Z}_2}$.
Note that the projections $\frac{(1+(-1)^{f'_L})(1+(-1)^{f'_R})}{4}$, $\frac{(1+(-1)^{f_L})(1+(-1)^{f'_R})}{4}$, $\frac{(1+(-1)^{f'_L})(1+(-1)^{f_R})}{4}$, and $\frac{(1+(-1)^{f_L+f_R'})(1+(-1)^{f_L'+f_R})}{4}$ produce analogous spectra, exchanging one or both chiralities of the spinor representations of $\mathfrak{so}(32)_L\oplus \mathfrak{so}(32)_R$.

The spectrum for the diagonal projection by  $\frac{(1+(-1)^f)(1+(-1)^{f'})}{4}$ consists of the rows from table \ref{tab:bosT16lightspectrum} marked with a dot, corresponding to the characters
\begin{align}
O_{32}\overline{O}_{32} + V_{32}\overline{V}_{32} + S_{32}\overline{S}_{32} + C_{32}\overline{C}_{32}\,,\end{align}
hence the allowed weight classes are exactly $(o,o)$, $(v,v)$, $(s,s)$, $(c,c)$. Therefore the diagonal center $\lbrace{(1,1),(z_v,z_v),(z_s,z_s),(z_c,z_c)\rbrace}$ acts trivially, giving the quotient of $Spin(32)\times Spin(32)$ by $\mathbb{Z}_2 \times \mathbb{Z}_2$, as in~\eqref{eq:gauge_group_bosonic_lattice}.

\bibliographystyle{utphys}
\bibliography{bibliography}

@article{Bedroya:2021fbu,
    author = "Bedroya, Alek and Hamada, Yuta and Montero, Miguel and Vafa, Cumrun",
    title = "{Compactness of brane moduli and the String Lamppost Principle in d \ensuremath{>} 6}",
    eprint = "2110.10157",
    archivePrefix = "arXiv",
    primaryClass = "hep-th",
    doi = "10.1007/JHEP02(2022)082",
    journal = "JHEP",
    volume = "02",
    pages = "082",
    year = "2022"
}

@article{Hamada:2021bbz,
    author = "Hamada, Yuta and Vafa, Cumrun",
    title = "{8d supergravity, reconstruction of internal geometry and the Swampland}",
    eprint = "2104.05724",
    archivePrefix = "arXiv",
    primaryClass = "hep-th",
    doi = "10.1007/JHEP06(2021)178",
    journal = "JHEP",
    volume = "06",
    pages = "178",
    year = "2021"
}

@article{Montero:2020icj,
    author = "Montero, Miguel and Vafa, Cumrun",
    title = "{Cobordism Conjecture, Anomalies, and the String Lamppost Principle}",
    eprint = "2008.11729",
    archivePrefix = "arXiv",
    primaryClass = "hep-th",
    doi = "10.1007/JHEP01(2021)063",
    journal = "JHEP",
    volume = "01",
    pages = "063",
    year = "2021"
}

@article{Kim:2019ths,
    author = "Kim, Hee-Cheol and Tarazi, Houri-Christina and Vafa, Cumrun",
    title = "{Four-dimensional $\mathbf{\mathcal{N}=4}$ SYM theory and the swampland}",
    eprint = "1912.06144",
    archivePrefix = "arXiv",
    primaryClass = "hep-th",
    doi = "10.1103/PhysRevD.102.026003",
    journal = "Phys. Rev. D",
    volume = "102",
    number = "2",
    pages = "026003",
    year = "2020"
}

@article{Adams:2010zy,
    author = "Adams, Allan and DeWolfe, Oliver and Taylor, Washington",
    title = "{String universality in ten dimensions}",
    eprint = "1006.1352",
    archivePrefix = "arXiv",
    primaryClass = "hep-th",
    reportNumber = "COLO-HEP-554, MIT-CTP-4155",
    doi = "10.1103/PhysRevLett.105.071601",
    journal = "Phys. Rev. Lett.",
    volume = "105",
    pages = "071601",
    year = "2010"
}

@article{Kumar:2009us,
    author = "Kumar, Vijay and Taylor, Washington",
    title = "{String Universality in Six Dimensions}",
    eprint = "0906.0987",
    archivePrefix = "arXiv",
    primaryClass = "hep-th",
    reportNumber = "MIT-CTP-4046",
    doi = "10.4310/ATMP.2011.v15.n2.a3",
    journal = "Adv. Theor. Math. Phys.",
    volume = "15",
    number = "2",
    pages = "325--353",
    year = "2011"
}

@article{Hamada:2023zol,
    author = "Hamada, Yuta and Loges, Gregory J.",
    title = "{Towards a complete classification of 6D supergravities}",
    eprint = "2311.00868",
    archivePrefix = "arXiv",
    primaryClass = "hep-th",
    reportNumber = "KEK-TH-2572",
    doi = "10.1007/JHEP02(2024)095",
    journal = "JHEP",
    volume = "02",
    pages = "095",
    year = "2024"
}

@phdthesis{Raucci:2024fnp,
    author = "Raucci, Salvatore",
    title = "{Spacetime aspects of non-supersymmetric strings}",
    eprint = "2409.19395",
    archivePrefix = "arXiv",
    primaryClass = "hep-th",
    school = "Pisa, Scuola Normale Superiore",
    month = "9",
    year = "2024"
}

@article{Basile:2021vxh,
    author = "Basile, Ivano",
    title = "{Supersymmetry breaking and stability in string vacua: Brane dynamics, bubbles and the swampland}",
    eprint = "2107.02814",
    archivePrefix = "arXiv",
    primaryClass = "hep-th",
    doi = "10.1007/s40766-021-00024-9",
    journal = "Riv. Nuovo Cim.",
    volume = "44",
    number = "10",
    pages = "499--596",
    year = "2021"
}

@article{Witten:1998cd,
    author = "Witten, Edward",
    title = "{D-branes and K-theory}",
    eprint = "hep-th/9810188",
    archivePrefix = "arXiv",
    reportNumber = "IASSNS-HEP-98-82",
    doi = "10.1088/1126-6708/1998/12/019",
    journal = "JHEP",
    volume = "12",
    pages = "019",
    year = "1998"
}

@article{Blum:1997gw,
    author = "Blum, Julie D. and Dienes, Keith R.",
    title = "{Strong / weak coupling duality relations for nonsupersymmetric string theories}",
    eprint = "hep-th/9707160",
    archivePrefix = "arXiv",
    reportNumber = "IASSNS-HEP-97-80",
    doi = "10.1016/S0550-3213(97)00803-1",
    journal = "Nucl. Phys. B",
    volume = "516",
    pages = "83--159",
    year = "1998"
}

@article{Blum:1997cs,
    author = "Blum, Julie D. and Dienes, Keith R.",
    title = "{Duality without supersymmetry: The Case of the SO(16) x SO(16) string}",
    eprint = "hep-th/9707148",
    archivePrefix = "arXiv",
    reportNumber = "IASSNS-HEP-97-67",
    doi = "10.1016/S0370-2693(97)01172-6",
    journal = "Phys. Lett. B",
    volume = "414",
    pages = "260--268",
    year = "1997"
}

@article{Witten:2023snr,
    author = "Witten, Edward",
    title = "{Anomalies and Nonsupersymmetric D-Branes}",
    eprint = "2305.01012",
    archivePrefix = "arXiv",
    primaryClass = "hep-th",
    month = "5",
    year = "2023"
}

@inproceedings{Sen:1999mg,
    author = "Sen, Ashoke",
    title = "{NonBPS states and Branes in string theory}",
    booktitle = "{Advanced School on Supersymmetry in the Theories of Fields, Strings and Branes}",
    eprint = "hep-th/9904207",
    archivePrefix = "arXiv",
    reportNumber = "MRI-PHY-P990411",
    pages = "187--234",
    month = "1",
    year = "1999"
}

@article{Sen:1998ki,
    author = "Sen, Ashoke",
    title = "{Type I D particle and its interactions}",
    eprint = "hep-th/9809111",
    archivePrefix = "arXiv",
    reportNumber = "MRI-PHY-P980960",
    doi = "10.1088/1126-6708/1998/10/021",
    journal = "JHEP",
    volume = "10",
    pages = "021",
    year = "1998"
}

@article{Sen:1998tt,
    author = "Sen, Ashoke",
    title = "{SO(32) spinors of type I and other solitons on brane - anti-brane pair}",
    eprint = "hep-th/9808141",
    archivePrefix = "arXiv",
    reportNumber = "MRI-PHY-P980858",
    doi = "10.1088/1126-6708/1998/09/023",
    journal = "JHEP",
    volume = "09",
    pages = "023",
    year = "1998"
}

@article{Polchinski:1995df,
    author = "Polchinski, Joseph and Witten, Edward",
    title = "{Evidence for heterotic - type I string duality}",
    eprint = "hep-th/9510169",
    archivePrefix = "arXiv",
    reportNumber = "IASSNS-HEP-95-81, NSF-ITP-95-135",
    doi = "10.1016/0550-3213(95)00614-1",
    journal = "Nucl. Phys. B",
    volume = "460",
    pages = "525--540",
    year = "1996"
}

@article{Fraiman:2023cpa,
    author = "Fraiman, Bernardo and Gra\~na, Mariana and Parra De Freitas, H\'ector and Sethi, Savdeep",
    title = "{Non-supersymmetric heterotic strings on a circle}",
    eprint = "2307.13745",
    archivePrefix = "arXiv",
    primaryClass = "hep-th",
    doi = "10.1007/JHEP12(2024)082",
    journal = "JHEP",
    volume = "12",
    pages = "082",
    year = "2024"
}

@article{Sugimoto:1999tx,
    author = "Sugimoto, Shigeki",
    title = "{Anomaly cancellations in type I D-9 - anti-D-9 system and the USp(32) string theory}",
    eprint = "hep-th/9905159",
    archivePrefix = "arXiv",
    reportNumber = "YITP-99-25",
    doi = "10.1143/PTP.102.685",
    journal = "Prog. Theor. Phys.",
    volume = "102",
    pages = "685--699",
    year = "1999"
}

@inproceedings{Sagnotti:1995ga,
	title        = {{Some properties of open string theories}},
	author       = {Sagnotti, Augusto},
	year         = 1995,
	booktitle    = {{Supersymmetry and unification of fundamental interactions. Proceedings, International Workshop, SUSY 95, Palaiseau, France, May 15-19}},
	pages        = {473--484},
	eprint       = {hep-th/9509080},
	archiveprefix = {arXiv},
	primaryclass = {hep-th},
	reportnumber = {ROM2F-95-18}
}

@article{Sagnotti:1996qj,
	title        = {{Surprises in open string perturbation theory}},
	author       = {Sagnotti, Augusto},
	year         = 1997,
	journal      = {Nucl. Phys. Proc. Suppl.},
	booktitle    = {{Theory of elementary particles. Proceedings, 30th International Symposium Ahrenshoop, Buckow, Germany, August 27-31, 1996}},
	volume       = {56B},
	pages        = {332--343},
	doi          = {10.1016/S0920-5632(97)00344-7},
	eprint       = {hep-th/9702093},
	archiveprefix = {arXiv},
	primaryclass = {hep-th},
	reportnumber = {ROM2F-97-4}
}

@article{Angelantonj:2002ct,
    author = "Angelantonj, Carlo and Sagnotti, Augusto",
    title = "{Open strings}",
    eprint = "hep-th/0204089",
    archivePrefix = "arXiv",
    reportNumber = "CERN-TH-2002-025, ROM2F-2002-08, LPTENS-02-14, CPHT-RR-020-0202, CPHT-RR-020.0202",
    doi = "10.1016/S0370-1573(02)00273-9",
    journal = "Phys. Rept.",
    volume = "371",
    pages = "1--150",
    year = "2002",
    note = "[Erratum: Phys.Rept. 376, 407 (2003)]"
}

@article{Dudas:2001wd,
    author = "Dudas, E. and Mourad, J. and Sagnotti, A.",
    title = "{Charged and uncharged D-branes in various string theories}",
    eprint = "hep-th/0107081",
    archivePrefix = "arXiv",
    reportNumber = "LPT-ORSAY-01-56, ROM2F-01-18",
    doi = "10.1016/S0550-3213(01)00552-1",
    journal = "Nucl. Phys. B",
    volume = "620",
    pages = "109--151",
    year = "2002"
}

@article{Kim:2019vuc,
    author = "Kim, Hee-Cheol and Shiu, Gary and Vafa, Cumrun",
    title = "{Branes and the Swampland}",
    eprint = "1905.08261",
    archivePrefix = "arXiv",
    primaryClass = "hep-th",
    doi = "10.1103/PhysRevD.100.066006",
    journal = "Phys. Rev. D",
    volume = "100",
    number = "6",
    pages = "066006",
    year = "2019"
}

@article{Cvetic:2020kuw,
    author = "Cveti\v{c}, Mirjam and Dierigl, Markus and Lin, Ling and Zhang, Hao Y.",
    title = "{String Universality and Non-Simply-Connected Gauge Groups in 8d}",
    eprint = "2008.10605",
    archivePrefix = "arXiv",
    primaryClass = "hep-th",
    reportNumber = "CERN-TH-2020-138, UPR-1306-T",
    doi = "10.1103/PhysRevLett.125.211602",
    journal = "Phys. Rev. Lett.",
    volume = "125",
    number = "21",
    pages = "211602",
    year = "2020"
}

@article{Angelantonj:2007ts,
    author = "Angelantonj, Carlo and Dudas, Emilian",
    title = "{Metastable string vacua}",
    eprint = "0704.2553",
    archivePrefix = "arXiv",
    primaryClass = "hep-th",
    reportNumber = "CPHT-RR-017.0417, DFTT-2007-5, LPT-ORSAY-07-23",
    doi = "10.1016/j.physletb.2007.06.031",
    journal = "Phys. Lett. B",
    volume = "651",
    pages = "239--245",
    year = "2007"
}

@article{Larotonda:2024thv,
    author = "Larotonda, Vittorio and Lin, Ling",
    title = "{Anomaly inflow and gauge group topology in the 10d Sugimoto string theory}",
    eprint = "2412.17894",
    archivePrefix = "arXiv",
    primaryClass = "hep-th",
    doi = "10.1007/JHEP06(2025)136",
    journal = "JHEP",
    volume = "06",
    pages = "136",
    year = "2025"
}

@article{Fraiman:2025yrx,
    author = "Fraiman, Bernardo and Parra de Freitas, H{\'e}ctor",
    title = "{Symmetries and dualities in non-supersymmetric CHL strings}",
    eprint = "2511.01674",
    archivePrefix = "arXiv",
    primaryClass = "hep-th",
    reportNumber = "MPP-2025-79; IFT-UAM/CSIC-25-111",
    month = "11",
    year = "2025"
}

@article{Bergman:1997rf,
    author = "Bergman, Oren and Gaberdiel, Matthias R.",
    title = "{A Nonsupersymmetric open string theory and S duality}",
    eprint = "hep-th/9701137",
    archivePrefix = "arXiv",
    reportNumber = "HUTP-97-A003, BRX-TH-402",
    doi = "10.1016/S0550-3213(97)00309-X",
    journal = "Nucl. Phys. B",
    volume = "499",
    pages = "183--204",
    year = "1997"
}

@article{Blumenhagen:1999ad,
    author = "Blumenhagen, Ralph and Kumar, Alok",
    title = "{A Note on orientifolds and dualities of type 0B string theory}",
    eprint = "hep-th/9906234",
    archivePrefix = "arXiv",
    reportNumber = "HUB-EP-99-29, CERN-TH-99-190",
    doi = "10.1016/S0370-2693(99)01002-3",
    journal = "Phys. Lett. B",
    volume = "464",
    pages = "46--52",
    year = "1999"
}

@article{Bianchi:1991eu,
    author = "Bianchi, M. and Pradisi, G. and Sagnotti, A.",
    title = "{Toroidal compactification and symmetry breaking in open string theories}",
    reportNumber = "ROM2F-91-15",
    doi = "10.1016/0550-3213(92)90129-Y",
    journal = "Nucl. Phys. B",
    volume = "376",
    pages = "365--386",
    year = "1992"
}

@article{Montonen:1977sn,
    author = "Montonen, C. and Olive, David I.",
    title = "{Magnetic Monopoles as Gauge Particles?}",
    reportNumber = "CERN-TH-2391",
    doi = "10.1016/0370-2693(77)90076-4",
    journal = "Phys. Lett. B",
    volume = "72",
    pages = "117--120",
    year = "1977"
}

@article{Osborn:1979tq,
    author = "Osborn, Hugh",
    title = "{Topological Charges for N=4 Supersymmetric Gauge Theories and Monopoles of Spin 1}",
    reportNumber = "DAMTP 79/4",
    doi = "10.1016/0370-2693(79)91118-3",
    journal = "Phys. Lett. B",
    volume = "83",
    pages = "321--326",
    year = "1979"
}

@article{Goddard:1976qe,
    author = "Goddard, P. and Nuyts, J. and Olive, David I.",
    title = "{Gauge Theories and Magnetic Charge}",
    reportNumber = "CERN-TH-2255",
    doi = "10.1016/0550-3213(77)90221-8",
    journal = "Nucl. Phys. B",
    volume = "125",
    pages = "1--28",
    year = "1977"
}

@book{Angelantonj:2024tns,
    author = "Angelantonj, Carlo and Florakis, Ioannis",
    title = "{A Lightning Introduction to String Theory}",
    eprint = "2406.09508",
    archivePrefix = "arXiv",
    primaryClass = "hep-th",
    doi = "10.1007/978-981-19-3079-9_53-1",
    year = "2024",
    publisher = "Handbook of Quantum Gravity"
}

@article{Leone:2025mwo,
    author = "Leone, Giorgio and Raucci, Salvatore",
    title = "{Aspects of strings without spacetime supersymmetry}",
    eprint = "2509.24703",
    archivePrefix = "arXiv",
    primaryClass = "hep-th",
    reportNumber = "IFT-UAM/CSIC-25-100",
    doi = "10.1007/s40766-025-00078-z",
    journal = "Riv. Nuovo Cim.",
    volume = "49",
    number = "3",
    pages = "75--136",
    year = "2026"
}

@article{Dudas:2025ubq,
    author = "Dudas, E. and Mourad, J. and Sagnotti, A.",
    title = "{Supersymmetry breaking with fields, strings and branes}",
    eprint = "2511.04367",
    archivePrefix = "arXiv",
    primaryClass = "hep-th",
    doi = "10.1016/j.physrep.2026.02.005",
    journal = "Phys. Rept.",
    volume = "1175",
    pages = "1--256",
    year = "2026"
}

@article{Bergman:1999km,
    author = "Bergman, Oren and Gaberdiel, Matthias R.",
    title = "{Dualities of type 0 strings}",
    eprint = "hep-th/9906055",
    archivePrefix = "arXiv",
    reportNumber = "CALT-68-2228, DAMTP-1999-74",
    doi = "10.1088/1126-6708/1999/07/022",
    journal = "JHEP",
    volume = "07",
    pages = "022",
    year = "1999"
}

@article{Faraggi:2007tj,
    author = "Faraggi, Alon E. and Tsulaia, Mirian",
    title = "{On the Low Energy Spectra of the Nonsupersymmetric Heterotic String Theories}",
    eprint = "0706.1649",
    archivePrefix = "arXiv",
    primaryClass = "hep-th",
    reportNumber = "LTH-746",
    doi = "10.1140/epjc/s10052-008-0545-2",
    journal = "Eur. Phys. J. C",
    volume = "54",
    pages = "495--500",
    year = "2008"
}

@article{Acharya:2022shu,
    author = "Acharya, Bobby Samir and Aldazabal, Gerardo and Font, Anamar{\'\i}a and Narain, Kumar and Zadeh, Ida G.",
    title = "{Heterotic strings on $ \mathbb{T}^{3}$/${\mathbb{Z}}_{2}$, Nikulin involutions and M-theory}",
    eprint = "2205.09764",
    archivePrefix = "arXiv",
    primaryClass = "hep-th",
    doi = "10.1007/JHEP09(2022)209",
    journal = "JHEP",
    volume = "09",
    pages = "209",
    year = "2022"
}

@article{Baykara:2026gem,
    author = "Baykara, Zihni Kaan and Dudas, Emilian and Vafa, Cumrun",
    title = "{M-theory on $S^1\vee S^1$ as Type 0A}",
    eprint = "2603.13468",
    archivePrefix = "arXiv",
    primaryClass = "hep-th",
    month = "3",
    year = "2026"
}

@article{Altavista:2026evd,
    author = "Altavista, Chiara and Anastasi, Edoardo and Raucci, Salvatore and Uranga, Angel M. and Wang, Chuying",
    title = "{Ho{\v{r}}ava-Witten theory on ${\mathbf{S}}^1\vee{\mathbf{S}}^1$ as type 0 orientifold}",
    eprint = "2603.25786",
    archivePrefix = "arXiv",
    primaryClass = "hep-th",
    reportNumber = "IFT-UAM/CSIC-26-39",
    month = "3",
    year = "2026"
}

@article{Baykara:2026vdc,
    author = "Baykara, Zihni Kaan and Delgado, Matilda and Dudas, Emilian and De Freitas, Hector Parra and Vafa, Cumrun",
    title = "{A Duality Web for Non-Supersymmetric Strings}",
    eprint = "2604.07433",
    archivePrefix = "arXiv",
    primaryClass = "hep-th",
    month = "4",
    year = "2026"
}

@article{Altavista:2026brr,
    author = "Altavista, Chiara and Raucci, Salvatore and Uranga, Angel M. and Wang, Chuying",
    title = "{Heterotic Ouroboros}",
    eprint = "2604.22915",
    archivePrefix = "arXiv",
    primaryClass = "hep-th",
    reportNumber = "IFT-UAM/CSIC-26-53",
    month = "4",
    year = "2026"
}

@article{Dasgupta:2026maq,
    author = "Dasgupta, Keshav and Tatar, Radu",
    title = "{Towards Wedge Construction of Four-Dimensional Non-Supersymmetric Theories and Torsion Classes}",
    eprint = "2605.05333",
    archivePrefix = "arXiv",
    primaryClass = "hep-th",
    month = "5",
    year = "2026"
}

@article{Basile:2026trt,
    author = {Basile, Ivano and L{\"u}st, Dieter},
    title = "{String dualities and wedge singularities}",
    eprint = "2606.05287",
    archivePrefix = "arXiv",
    primaryClass = "hep-th",
    reportNumber = "MPP-2026-102",
    month = "6",
    year = "2026"
}

@article{Kamal:2026msr,
    author = "Kamal, Ahmed Rakin",
    title = "{A Circle That Won't Return: The Fate of RR Fluxes and D-branes in Type 0A Tachyon Condensation}",
    eprint = "2606.26280",
    archivePrefix = "arXiv",
    primaryClass = "hep-th",
    month = "6",
    year = "2026"
}

@inproceedings{Sagnotti:1987tw,
    author = "Sagnotti, Augusto",
    title = "{Open Strings and their Symmetry Groups}",
    booktitle = "{NATO Advanced Summer Institute on Nonperturbative Quantum Field Theory (Cargese Summer Institute)}",
    eprint = "hep-th/0208020",
    archivePrefix = "arXiv",
    reportNumber = "ROM2F-87-25",
    month = "9",
    year = "1987"
}

@article{Pradisi:1988xd,
    author = "Pradisi, Gianfranco and Sagnotti, Augusto",
    title = "{Open String Orbifolds}",
    reportNumber = "ROM2F-88-16",
    doi = "10.1016/0370-2693(89)91369-5",
    journal = "Phys. Lett. B",
    volume = "216",
    pages = "59--67",
    year = "1989"
}

@article{Horava:1989vt,
    author = "Horava, Petr",
    title = "{Strings on World Sheet Orbifolds}",
    reportNumber = "PRA-HEP-89/1",
    doi = "10.1016/0550-3213(89)90279-4",
    journal = "Nucl. Phys. B",
    volume = "327",
    pages = "461--484",
    year = "1989"
}

@article{Bianchi:1990yu,
    author = "Bianchi, Massimo and Sagnotti, Augusto",
    title = "{On the systematics of open string theories}",
    reportNumber = "ROM2F-90-20",
    doi = "10.1016/0370-2693(90)91894-H",
    journal = "Phys. Lett. B",
    volume = "247",
    pages = "517--524",
    year = "1990"
}

@article{Bianchi:1990tb,
    author = "Bianchi, Massimo and Sagnotti, Augusto",
    title = "{Twist symmetry and open string Wilson lines}",
    reportNumber = "ROM2F-90-28",
    doi = "10.1016/0550-3213(91)90271-X",
    journal = "Nucl. Phys. B",
    volume = "361",
    pages = "519--538",
    year = "1991"
}

@article{Fioravanti:1993hf,
    author = "Fioravanti, D. and Pradisi, G. and Sagnotti, A.",
    title = "{Sewing constraints and nonorientable open strings}",
    eprint = "hep-th/9311183",
    archivePrefix = "arXiv",
    reportNumber = "ROM2F-93-33",
    doi = "10.1016/0370-2693(94)90255-0",
    journal = "Phys. Lett. B",
    volume = "321",
    pages = "349--354",
    year = "1994"
}

@article{Pradisi:1995pp,
    author = "Pradisi, G. and Sagnotti, A. and Stanev, Ya. S.",
    title = "{The Open descendants of nondiagonal SU(2) WZW models}",
    eprint = "hep-th/9506014",
    archivePrefix = "arXiv",
    reportNumber = "ROM2F-95-09, CPTH-RR-358-0695",
    doi = "10.1016/0370-2693(95)00840-H",
    journal = "Phys. Lett. B",
    volume = "356",
    pages = "230--238",
    year = "1995"
}

@article{Ginsparg:1986wr,
    author = "Ginsparg, Paul H. and Vafa, C.",
    title = "{Toroidal Compactification of Nonsupersymmetric Heterotic Strings}",
    reportNumber = "HUTP-86-A064A, HUTP-86-A064",
    doi = "10.1016/0550-3213(87)90387-7",
    journal = "Nucl. Phys. B",
    volume = "289",
    pages = "414",
    year = "1987"
}

@article{Kaidi:2019tyf,
    author = "Kaidi, Justin and Parra-Martinez, Julio and Tachikawa, Yuji",
    title = "{Topological Superconductors on Superstring Worldsheets}",
    eprint = "1911.11780",
    archivePrefix = "arXiv",
    primaryClass = "hep-th",
    reportNumber = "IPMU-19-0164, UCLA/TEP/2019/106",
    doi = "10.21468/SciPostPhys.9.1.010",
    journal = "SciPost Phys.",
    volume = "9",
    pages = "10",
    year = "2020"
}

@article{Michishita:1999it,
    author = "Michishita, Yoji",
    title = "{D0-branes in SO(32) x SO(32) open type 0 string theory}",
    eprint = "hep-th/9907094",
    archivePrefix = "arXiv",
    reportNumber = "KUNS-1586",
    doi = "10.1016/S0370-2693(99)01123-5",
    journal = "Phys. Lett. B",
    volume = "466",
    pages = "161",
    year = "1999"
}

@article{Mourad:2017rrl,
    author = "Mourad, J. and Sagnotti, A.",
    title = "{An Update on Brane Supersymmetry Breaking}",
    eprint = "1711.11494",
    archivePrefix = "arXiv",
    primaryClass = "hep-th",
    month = "11",
    year = "2017"
}

@article{Dabholkar:1995ep,
    author = "Dabholkar, Atish",
    title = "{Ten-dimensional heterotic string as a soliton}",
    eprint = "hep-th/9506160",
    archivePrefix = "arXiv",
    reportNumber = "CALT-68-2002",
    doi = "10.1016/0370-2693(95)00949-L",
    journal = "Phys. Lett. B",
    volume = "357",
    pages = "307--312",
    year = "1995"
}

@article{Hull:1995nu,
    author = "Hull, C. M.",
    title = "{String-string duality in ten-dimensions}",
    eprint = "hep-th/9506194",
    archivePrefix = "arXiv",
    reportNumber = "QMW-TH-95-25, QMW-95-25",
    doi = "10.1016/0370-2693(95)01000-G",
    journal = "Phys. Lett. B",
    volume = "357",
    pages = "545--551",
    year = "1995"
}

\end{document}